\documentclass[useAMS,usenatbib]{mn2e}
\usepackage{amsmath}
\usepackage{amssymb}
\usepackage[dvips]{graphicx}
\usepackage{mathrsfs}

\begin{document}

\title[Maximal spin in a symbiotic system of BH, disk and jet]{Maximal spin
and energy conversion efficiency \\
in a symbiotic system of black hole, disk and jet}
\author[Z. Kov\'{a}cs, L. \'{A}. Gergely and P. L. Biermann ]{Zolt\'{a}n Kov%
\'{a}cs$^{1\star }$, L\'{a}szl\'{o} \'{A}. Gergely$^{2,3\ddag }$, Peter L.
Biermann$^{4,5,6,7,8\dag }$ \\
$^{1}~$ Department of Physics and Center for Theoretical and Computational
Physics, The University of Hong\\
Kong, Pok Fu Lam Road, Hong Kong\\
$^{2}$ ~Department of Theoretical Physics, University of Szeged, Tisza Lajos
krt. 84-86, Szeged 6720, Hungary\\
$^{3}~$ Department of Experimental Physics, University of Szeged, 6720
Szeged, D\'{o}m t\'{e}r 9, Hungary\\
$^{4}~$ Max-Planck-Institut f{\"{u}}r Radioastronomie, Auf dem H{\"{u}}gel
69, Bonn, Germany\\
$^{5}~$ FZ Karlsruhe, and Department of Physics, University of Karlsruhe,
Germany\\
$^{6}~$ Department of Physics and Astronomy, University of Alabama,
Tuscaloosa, 35487 AL, USA \\
$^{7}~$ Department of Physics, University of Alabama at Huntsville, 35899 AL, USA\\
$^{8}~$ Department of Physics and Astronomy, University of Bonn, Germany\\
\\
{\small {$^{\star }$E-mail:zkovacs@hku.hk\qquad $^{\ddag }$%
E-mail:gergely@physx.u-szeged.hu\qquad $^{\dag }$%
E-mail:plbiermann@mpifr-bonn.mpg.de\qquad\ } }}

\date{\today }

\maketitle

\begin{abstract}
We study a combined model of black hole - accretion disk - magnetosphere -
jet symbiosis, applicable for supermassive black holes. We quantify the mass
and spin evolution and we analyze how the limiting value of the spin
parameter and the conversion efficiency of accreted mass into radiation
depend on the interplay of electromagnetic radiation reaction, magnetosphere
characteristics and truncation radius of radiation. The dominant effect
comes from the closed magnetic field line region, which reduces the spin
limit to values $\sim$0.89 (instead $\sim$0.99 in its absence). Therefore
observations on black hole spins could favour or disfavour the existence of
the closed magnetic field line region (or its coupling to the disk). We also
find that the suppression of radiation from the innermost part of the
accretion disk, inferred from observations, and a collimated jet both
increase the spin limit and the energy conversion efficiency.
\end{abstract}

\section{INTRODUCTION}

Ever since it was shown, that almost all galaxies have a central
super-massive black hole (Kormendy \& Richstone 1995, Faber et al. 1997), it
has become of renewed interest to understand the life-cycle of black holes.
Is it driven by accretion from an accretion disk (Shakura 1972, Shakura \&
Sunyaev 1973, Novikov \& Thorne 1973), mergers with other black holes
(Hughes \& Blandford 2003), an alternating sequence of the two (Berti \&
Volonteri 2008), eating stars whole, spin-down driving a relativistic jet,
or some other not yet understood process? Early statistics have already showed
that existing black holes are active in the sense to have an accretion disk
shine prominently only a small fraction of the time, of order one percent or
so.

On the other hand, digging deep with observations in the radio regime shows
that even very quiet black holes are all active in the sense of showing
compact radio emission. Very long baseline (usually intercontinental) radio
interferometry (VLBI) shows that in all cases, when the data are
sufficiently deep, the activity is driving a relativistic jet (Chini et al.
1989\textbf{,} Nagar et al. 2000, 2001, 2002, 2005, Falcke et al. 2000).
This emission is strong, when the jet happens to point in our direction, and
so is strongly enhanced by re\allowbreak la\allowbreak tivistic boosting.
And yet, the power is far sub-Eddington in many such cases. Therefore it is
hard to accept that accretion could provide the power. For jets pointing
elsewhere these cases populate the low end of the radio luminosity function
(Windhorst et al. 1984, 1985, 1990, Perez-Fournon \& Biermann 1984). In such
cases spin-down almost certainly powers the activity (Blandford 1976).

There are several proposals on the ways of interpreting the compact radio emission,
some with emission from the disk itself, and some with emission from a
relativistic jet. Since we can directly observe a relativistic jet in some
cases, we follow Occam's razor, and adopt the point of view, that all such
cases have the same mechanism. Therefore we interpret the compact radio
emission as arising from a relativistic jet in all cases (Falcke, Malkan \& Biermann 1995; Falcke \& Biermann 1995; Falcke 1996; Falcke \& Biermann 1996; Falcke \& Biermann 1999; Markoff, Falcke \& Fender 2001; Yuan, Markoff \& Falcke 2002a; Yuan et al. 2002b).

It had been noted early on, that the low radio frequency spectrum suggested
a low energy cutoff in the electron energy distribution, such as would
naturally occur from the decay of charged pions, created in hadronic
collisions (Falcke et al. 1995a, 1995b, 1996,\textbf{\ }Gopal-Krishna et al.
2004). The site to do this would be the inner accretion disks in a low
accretion mode, when the disk turns into a very hot Advection Dominated
Accretion Flow (ADAF). For a dimensionless spin-parameter $a_{\ast }>0.95$
the temperature reaches to relativistic levels (Mahadevan 1998), and the
thermal collisions become hadronic, giving rise to pion production. As a
consequence such radio spectra suggest that the spin is high. However, in
the case, that the jets start purely electrodynamically, as Poynting flux
jets, the charged pion production would be in the shocks around stars or
compact clouds entering the jets sideways (e.g., Gamma rays from cloud
penetration at the base of AGN jets, Araudo et al. (2010)).

As almost all otherwise quiescent super-massive black holes are active in
such a sense, it implies that they maintain their spin at a high level,
compatible with the suggestion of Blandford (1976), that the decay time of
the black hole magnetic fields is very long; this is feasible as long as the
power output used to drive the jet is a small fraction of the energy
reservoir, the rotation of the black hole. The power output estimated by
Whysong \& Antonucci (2003) for the radio galaxies M87 (= NGC4486 = Vir A)
and NGC5128 (=Cen A) suggests that indeed the power output by the jets is of
order $10^{-2.5}$ of the limiting Eddington luminosity, or even less,
allowing a long life time. This limit implies that no matter is supplied
from the disk to the jet at all, which means that the jet starts as a
Poynting flux dominated jet (Sikora et al. 2005). Armitage \& Natarajan
(1999) confirmed that a fully rotating black hole can power a Poynting flux
dominated jet for the very long lifetime of AGNs.

Thus it is of tantamount importance to understand, which process allows the
super-massive black holes to attain such a high spin state.

Due to accretion, black holes spin up to the maximal spin limit, even in the
absence of initial rotation (Bardeen 1970). For the maximally spinning
(extreme) Kerr black holes the efficiency of accreted rest mass conversion
into outgoing electromagnetic radiation is $42.3\%$. A refined accretion
model, assuming a geometrically thin, but optically thick steady state
accretion disk (composed of rotating plasma with high opacity) was
introduced by Shakura \& Sunyaev (1973) and Novikov \& Thorne (1973). This
model takes into account the additional effect of the photons emitted from
the disk and captured by the black hole, modifying both the energy and
angular momentum transport from the disk into the black hole (Page \& Thorne
1974). Therefore the spin evolution and mass growth of black holes is
slightly changed as compared to Bardeen's model. The spin limit is reduced
to $0.9982$ (compared to $1$ in the Bardeen approach) for an electron
scattering atmosphere; also the conversion efficiency diminishes to $30\%$
(Thorne 1974). The black holes with radiating accretion disks leading to
photon capture are referred to as \textit{canonical black holes}, while the
accretion disk is discussed in the framework of a \textit{hydrodynamic}
model with \textit{steady-state accretion}.

In \textit{magnetohydrodynamic} (MHD) models, magnetic stresses in the disk
modify the dynamics of the accretion disk. The study of MHD accretion flows
for various magnetosphere models lead to a modified energy and angular
momentum transport (Camenzind 1986, 1987, Takahashi et al. 1990, Nitta,
Takahashi \& Tomimatsu 1991, Hirotani et al. 1992). Blandford \& Znajek
(1977) have found a mechanism (BZ) of extracting both energy and angular
momentum from the rotating black hole via open poloidal magnetic field lines
emanating from its event horizon (such that the total magnetic flux over the
horizon vanishes). In another configuration of closed poloidal magnetic
field lines, also emanating from the event horizon, but anchored in the
disk, the energy and angular momentum are exchanged between the rapidly
rotating black hole and the slowly rotating disk (Li 2000, Wang, Xiao \& Lei
2002).

The BZ mechanism is nowadays regarded as a possible source of powering jets
in AGN's and gamma-rray bursts (GRB's), where the angular momentum is carried away by outflowing
jets. Jets existing in symbiosis with accretion disks and black holes
represent an efficient mechanism for extracting energy and angular momentum
from the disk and black hole. Based on Blandford \& K\"{o}nigl (1979)
arguments, Falcke \& Biermann (1996) developed a simple model where the
plasma jet emanating from the inner region of the disk forms a \textit{%
symbiotic} system with the disk and the black hole. This model, in
conjunction with the BZ mechanism describes the rest mass, angular momentum
and energy transport between the black hole and its environment in the
presence of magnetic field and jets. Focusing on the role of jets (by
switching off the magnetic field) this symbiotic model was discussed in
detail by Donea \& Biermann (1996). An alternative model for launching
relativistic jets in active galactic nuclei (AGN) from an accreting Kerr
black hole by converting the accretion disc energy into jet energy, was
recently advanced by Du\c{t}an (2010).

\textit{Composite} models of accretion disks involve a radial transition
between a Shakura-Sunyaev disk on the outside to an ADAF on the inside. In
an ADAF the heat generated via viscosity is advected inward rather than
radiated away. Some composite models apply a geometrically thick and
optically thin hot corona (Mannheim et al. 1995) positioned between the
marginally stable orbit and the inner edge of the geometrically thin and
optically thick accretion disk, lying at a few gravitational radii (Thorne
\& Price 1975, Shapiro, Lightman \& Eardley 1976). In composite models the
soft and hard components of the broad band X-ray spectra of galactic black
holes were attributed to the thermal radiation of the accretion disk and the
emission mechanism in the corona, respectively. In another type of composite
models the corona lies above and under the accretion disk and the soft
photons arriving from the disk produce a hard emission via their inverse
comptonization by the thermally hot electrons in the corona (Liang \& Price
1977). In both configurations the soft photon flux of the disk and, in turn,
the photon capture effect on the mass-energy growth of the black hole are
reduced. The spin evolution and its limiting value are therefore closer to
the predictions of Bardeen than for canonical black holes.

We picture the evolution of an activity event of a black hole as follows.
Two galaxies interact and feed low angular momentum material into the
centers (Toomre \& Toomre 1972). In case both have a central black hole,
these black holes merge, and typically produce a merged black hole with
relatively high spin, of order $0.7$ (Berti \& Volonteri 2008). Then
accretion near maximum levels pushes up the spin to near maximum (Bardeen
1970), and establishes a maximal magnetic field, and a jet, initially weak.
When accretion slows down, the jet becomes strong, and the emission from the
inner disk dies away. In the following long phase the black hole powers the
jet from its rotational energy alone. In the present work we study \textit{%
the evolution of the spin and mass} of the black hole coexisting in \textit{%
a symbiotic system with a radiatively truncated accretion disk, a pair of
jets, and a magnetosphere} containing both open and closed field line
regions. As no matter is supplied from the disk to the Poynting flux
dominated jet, its effect appears only in the suppression of the radiation
from the accretion disk in its innermost region. For simplicity, we do not
include possible contributions of hard photons radiated away by a hot
corona, neither the ADAF mechanism. An ionization instability may also
influence the lifetime cycle and overall complexity in the supermassive
black hole environment (Janiuk \& Czerny 2011). Despite the simplifications,
we present the most complete symbiotic model up to date, and we employ it
for the study of spin and mass evolution.

In Section 2 we briefly review the elements of the symbiotic model. In
Section 3 we discuss the structure equations of the disk, as derived from
conservation laws and give a detailed presentation on the effects of the
magnetic fields with coexisting open and closed field line topologies. We
also discuss the effects on the radial profiles of the flux emitted by the
disk and write up a generic flux formula. Section 4 presents study of the
evolution of the black hole spin and mass and the modifications induced in
the energy conversion efficiency by magnetic fields and jets. Section 5
contains the concluding remarks.

\textit{Notation.} 4-vectors and tensors are denoted by bold face italic
letters. The only exceptions are the Killing vectors $\partial /\partial t$
and $\partial /\partial \phi $. The scalar product of 4-vectors and the
contraction of 4-vectors with symmetric tensors are denoted by a dot, i.e., $%
\boldsymbol{q}\cdot \boldsymbol{u}=q^{\mu }u_{\mu }$ and $\boldsymbol{T}%
\cdot \partial /\partial \phi =T^{\mu \nu }(\partial /\partial \phi )_{\mu }$%
, where $T^{\mu \nu }=T^{\nu \mu }$. We use geometrical units, with $G=1$
and $c=1$.

\section{ELEMENTS OF THE SYMBIOTIC MODEL}

In subsection 2.1 we review the dynamics of test particles in the rotating
black hole geometry. In subsection 2.2 we summarize the relevant results on
accretion disk physics from Shakura \& Sunyaev (1973), Novikov \& Thorne
(1973), Page \& Thorne (1974), for later
generalization. The presented assumptions regarding geometry and matter
stand at the basis of the steady-state, thin accretion disk model. Then in
subsection 2.3 we discuss the magnetic field and in subsection 2.4 the
characteristics of the jets.

\subsection{The geometry}

The line element in a stationary, axisymmetric configuration is 
\begin{equation}
ds^{2}=g_{tt}dt^{2}+2g_{t\phi }{d}t{d}\phi +g_{\phi \phi }d\phi
^{2}+g_{11}d(x^{1})^{2}+g_{22}d(x^{2})^{2}~,  \label{ds2gen}
\end{equation}%
with $g_{\mu \nu }\equiv \boldsymbol{g}(\partial /\partial x^{\mu },\partial
/\partial x^{\nu })=(\partial /\partial x^{\mu })\cdot (\partial /\partial
x^{\nu })$ the components of the metric $\boldsymbol{g}$. The coordinates $%
x^{\mu }=(t,x^{1},x^{2},\phi )$ are adapted to the time-like and rotational
Killing vector fields $\partial /\partial t$ and $\partial /\partial \phi $.
In addition we assume asymptotic flatness and reflection symmetry about the
equatorial plane.

Geodesic circular orbits in the equatorial plane are characterized by their
angular velocity $\Omega \equiv d\phi /dt$, specific (normalized to mass)
energy $\widetilde{E}$ and specific angular momentum $\widetilde{L}$,
measured at infinity (Shibata \& Sasaki 1998): 
\begin{eqnarray}
\Omega &=&\frac{-g_{t\phi ,r}+\sqrt{(g_{t\phi ,r})^{2}-g_{tt,r}g_{\phi \phi
,r}}}{g_{\phi \phi ,r}}\;,  \label{Omegagen} \\
\widetilde{E} &=&-\boldsymbol{u}\cdot \partial /\partial t=-(g_{tt}+\Omega
g_{t\phi })/|\boldsymbol{k}|\;,  \label{tildeEgen} \\
\widetilde{L} &=&\boldsymbol{u}\cdot \partial /\partial \phi =(g_{t\phi
}+\Omega g_{\phi \phi })/|\boldsymbol{k}|\;.  \label{tildeLgen}
\end{eqnarray}%
Here the 4-velocity $\boldsymbol{u}=\boldsymbol{k}/|\boldsymbol{k}|$ of a
stationary observer is determined by 
\begin{equation}
\boldsymbol{k}=\partial /\partial t+\Omega \partial /\partial \phi ~,
\label{k}
\end{equation}%
with norm 
\begin{equation}
|\boldsymbol{k}|=\sqrt{-g_{tt}-\Omega g_{t\phi }-\Omega ^{2}g_{\phi \phi }}%
\;.
\end{equation}%
As $\Omega $ does not depend on either $t$ or $\phi $, it is a constant on
the transitivity surfaces of the two Killing vectors and therefore $%
\boldsymbol{k}$ is also a Killing vector.

The back-reaction of orbiting particles to the geometry (\ref{ds2gen}) is
neglected, considering effectively a vacuum. The vacuum space-time obeying
axial symmetry, stationarity and asymptotic flatness is the Kerr space-time,
characterized by two constants, its mass $M$ and angular momentum $J$\textbf{%
,} given by%
\begin{equation}
ds^{2}=-\frac{\Sigma \Delta }{A}dt^{2}+\widetilde{\omega }^{2}({d}\phi
-\omega {d}t)^{2}+\frac{\Sigma }{\Delta }dr^{2}+\Sigma d\theta ^{2}\;,
\label{ds2Kerr}
\end{equation}%
with the metric functions 
\begin{eqnarray}
\Sigma &=&r^{2}+a^{2}\cos ^{2}\theta \;, \\
\Delta &=&r^{2}+a^{2}-2Mr\;,  \label{Delta} \\
A &=&(r^{2}+a^{2})^{2}-a^{2}\Delta \sin ^{2}\theta \;, \\
\widetilde{\omega }^{2} &=&A\sin ^{2}\theta /\Sigma \;.  \label{tildeomega}
\end{eqnarray}%
The Kerr black hole has two horizons (the roots of $\Delta =0$) at $r_{\pm
}=M(1\pm \sqrt{1-a_{\ast }^{2}})$, where $a=J/M$ and $a_{\ast }=a/M=J/M^{2}$
are the dimensional and dimensionless spin parameters, respectively.

In the vicinity of the equatorial plane the coordinate $z=r\cos \theta $ is
suitable to replace $\theta $ such that the Kerr metric is approximated as%
\begin{equation}
ds^{2}=-\mathscr D\mathscr A^{-1}dt^{2}+r^{2}\mathscr A({d}\phi -\omega {d}%
t)^{2}+\mathscr D^{-1}dr^{2}+dz^{2}\;\;.  \label{ds2}
\end{equation}%
Eqs. (\ref{Omegagen})-(\ref{tildeLgen}) then become 
\begin{eqnarray}
\Omega &=&M^{-1}x^{-3}\mathscr B^{-1}\;,  \label{Omega} \\
\widetilde{E} &=&\mathscr C^{-1/2}\mathscr G\;,  \label{tildeE} \\
\widetilde{L} &=&Mx\mathscr C^{-1/2}\mathscr F\;,  \label{tildeL}
\end{eqnarray}%
with%
\begin{eqnarray}
\mathscr A &=&1+a_{\ast }^{2}x^{-4}+2a_{\ast }^{2}x^{-6}\;,  \label{calA} \\
\mathscr B &=&\pm (1\pm a_{\ast }x^{-3})\;,  \label{calB} \\
\mathscr C &=&1-3x^{-2}\pm 2a_{\ast }x^{-3}\;,  \label{calC} \\
\mathscr D &=&1-2x^{-2}+a_{\ast }^{2}x^{-4}\;,  \label{calD} \\
\mathscr G &=&1-2x^{-2}\pm a_{\ast }x^{-3}\;,  \label{calG} \\
\mathscr F &=&\pm (1\mp 2a_{\ast }x^{-3}+a_{\ast }^{2}x^{-4})  \label{calF}
\end{eqnarray}%
and 
\begin{equation}
\omega =2M^{-1}a_{\ast }x^{-6}\mathscr A^{-1}~.
\end{equation}%
Here $x=\sqrt{r/M}$ (Novikov \& Thorne, Page \& Thorne 1974). In Eqs. (\ref%
{calA})-(\ref{calF}) the upper (lower) signs refer to corotating
(retrograde) orbits.\footnote{%
In SI units distances are conveniently expressed in units of $GM/c^{2}$ (the
gravitational radius). In our chosen units then distances can be expressed
in units of $M$.}

A particle with zero orbital angular velocity falling into the Kerr black
hole will be subject to the frame dragging effect, and as a result will
acquire an angular velocity $\Omega ^{H}\equiv \Omega \left( r_{+}\right)
=a_{\ast }/2r_{+}$, known as the horizon rotational velocity.\footnote{%
All indices $^{H}$ refer to the outer horizon of the Kerr black hole.}

\begin{figure}
\centering
\includegraphics[width=.48\textwidth]{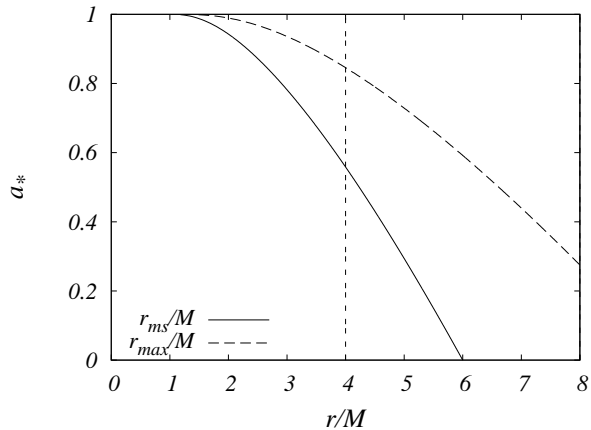}
\caption{The relation between the spin parameter $a_{\ast }$ and the
marginally stable orbit $r_{ms}$ (solid curve), as given by Eq. (\protect\ref%
{aux2}). At $r=4M$ (vertical dashed line) the radiative truncation is
possible only for fairly rapidly rotating black holes with $0.56\leq a_{\ast
}\leq 1$. For $r>6M$ the radiative truncation is possible at any $a_{\ast }$%
. The dashed curve shows for each $a_{\ast }\,$the value of the radius,
where the flux $F$ emitted by the accretion disk is maximal, obtained by
extremizing Eq. (\protect\ref{F}) with respect to $r/M$.}
\label{r_ms}
\end{figure}

The innermost marginally stable circular geodesic orbit, with coordinate
radius $r_{ms}$ is given by (Bardeen, Press \& Teukolsky 1972), which
rewritten in the dimensionless variable $x_{ms}=\sqrt{r_{ms}/M}$ reads: 
\begin{equation}
x_{ms}^{4}-6x_{ms}^{2}\pm 8a_{\ast }x_{ms}-3a_{\ast }^{2}=0~.  \label{aux2}
\end{equation}%
Special solutions include the Schwarzschild limit $a_{\ast }=0$, in which we
recover $r_{ms}=6M$; and the extreme Kerr limit $a_{\ast }=1$ for which we
get $r_{ms}=M$ for corotating and $r_{ms}=9M$ for counterrotating particles.
The generic dependence of $r_{ms}$ on $a_{\ast }\,$ is represented on Fig. %
\ref{r_ms}. 

\subsection{The accretion disk}

Particles on circular geodesic orbits in the equatorial plane of the black
hole form an accretion disk of plasma. Novikov \& Thorne (1973) have
introduced the hydrodynamic model of an anisotropic fluid for the disk with
the energy-momentum tensor%
\begin{equation}
\boldsymbol{T}^{HD}=\rho _{0}\boldsymbol{u}\otimes \boldsymbol{u}+%
\boldsymbol{u}\otimes \boldsymbol{q}+\boldsymbol{q}\otimes \boldsymbol{u}+%
\boldsymbol{t}\;.  \label{THD}
\end{equation}%
Here $\boldsymbol{u}$ is the 4-velocity, $\rho _{0}$ the density, $%
\boldsymbol{q}$ the energy flow vector and $\boldsymbol{t}$ the stress
tensor in the averaged rest-frame, obeying $\boldsymbol{u}\cdot \boldsymbol{q%
}=0=\boldsymbol{u}\cdot \boldsymbol{t}$. The matter variables $\rho _{0},~%
\boldsymbol{q}$\ and $\boldsymbol{t}$\ are local, unaveraged quantities. A
generalization of the above energy-momentum tensor would be to add the
specific internal energy $\Pi \boldsymbol{u}\otimes \boldsymbol{u}$, however
Page \& Thorne (1974) justifiably assume that this is negligible compared to
the gravitational potential energy.

Due to the axial and stationary symmetry of the geometry it is practical to
introduce averages over $\phi $ and $t$ for any (tensorial) matter quantity $%
\Psi $ as%
\begin{equation}
\left\langle \Psi (z,r)\right\rangle =\frac{1}{2\pi \Delta t}%
\int_{t}^{t+\Delta t}\int_{0}^{2\pi }\Psi (t,r,z,\phi )dtd\phi ~,
\end{equation}%
where $\Psi $ is Lie dragged along $\partial /\partial t$ and $\partial
/\partial \phi $ and $\Delta t$ is a characteristic time scale.

\textbf{\ }The maximum half thickness $H_{disk}$ of the disk being much
less, than its characteristic radial scale, the disk is considered
geometrically thin. Optically however, the disk is thick, thus it has high
opacity. Its surface energy density is 
\begin{equation}
\Sigma _{disk}\left( r\right) =\int_{-H_{disk}}^{H_{disk}}\left\langle \rho
_{0}\right\rangle dz~,
\end{equation}%
where 
\begin{equation}
u_{disk}^{r}=\Sigma _{disk}^{-1}\int_{-H_{disk}}^{H_{disk}}\left\langle \rho
_{0}u^{r}\right\rangle dz~
\end{equation}%
represents the radial velocity of the local rest frame of the baryons. Then
the radial rest mass flow through the ring is defined as 
\begin{equation}
\dot{M}_{0}=-2\pi r\Sigma _{disk}u_{disk}^{r}  \label{M0dot}
\end{equation}%
(the minus sign arises from the fact that $u_{disk}^{r}$ point outwards,
while the mass flow is inwards).

Following Page \& Thorne 1974 we assume that there is no energy flow in the
disk ($q^{r}=0=q^{\phi }$), therefore any energy flow is entirely in the $z$%
-direction. We also define the average of the energy flow vector $%
\boldsymbol{q}$ as 
\begin{equation}
F=\langle q^{z}(H_{disk})\rangle =-\langle q^{z}(-H_{disk})\rangle ~,
\end{equation}%
(the last equality coming from the reflection symmetry over the equatorial
plane, thus $F$ will have opposed signs if evaluated at the upper or lower
disk boundary) and $W_{\phi }{}^{r}$ as the $r\phi $-component of the
averaged stress tensor $\boldsymbol{t}$, integrated over the thickness of
the disk%
\begin{equation}
W_{\phi }{}^{r}=\int_{-H_{disk}}^{H_{disk}}\left\langle t_{\phi
}^{~r}\right\rangle dz~.
\end{equation}

This hydrodynamic model is useful as long as the effect of the magnetic
fields on the geometry and null geodesics is at most of the order of the
Larmor radius (small compared to the gravitational radius).

\subsection{The magnetic field}

The thin disk models can be generalized by considering external magnetic
fields around the accretion disk - black hole system. The topology of the
black hole magnetosphere consists of open and closed flux lines. Open flux
lines emanate from the event horizon or the disk and extend to infinity
forming a jet, whereas closed ones connect the horizon and the accretion
disk (Macdonald \& Thorne 1982). Open field lines also could emanate from
the outer disk region, outside the maximal extension of the closed line
region (Uzdensky 2005).

The magnetic field geometry in accretion flows in the relativistic case was
studied by many, especially by Hirose et al. (2004) and McKinney (2005),
building on earlier work. Our approximation constitutes a subset of their
configurations, and one of their results is that the configuration
connecting black hole and disk is more important for thin disks and does not
transfer significant amounts of angular momentum. However, since we argue
for a radiative inner edge of the accretion disk to be possibly far outside
the inner-most stable orbit, a direct comparison is difficult.

The open field line topology emanating from the horizon provides the basic
idea for the Blandford-Znajek (BZ) mechanism, which describes how rotational
energy and angular momentum can be extracted electromagnetically from the
spinning hole and transported to large distances in the form of a Poyn\-ting
flux jet (Blandford \& Znajek, Znajek 1977, Hirose et al. 2004, McKinney
2005). Let $\Omega ^{F}$ denote the rotational velocity of the open magnetic
field lines. We assume the maximally efficient BZ mechanism, which implies
that at the event horizon $\Omega ^{F}\left( r_{H}\right) =\Omega ^{H}/2$.

The magnetic field at the horizon is given by the history of accretion to
the black hole (Blandford \& Znajek 1977). The strength of magnetic field at the horizon can be written
in the term of the accretion rate of the rest mass (Moderski \& Sikora 1996)
as%
\begin{equation}
B_{H}^{2}=\frac{2\dot{M}_{0}}{r_{H}^{2}}\;.  \label{B2H}
\end{equation}%
We assume Eq. (\ref{B2H}) to hold in our model. This approximation was also
found in considerations by Krolik (1999) to be generally valid.

\begin{figure}
\centering\includegraphics[width=.4\textwidth]{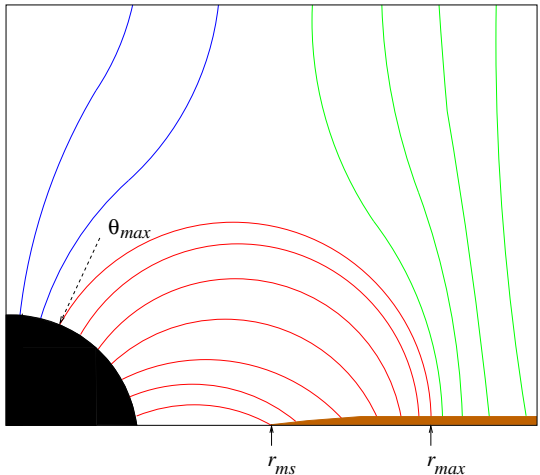}
\caption{The open and closed magnetic field line topology.}
\label{bhm}
\end{figure}

In addition to the parameters $B_{H}$ and $\Omega ^{F}$ of the open magnetic
field, we introduce the opening angle $\theta _{max}$ of the cone separating
open and closed field lines in each hemisphere. The open magnetic field
lines emanate from the horizon for $0<\theta <\theta _{max}$, whereas the
closed ones connect the horizon to the disk between $\theta _{max}$ and the
equatorial plane. The maximal extension of the closed magnetic field on the
disk is at $r_{\max }$, outside which another open field line system begins.
We sketch this configuration on Fig. \ref{bhm}. 

For the closed magnetic field, Wang, Xiao \& Lei (2002) employed the power
law distribution $B_{D}\propto r^{-n}$, $0<n<3$ given by Blandford (1976)
and defined the poloidal component $B_{D}$ on the disk as 
\begin{equation}
\frac{B_{D}}{B_{H}}=\frac{r_{H}}{\widetilde{\omega }(r_{ms})}\left( \frac{r}{%
r_{ms}}\right) ^{-n}~.  \label{BDBH}
\end{equation}%
They assumed that the pressures due to the magnetic field and infalling
matter are in balance at the inner edge of the disk ($r_{ms}$) and the ratio 
$B_{H}/B_{D}$ varies from $1.8$ to $3$ for $0<a_{\ast }<1$.

As the closed field lines connect the event horizon to the accretion disk,
the rotating black hole exerts a magnetic torque on the disk, provided the
flux lines are frozen into the plasma (ideal MHD condition). This torque
transfers angular momentum and energy from the black hole to the disk ($%
\Omega <\Omega ^{H}$ is assumed), modifying the global energy and angular
momentum balance (Li 2000, 2003).

In the MHD simulations of Hirose et al. (2004) these magnetic field
configurations were found to be uncommon, connecting the black hole horizon
to the disk. However, since we include a radiative cutoff of the inner disk,
inspired by observations, our configuration is not readily compared to
theirs. Also, we have the additional free parameter of a maximum latitude $%
\theta _{max}$ of magnetic field lines connecting, which allows the black
hole to be stronger or weaker in its effect on the jet than given by the
current accretion rate as considered in Hirose et al. (2004). Simulations of
such configurations would be very useful, when the black hole has acquired
already its maximal magnetic field, and then the accretion rate dips below
its earlier rate; then the newly accreted field is weaker, and the
separatrix between disk connection and jet configuration could be different,
resulting in a different $\theta _{max}$.

By solving numerically the Grad-Shafranov (GS) equation, Uzdensky (2005)
studied the topology of a stationary axisymmetric force-free degenerate
magnetosphere of a Kerr black hole with closed field lines between the event
horizon and a thin, Keplerian, infinitely conducting accretion disk. He
assumed the radius $r_{\max }$ proportional to $\psi _{H}/\psi _{D}$, the
ratio of the total flux measured at the event horizon and on the disk at $%
r_{\max }$ (his Eq. (42), plausible, but labelled by him as arbitrary). This
is a modified counterpart of the Blandford power law given by Eq. (\ref{BDBH}%
), with qualitative agreement for the power $n=3$ (modulo a scaling factor).

The Uzdensky (2005) solution of the GS equation determines the radial
extension of the field lines anchored in the disk as a function of the spin.
In the limit of slow rotation the size of the magnetically connected inner
part of the disk decreases with the increasing black hole spin. We stress
here that the model we adapt (based on the Blandford power law) also
predicts this shrinking of the closed magnetic field region with increasing
spin, as illustrated on Fig. \ref{r_max}.

Uzdensky did not include any magnetic field lines running from the black
hole to infinity, and so in our language his $\theta _{max}=0$. Our model
includes the jet, and hence the open magnetic field lines emanating from the
horizon, thus $\theta _{max}$ is a free parameter. Uzdensky solves a
different problem, the one for a jet-less black hole, and so it is not
immediately obvious how to directly use his results for the more general
setup we discuss. Therefore the application of the Blandford power law for
the ratio $B^{H}/B^{D}$ seems to us to be the most straightforward way to
compare the effect of the truncation of the radiation on the spin evolution
with previous results.

\subsection{The jet}

Numerical simulations of rotating black holes with a magnetosphere and a
thin accretion disk produced realistic jet profiles which have a complicated
double layer structure with an inner gas pressure-driven and an outer
magnetically driven jet (Mizuno et al. 2006). For the magnetically driven
jets the open field lines may be connected to the disk, without touching the
event horizon. These magnetohydrodynamic mechanisms dissipate energy and
pump angular momentum from the disk into the jet. The black hole spin
evolution is driven by spin-down to drive the jet and independently by
spin-up by accretion from the disk.

With a non-zero $4$-velocity component $u^{\phi }$, the plasma forms a jet
extracting rest mass, angular momentum and energy from the disk. The loss of
matter by the jets is specified by the parameter%
\begin{equation}
q_{jet}=\frac{\dot{M}_{jet}}{\dot{M}_{0}}\in \lbrack 0,1)\;,  \label{qm}
\end{equation}%
where $0\leq q_{jet}<1$ (vanishing in our model) and $\dot{M}_{jet}$ is the
rate of the rest mass flow into the jet (Falcke \& Biermann 1995). As the
geometry is reflection symmetric with respect to the equatorial plane, $\dot{%
M}_{jet}$ is composed of two equal contributions, of the upstream and
downstream jets.

The balance equations for mass, angular momentum and energy in the presence
of the jets were derived in Donea \& Biermann (1996), with parameters $%
q_{jet}$ and $r_{in}$ (the cut-off radius of the radiating disk) describing
the global properties of the jet.

We adopt the simplified approach as in Falcke \& Biermann (1995). The
differential rotation of the orbiting plasma winds up the radial magnetic
field lines acting as a dynamo which generates a poloidal magnetic field.
The latter produces a vertical Poynting flux jet, containing negligible
baryonic matter. The jet being in a Poynting flux limit implies $q_{jet}=0$.
This assumption decouples the energy and angular momentum transport in the
jet from the accretion disk properties.

Almost all supermassive black holes show signs of relativistic jets
(Perez-Fournon \& Biermann 1984, Chini et al. 1989, Falcke et al. 2000%
\textbf{,} Nagar et al. 2000, 2001, 2002, 2005), exhibiting compact radio
emission and extreme variability in this emission (the explanation of this
variability needing relativistic motion). The jet activity is correlated
with the high spin. Should this really be true for the vast majority of
otherwise quiescent black holes, this would be consistent with the
hypothesis, that these black holes are in a high spin state. Theoretical
estimates are consistent with assuming that the intrinsic Poynting flux of a
black hole in a high spin state is of order of $10^{-2.5}$ of the Eddington
luminosity (Armitage \& Natarian 1999), also consistent with the data on
some nearby radio galaxies, Cyg A, M87 and Cen A (Whysong \& Antonucci
2003). This low value implies that the lifetime of activity in this state is
about the Hubble time. In other worlds the spin-down due to the jet is
inefficient, the jet driving can be maintained for extremely long times,
independently of any accretion. This low backreaction is the main reason we
will not include the jet contribution in the balance equations.

\subsection{The inner disk}

There is observational evidence from the source GRS1915+105 (Fender et al.
2003, Eikenberry 1998) that the innermost region of the disk does not
radiate when the jet is strong. We will therefore assume that unless in the
exterior luminous part of the disk, where energy is dissipated, below a
certain radius $r_{in}$ the disk does not radiate. Thus all energy and
angular momentum from this region is transported into the black hole, hence
powering up the jet. The jet is not connected to the disk directly, but only
to the black hole, allowing it to start as a pure Poynting flux. Baryonic
matter is added into the jet only later, far up from the black hole by
intersecting clouds, stars, stellar winds and from shear and boundary
instabilities (Araudo et al. 2010). This model allows the early Lorentz
factor of the jet flow to be very high, as recently suggested by various
observations (see e.g. Ghisellini \& Tavecchio 2008).

In Abramowicz et al. (1978) the equipotential surfaces in the absence of
magnetic fields were employed to derive the thickness of the disk near the
inner edge. It was shown that a non-radiating disk can no longer be
described as geometrically thin, its pressure becomes important for the
radial disk structure, and the inner edge is pushed in the region between
the marginally stable and the marginally bound circular orbits. For
Schwarzschild black hole the marginally stable orbit is at $6M$, while the
innermost bound circular orbit is at $4M$. With increasing black hole spin
the marginally stable orbit approaches the horizon, the marginally bound
orbit being trapped in between, their difference becoming very small.

We also argue that all angular momentum of the accretion disk goes into the
black hole from the interior zone, independent of whether the disk is thick
or thin, either by direct accretion or via the torques of the closed
magnetic field lines and the details of this transfer do not matter in a
quasi-stationary situation. In particular, the exact location of the exact
inner edge of the disk (below $r_{in}$) does not influence the angular
momentum transport towards the black hole; all goes in. These considerations
enable us to use as a good approximation for the inner edge of the disk the
marginally stable orbit, especially at the high spin values investigated in
this paper.

What we have then is a radiatively truncated thin disk, with the inner edge
of the radiation zone at $r_{in}$ (instead of $r_{ms}$ in the absence of the
jet). The truncated disk model may also be consistent with the composite
(ADAF) models where the inner edge of the thin disk under a hot corona can
be found at a radius estimated between $2M$ and $100M$ (Done, Madejski \& 
\.{Z}ycki, 2000). For our analysis we chose radiative truncation radii
between $4M$ and $12M$ which is twice the range of radii considered by Donea
\& Biermann (1996). With this larger range the radiative truncation effect
on the mass accretion becomes easier to study. Fig. \ref{r_ms} shows the
radius of the marginally stable orbit as function of the spin parameter, as
given by Novikov \& Thorne (1973) and Page \& Thorne (1974).

This approximation is different from what was considered by Gammie (1999),
Krolik (1999); and Agol \& Krolik (2000), all of whom considered the
conditions at the inner-most stable orbit. Our work is complementary to
theirs, but as noted earlier supported by observations of the source
GRS1915+105 as useful to consider (see also Donea \& Biermann (1996)).

For lower values of the spin parameter, $a_{\ast }\approx 0.56$; the
inner edge of the thin disk is located at higher radii than $4M$, so the\
truncation of the radiation at $4M$ is not possible. The radiative
truncation at any radius greater than $6M$ is however possible for the whole
range of the spin parameter.

\section{STRUCTURE EQUATIONS OF THE DISK\label{structure}}

For standard accretion disks Page \& Thorne (1974) derived the integral form
of the conservation laws of rest mass $M_{0}$, angular momentum and energy
of the the accretion disk and black hole.\ The global transport equations
were obtained by averaging the continuity equation and the total divergence
of the density-flux 4-vectors of total angular momentum $\boldsymbol{J}=%
\boldsymbol{T}\cdot \partial /\partial \phi $ and energy $\boldsymbol{E}=-%
\boldsymbol{T}\cdot \partial /\partial t$, respectively.

In subsection \ref{balance} we generalize this result by including the
effects of the magnetic field lines. We discuss the various source terms
appearing in the balance equations in the remaining subsections, also the
power and torque of the open magnetic field lines.

\subsection{Balance equations in the symbiotic model\label{balance}}

\subsubsection{Rest mass conservation}

Rest mass conservation implies that the time averaged rate of the accretion
of rest mass (the radial matter flow through a ring of thickness $\Delta r\,$%
), given by Eq. (\ref{M0dot}) is independent of the radius (Page \& Thorne
1974) 
\begin{equation}
\dot{M}_{0}=-2\pi r\Sigma _{disk}u_{disk}^{r}=const.
\end{equation}%
for any $r>r_{in}$.

\subsubsection{Angular-momentum conservation}

The angular momentum balance equation is found by repeating the derivation
(22)-(27) of Page \& Thorne (1974), and adding the magnetic contribution. It
is necessary to use $\left( -g\right) ^{1/2}=r$ emerging from Eq. (\ref{ds2}), that $%
u_{\phi }=\widetilde{L}$, employ the time and azimuthal averaging and the
reflection symmetry about the plane $z=0$ (such that $u_{jet}^{z}\left(
-z\right) =-u_{jet}^{z}\left( z\right) $). The angular momentum transport
equation becomes 
\begin{equation}
\lbrack \dot{M}_{0}\widetilde{L}-2\pi rW_{\phi }{}^{r}]_{,r}=4\pi rF%
\widetilde{L}-\left( \frac{d\Delta L^{D}}{dt}\right) _{,r}~.
\label{qmdotM0tildeL}
\end{equation}%
The torque $d\Delta L^{D}/dt$ exerted on the disk by the spinning black hole
via the closed magnetic field lines gives an additional term, which will be
calculated in subsection 3.3.

Eq. (\ref{qmdotM0tildeL}) represents the balance of four types of
contributions determining the angular momentum transport from the disk to
the black hole. On the left hand side there are the contributions of the
rest mass flow $\dot{M}_{0}$ and the torques in the disk, on the right hand
side the contributions of the flux $F$ of the radiation emitted from the
surface of the disk and the contribution of the closed magnetic field lines.

\subsubsection{Energy conservation}

Energy conservation arises from the balance of the energy transport in terms
of the rest mass accretion, torques in the disk (left hand side terms) and
the surface radiation and contribution of the closed magnetic field (right
hand side terms). Similar considerations as for the rest mass and angular
momentum conservations lead to 
\begin{equation}
\lbrack \dot{M}_{0}\widetilde{E}-2\pi r\Omega W^{r}{}_{\phi }]_{,r}=4\pi rF%
\widetilde{E}-P_{,r}^{D}~.  \label{qmdotM0tildeE}
\end{equation}%
The expression of the extracted power $P^{D}$ will be given in subsection %
\ref{torque_closed}. Again, only the closed field lines contribute.

The expressions (\ref{qmdotM0tildeE}) and (\ref{qmdotM0tildeL}) are the 
\textit{radial structure equations} of the accretion disk with external
magnetic fields and jets.

\subsection{Power and torque on open magnetic field lines}

Although not contributing to the structure equations of the disk, we present
here the energy and angular momentum transport due to the BZ mechanism, as
they will contribute to the mass and spin evolution of the black hole, to be
discussed in Section 4. These transport phenomena have been described in the
framework of a 3+1 decomposition of the Kerr geometry (Macdonald \& Thorne
1982, Thorne, Price \& Macdonald 1986). In the 3+1 formalism the torque
exerted on the open magnetic field lines by the rotating black hole can be
expressed as an the integral across a surface $\mathscr S$ intersecting all
open field lines emanating from the horizon and $d\psi $ represents an
element of the flux tube $\psi $ transverse to $\mathscr S$. 
\begin{equation}
-\frac{d\Delta L^{H}}{dt}=\int_{\mathscr S}\frac{\Omega ^{H}-\Omega ^{F}}{%
4\pi }\widetilde{\omega }^{2}B_{H}{d}\psi ~,  \label{dLHdt}
\end{equation}%
The power transmitted by this torque and pumped into the flux tube becomes%
\begin{equation}
P^{H}=\int_{\mathscr S}\Omega ^{F}\frac{\Omega ^{H}-\Omega ^{F}}{4\pi }%
\widetilde{\omega }^{2}B_{H}{d}\psi \;,  \label{P}
\end{equation}%
where $\widetilde{\omega }$ is the cylindrical radius of flux tube at the
event horizon, given by Eq. (\ref{tildeomega}), and $B_{H}$ is the strength (%
\ref{B2H}) of the magnetic field normal to the horizon.

In order to determine these quantities in the BZ process, we integrate over
the open field lines spanning between $\theta =0$ and $\theta _{max}$, in
both hemispheres. Evaluating the integrands in Eqs. (\ref{dLHdt}) and (\ref%
{P}) on the horizon and expressing the magnetic flux $\psi $ as a function
of $\theta $, we obtain the net torque and the extracted power due to the BZ
process as%
\begin{equation}
-\frac{d\Delta L^{H}}{dt}=8M^{3}B_{H}^{2}\int_{0}^{\theta _{max}}\frac{%
(\Omega ^{H}-\Omega ^{F})r_{+}^{2}\sin ^{3}\theta }{2M-r_{-}\sin ^{2}\theta }%
{d}\theta  \label{dLHdtexp}
\end{equation}%
and 
\begin{equation}
P^{H}=8M^{3}B_{H}^{2}\int_{0}^{\theta _{max}}\frac{\Omega ^{F}(\Omega
^{H}-\Omega ^{F})r_{+}^{2}\sin ^{3}\theta }{2M-r_{-}\sin ^{2}\theta }{d}%
\theta ~.  \label{Pexp}
\end{equation}

In the particular case of $\theta _{max}=\pi /4$ Eqs. (\ref{dLHdtexp}) and (%
\ref{Pexp}) give values close to those of the prevalent approximations (e.g.
Park \& Vishniac 1988 and Moderski \& Sikora 1996) 
\begin{eqnarray}
-\frac{d\Delta L^{H}}{dt} &\simeq &\frac{1}{8}B_{H}^{2}r_{H}^{4}(\Omega
^{H}-\Omega ^{F})\;, \\
P^{H} &\simeq &\frac{1}{8}B_{H}^{2}r_{H}^{4}\Omega ^{F}(\Omega ^{H}-\Omega
^{F})
\end{eqnarray}%
for rapidly rotating black holes with $a_{\ast }\simeq 0.98$. Since we
assumed the rotation frequency $\Omega ^{F}=\Omega ^{H}/2$ at the event
horizon (maximally efficient BZ mechanism), Eqs. (\ref{dLHdtexp}) and (\ref%
{Pexp}) can be integrated analytically. For $a_{\ast }>0$ the result is 
\begin{eqnarray}
-\frac{d\Delta L^{H}}{dt} &=&8M^{3}B_{H}^{2}(\Omega ^{H}-\Omega ^{F})\frac{%
r_{+}^{2}}{r_{-}}\left[ \cos \theta _{max}-1\frac{{}}{{}}\right.  \notag \\
&&-\frac{2}{a_{\ast }}\tan ^{-1}\left( \frac{r_{-}}{Ma_{\ast }}\cos \theta
_{max}\right)  \notag \\
&&\left. +\frac{2}{a_{\ast }}\tan ^{-1}\left( \frac{r_{-}}{Ma_{\ast }}%
\right) \right]  \label{dLHdtanal}
\end{eqnarray}%
and 
\begin{equation}
P^{H}=-\Omega ^{F}d\Delta L^{H}/dt\;.  \label{Panal}
\end{equation}

\subsection{Power and torque on closed magnetic field lines\label%
{torque_closed}}

\begin{figure}
\centering
\includegraphics[width=.48\textwidth]{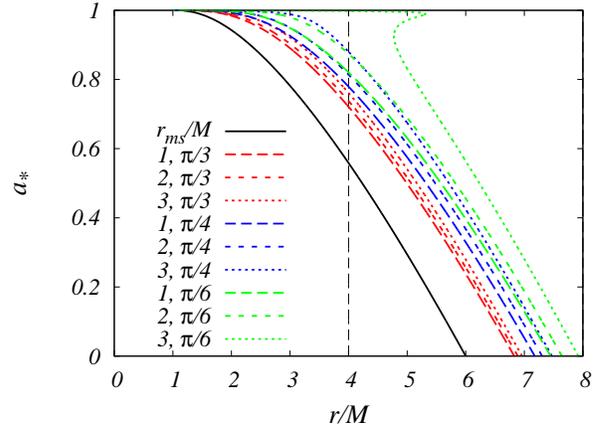}
\caption{The spin parameter $a_{\ast }$ as function of the maximal radius $%
r_{max}$ (representing the maximal extension of the closed magnetic field
lines emanating from the horizon), for various parameter values $n=1$, $2$, $%
3$ and $\protect\theta _{max}=\protect\pi /3$, $\protect\pi /4$, $\protect%
\pi /6$. The solid (black) curve indicates the value of $r_{ms}$ for each $%
a_{\ast }$ (representing the minimal extension of the closed magnetic field
lines). The radius $4M$ (the minimal radiative truncation radius) is
indicated by the vertical dashed line.}
\label{r_max}
\end{figure}

\begin{figure}
\centering\includegraphics[width=.48\textwidth]{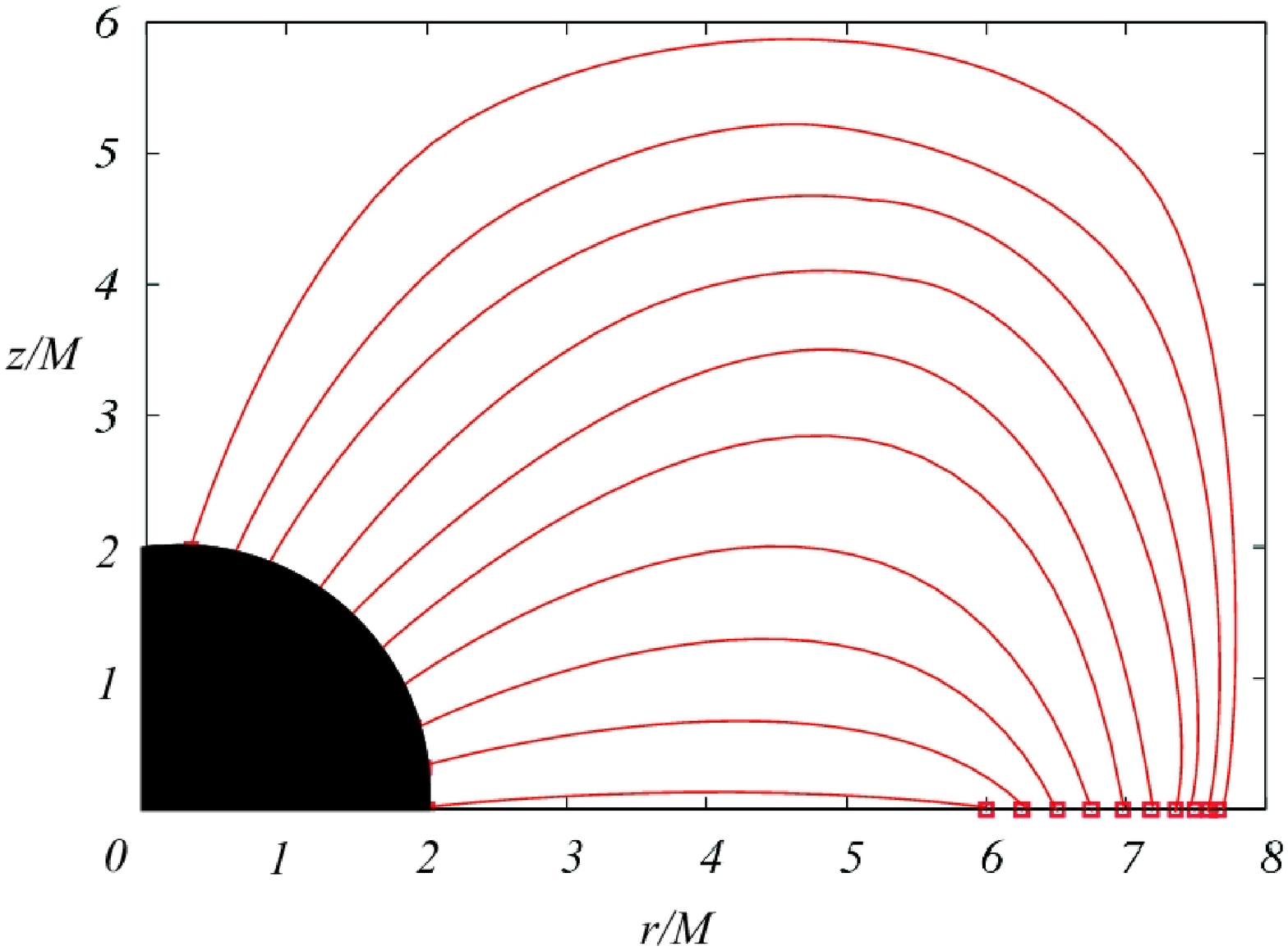} %
\includegraphics[width=.48\textwidth]{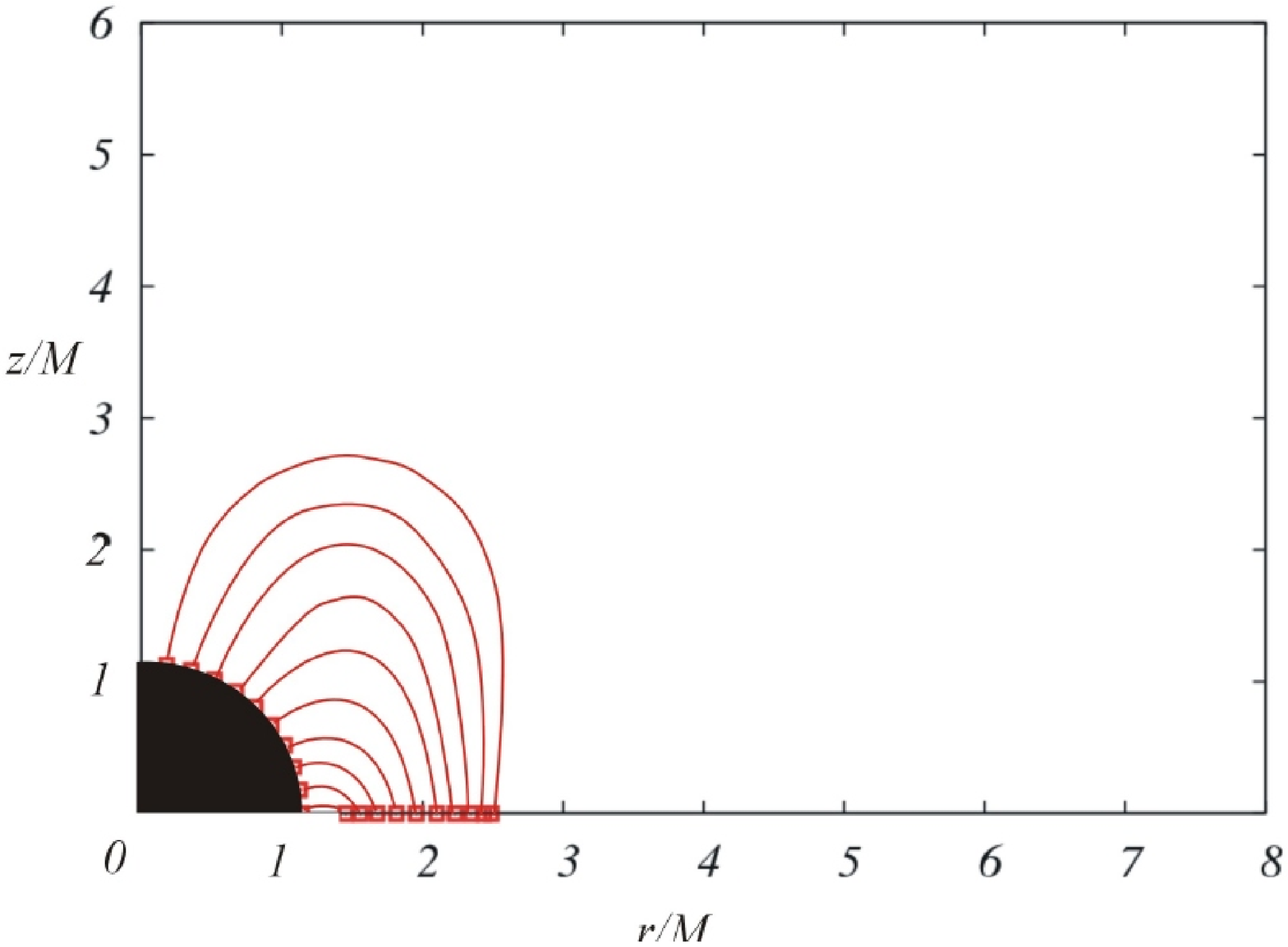}
\caption{The closed magnetic field configuration for the Schwarzschild black
hole (upper sketch) and a rapidly rotating, $a_{\ast }=0.99$ black hole
(lower sketch). The endpoints of the magnetic field lines were determined by
Eq. (\protect\ref{sin2theta}) for $n=1$.}
\label{theta_r}
\end{figure}

The effects of the closed magnetic field lines on the disk arise in a
similar fashion. The torque and the power emerging from the magnetic
coupling between the black hole and the disk was derived by Wang, Xiao \&
Lei (2002), applying the model developed by Li (2000). The torque is: 
\begin{equation}
-\frac{d\Delta L^{D}}{dt}=\int_{\mathscr S}\left( \frac{\Omega ^{H}-\Omega }{%
4\pi ^{2}}\right) \frac{{d}\psi }{{d}Z^{H}}{d}\psi \;,  \label{dLDdt}
\end{equation}%
where ${d}Z^{H}$ is the total impedance of the event horizon over the
surface $\mathscr S$ threaded by all closed field lines. Similarly, the
power transferred by the magnetic torque can be written as 
\begin{equation}
P^{D}=\int_{\mathscr S}\Omega \left( \frac{\Omega ^{H}-\Omega }{4\pi ^{2}}%
\right) \frac{{d}\psi }{{d}Z^{H}}{d}\psi \;.  \label{PD}
\end{equation}%
Employing the flux conservation in an arbitrary flux tube between the
horizon and the disk, the integrals from Eqs. (\ref{dLDdt}) and (\ref{PD})
can be rewritten as integrals over the radial coordinate. By virtue of Eq. (%
\ref{BDBH}), the equations (\ref{dLDdt}) and (\ref{PD}) lead to (Wang et al.
2003): 
\begin{eqnarray}
-\frac{d\Delta L^{D}}{dt} &=&\frac{4M^{2}B_{H}^{2}}{\sqrt{\mathscr A(x_{ms})}%
}\int_{r_{ms}}^{r_{max}}(\Omega ^{H}-\Omega )\left( \frac{r}{r_{ms}}\right)
^{1-n}  \notag \\
&&\times \sqrt{\frac{\mathscr A}{\mathscr D}}\frac{r_{+}^{2}\sin ^{2}\theta 
}{2M-r_{-}\sin ^{2}\theta }{d}r  \label{dLDdt2}
\end{eqnarray}%
and%
\begin{eqnarray}
P^{D} &=&\frac{4M^{2}B_{H}^{2}}{\sqrt{\mathscr A(x_{ms})}}%
\int_{r_{ms}}^{r_{max}}\Omega (\Omega ^{H}-\Omega )\left( \frac{r}{r_{ms}}%
\right) ^{1-n}  \notag \\
&&\times \sqrt{\frac{\mathscr A}{\mathscr D}}\frac{r_{+}^{2}\sin ^{2}\theta 
}{2M-r_{-}\sin ^{2}\theta }{d}r\;,  \label{PD2}
\end{eqnarray}%
where 
\begin{equation}
\sin ^{2}\theta =1-\left[ \frac{1}{2M\sqrt{\mathscr A(x_{ms})}}%
\int_{r_{ms}}^{r}\left( \frac{r}{r_{ms}}\right) ^{1-n}\sqrt{\frac{\mathscr A%
}{\mathscr D}}{d}r\right] ^{2}\;.  \label{sin2theta}
\end{equation}%
The radial coordinate $r_{max}$ of the outermost closed flux line emanating
from the horizon at the angle $\theta _{max}$ can be determined by
substituting $\theta _{max}$ in Eq. (\ref{sin2theta}). In the prefactor of
the integrals the function $\mathscr A$ is evaluated at the marginally
stable orbit $x_{ms}$. These expressions are sources in the global angular
momentum and energy balance equations. Fig. \ref{r_max} shows the radii $%
r_{ms}$ and $r_{max}$ for the power law exponent $n=1$, $2$ and $3$ and the
boundary angle $\theta _{max}=\pi /3$, $\pi /4$ and $\pi /6$ in the full
range of the spin parameter. The closed flux lines connect the event horizon
to the disk, ending at radial coordinates contained between $r_{ms}$ and $%
r_{max}$ for each configuration of $a_{\ast }$, $n$ and $\theta _{max}$.
When the radiation from the innermost part of the disk is suppressed, $%
r_{ms} $\ has to be replaced with $r_{in}$. 
The closed magnetic field line configuration is determined by Eq. (\ref%
{sin2theta}). We illustrate this for both the Schwarzschild black hole and a
high spin $a_{\ast }=0.99$ black hole on Fig. \ref{theta_r}. For each value
of the polar angle $\theta $ under which the open lines emanate from the
horizon there is a well-defined radius $r$ where they couple to the disk. We
recover that $r_{\max }$ is decreasing with increasing spin, a property also
shown by Uzdensky (2005). 

\subsection{The flux of the radiation emitted from the disk}

\subsubsection{Standard accretion disk}

By eliminating the torque term from the conservation laws, then employing
the energy-angular momentum relation $\widetilde{E}_{,r}=\Omega \widetilde{L}%
_{,r}$ for circular geodesic orbits, Page \& Thorne (1974) obtained the
radial distribution of the time-averaged radiation emitted from the surface
of standard accretion disk as 
\begin{equation}
F(r)=-\frac{\dot{M}_{0}}{4\pi r}\frac{\Omega _{,r}}{(\widetilde{E}-\Omega 
\widetilde{L})^{2}}\int_{r_{ms}}^{r}(\widetilde{E}-\Omega \widetilde{L})%
\widetilde{L}_{,r}dr\;.  \label{F}
\end{equation}%
Since the inner edge of the disk is located at the marginally stable orbit $%
r_{ms}$ they set the lover boundary of the flux integral to this radius.
They also prescribed {}the \textit{no-torque\ inner boundary condition} at $%
r_{ms}$ (which sets the integration constant to zero), by arguing that
accreting matter at the marginally stable orbit falls freely into the black
hole, unable to exert considerable torque on the disk. The explicit analytic
expression of $F$ is given as Eq. (\ref{Fanal}) of Appendix A. Remarkably,
with $r=Mx^{2}$ the flux $F$ becomes manifestly proportional to $\dot{M}%
_{0}/M^{2}$, the proportionality factor depending only on $a_{\ast }$ and $x$%
.

Although the no-torque inner boundary condition might not hold if closed
magnetic field lines connecting the inner edge of the disk and the event
horizon convey torque from the hole to the disk (Krolik 1999, Gammie 1999,
Agol \& Krolik 2000), Afshordi and Paczy\'{n}ski (2003) have proven that the
torque at the inner edge can be neglected in the \textit{absence} of \textit{%
large scale } radial magnetic fields in the plunging region. In another
analytic model developed by Li (2003) the magnetic field lines connect the
surface of the disk to higher latitudes of the event horizon. As a
consequence, the torque is transported outwards in the disk and it vanishes
at the inner edge. Following this model in our symbiotic scheme the magnetic
field will not produce considerable effects in the \textit{plunging} region
and we can also impose the no-torque boundary condition.

\subsubsection{Symbiotic model}

The torque term from Eqs. (\ref{qmdotM0tildeL}) and (\ref{qmdotM0tildeE})
can be eliminated in the generic case either. By integrating the result, the
flux of the disk is obtained as: 
\begin{eqnarray}
F_{tot} &=&F+\frac{\dot{M}_{0}}{4\pi r}\frac{\Omega _{,r}}{(\widetilde{E}%
-\Omega \widetilde{L})^{2}}\int_{r_{in}}^{r}(\widetilde{E}-\Omega \widetilde{%
L})  \notag \\
&&\times \left[ c_{L}-\left( \frac{c_{E}-\Omega c_{L}}{\Omega _{,r}}\right)
_{,r}\right] dr\;.  \label{ftot0}
\end{eqnarray}%
with 
\begin{eqnarray}
c_{L} &\equiv &-\dot{M}_{0}^{-1}(d\Delta L^{D}/dt)_{,r}\;,  \label{cL} \\
c_{E} &\equiv &-\dot{M}_{0}^{-1}P_{,r}^{D}\;.  \label{cE}
\end{eqnarray}%
Here $F$ is given in Eq. (\ref{F}), with the modification that its lower
integration limit should be $r_{in}$, rather than $r_{ms}$ (provided $%
r_{in}>r_{ms}$, otherwise $r_{ms}$). Integrating by parts the second term on
the right hand side of Eq. (\ref{ftot0}) we obtain 
\begin{eqnarray}
F_{tot} &=&F+\frac{\dot{M}_{0}}{4\pi r}\frac{\Omega _{,r}}{(\widetilde{E}%
-\Omega \widetilde{L})^{2}}\left\{ \left[ \frac{(\widetilde{E}-\Omega 
\widetilde{L})}{-\Omega _{,r}}(c_{E}-\Omega c_{L})\right] _{r_{in}}^{r}%
\right.  \notag \\
&&+\left. \frac{\dot{M}_{0}}{4\pi r}\int_{r_{in}}^{r}\frac{(\widetilde{E}%
-\Omega \widetilde{L})}{\Omega }[hc_{E}+(1-h)\Omega c_{L}]dr\right\} \;. 
\notag \\
&&  \label{ftot}
\end{eqnarray}%
Here we introduced the shorthand notation 
\begin{equation}
h(x,a_{\ast })\equiv \frac{\lbrack \ln (\widetilde{E}-\Omega \widetilde{L}%
)]_{,r}}{[\ln \Omega ]_{,r}}=-\frac{\mathscr{F}}{x^{2}\mathscr{C}}  \notag
\end{equation}%
with the functions $\mathscr{C}$ and $\mathscr{F}$ given by Eqs. (\ref{calC}%
) and (\ref{calF}).

The photon flux $F_{tot}$ emitted by the accretion disk depends on the
parameters introduced in the global description of the unified model. For a
given rotation frequency $\Omega ^{F}$ of the magnetic field, the topology
of magnetic field is specified by the exponent $n$ and the separation angle $%
\theta _{max}$, whereas the radiation from the disk is characterized by the
cutoff radius $r_{in}$. By specifying these parameters we have fully
determined the structure equations for the thin accretion disk, rotating in
the black hole magnetosphere.

\subsubsection{Special case: jet with no magnetosphere}

Here we consider the radiatively truncated thin disk scheme introduced in
the disk-jet-hole symbiotic model without magnetosphere. If no torque is
exerted on a Keplerian disk by the rotating hole we have $c_{L}=0$ and $%
c_{E}=0$. Then the flux integral (\ref{ftot}) simplifies to 
\begin{eqnarray}
F_{tot} &=&F=-\frac{\dot{M}_{0}}{4\pi r}\frac{\Omega _{,r}}{(\widetilde{E}%
-\Omega \widetilde{L})^{2}}\int_{\mathrm{max}(r_{ms},r_{in})}^{r}(\widetilde{%
E}-\Omega \widetilde{L})\widetilde{L}_{,r}dr\;,  \notag \\
&&  \label{ftotjet}
\end{eqnarray}%
where $F$ is given by Eq. (\ref{Fanal}) again, but with the substitution $%
x_{ms}^{2}\rightarrow \mathrm{max}(r_{ms},r_{in})/M$.

\begin{figure}
\centering
\includegraphics[width=.48\textwidth]{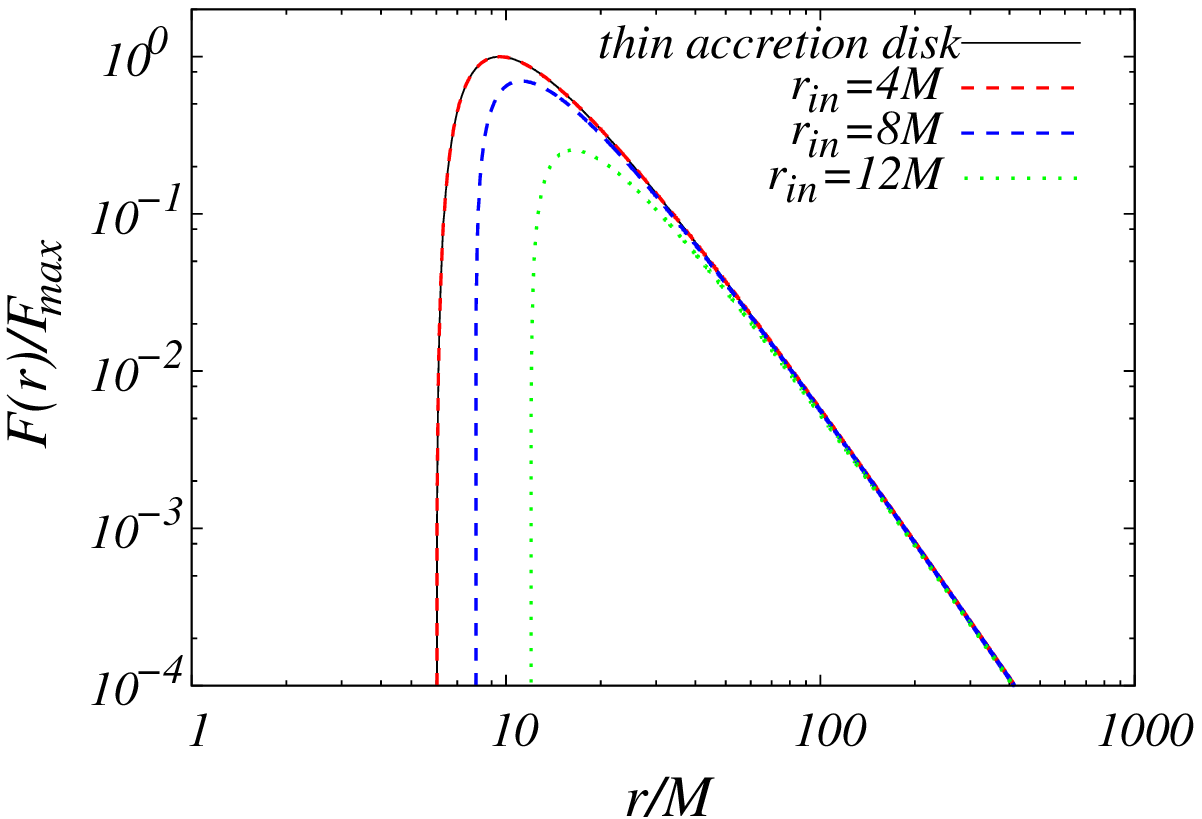} %
\includegraphics[width=.48\textwidth]{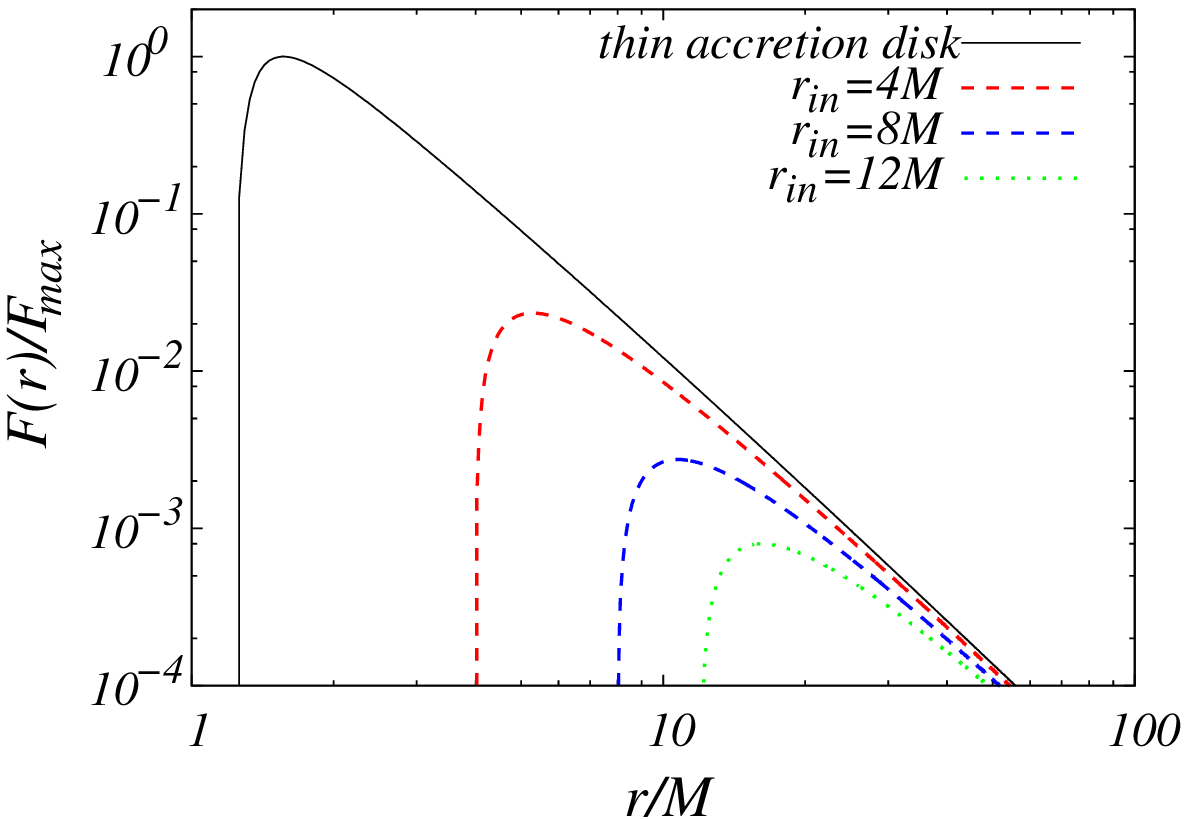}
\caption{The time-averaged flux $F(r)$ radiated by a steady-state thin
accretion disk about a Kerr black hole and by various truncated disk models
with the nozzle radii $r_{in}=4M$, $8M$ and $12M$. The upper plot is for
zero spin, the lower plot for the maximally allowed spin $a_{\ast }=0.9982$
of the canonical black hole. The flux is normalized to the maximal flux
emitted from the thin accretion disk $F_{max}=1.37\times 10^{-5}\dot{M}%
_{0}/M^{2}$ for the static case and $F_{max}=0.00498\dot{M}_{0}/M^{2}$ for
the rotating black hole. }
\label{F_td}
\end{figure}

Falcke \& Biermann (1995) have considered a direct connection between disk
and jet, truncating the disk at values between $4M$ and $12M$. In our model
the accretion disk does not power the jet directly, thus the inner edge of
the accretion disk is at $r_{ms}$, but we suppress the radiation below $%
r_{in}$. Fig. \ref{F_td} shows the radial distribution of the photon flux
emitted from the surface of the disk with the radiative truncation at $%
r_{in}=4M$, $8M$ and $12M$, for $a_{\ast }=0$ (static case) and $a_{\ast
}=0.9982$ (at the spin limit of the canonical black hole). For rapid
rotation the suppression of the radiation below $r_{in}$ decreases the
integrated flux considerably, provided $r_{in}>r_{ms}$. At low spin $r_{ms}$
is larger than $4M$ and the radiative truncation only at $8M$ decreases the
maximal value of $F(r)$. This behaviour is illustrated in Fig. \ref{F_td}. 
\begin{figure}
\centering
\includegraphics[width=.48\textwidth]{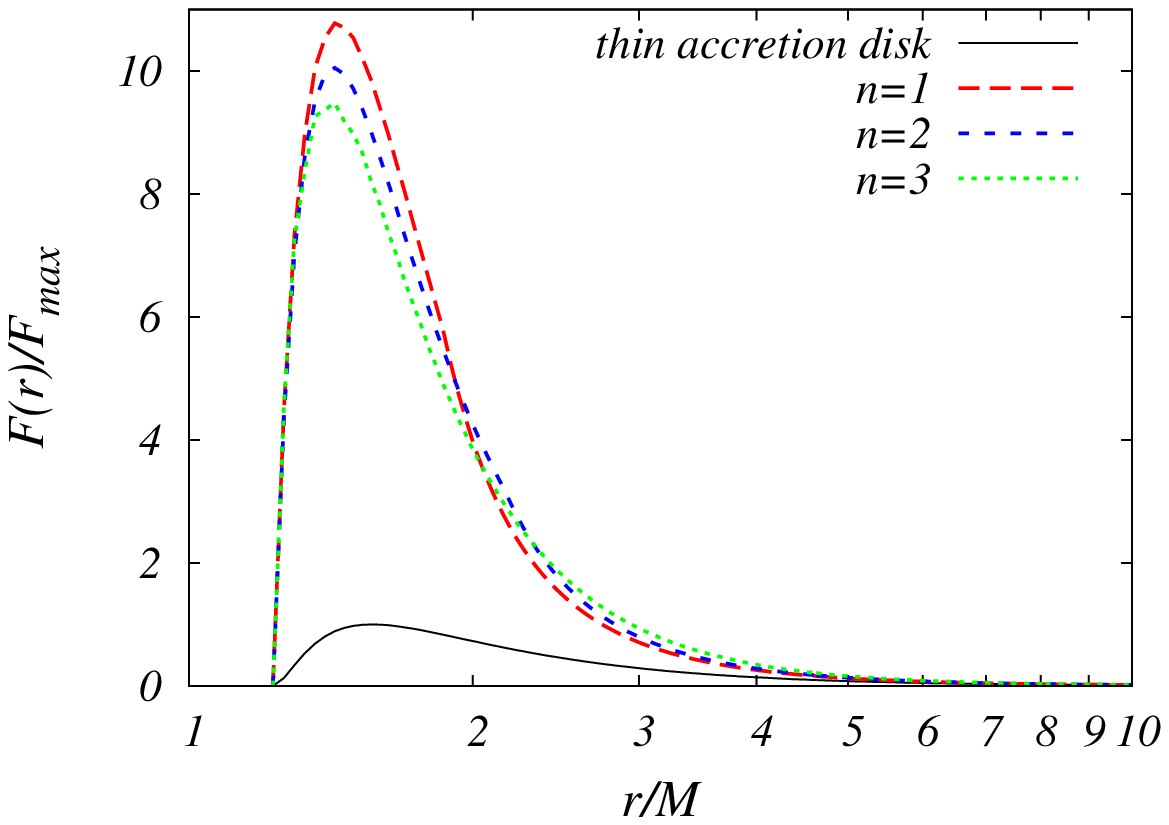} %
\includegraphics[width=.48\textwidth]{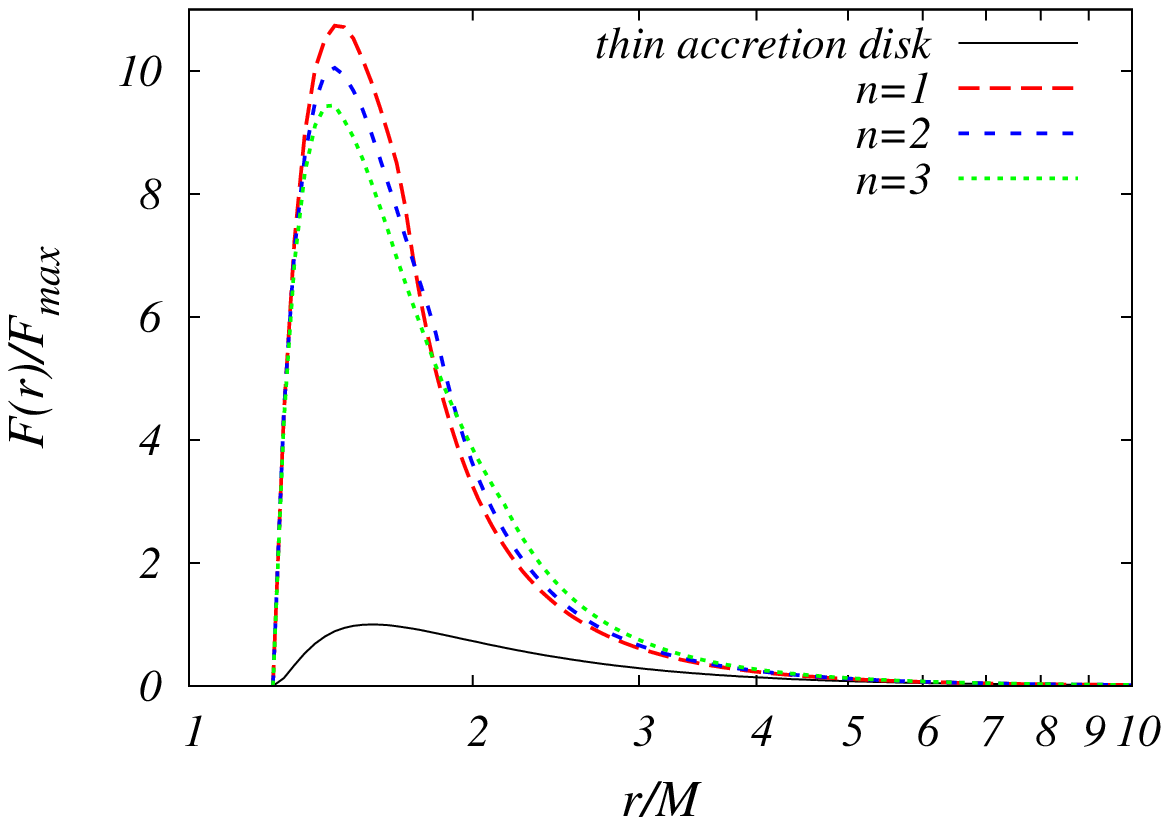} %
\includegraphics[width=.48\textwidth]{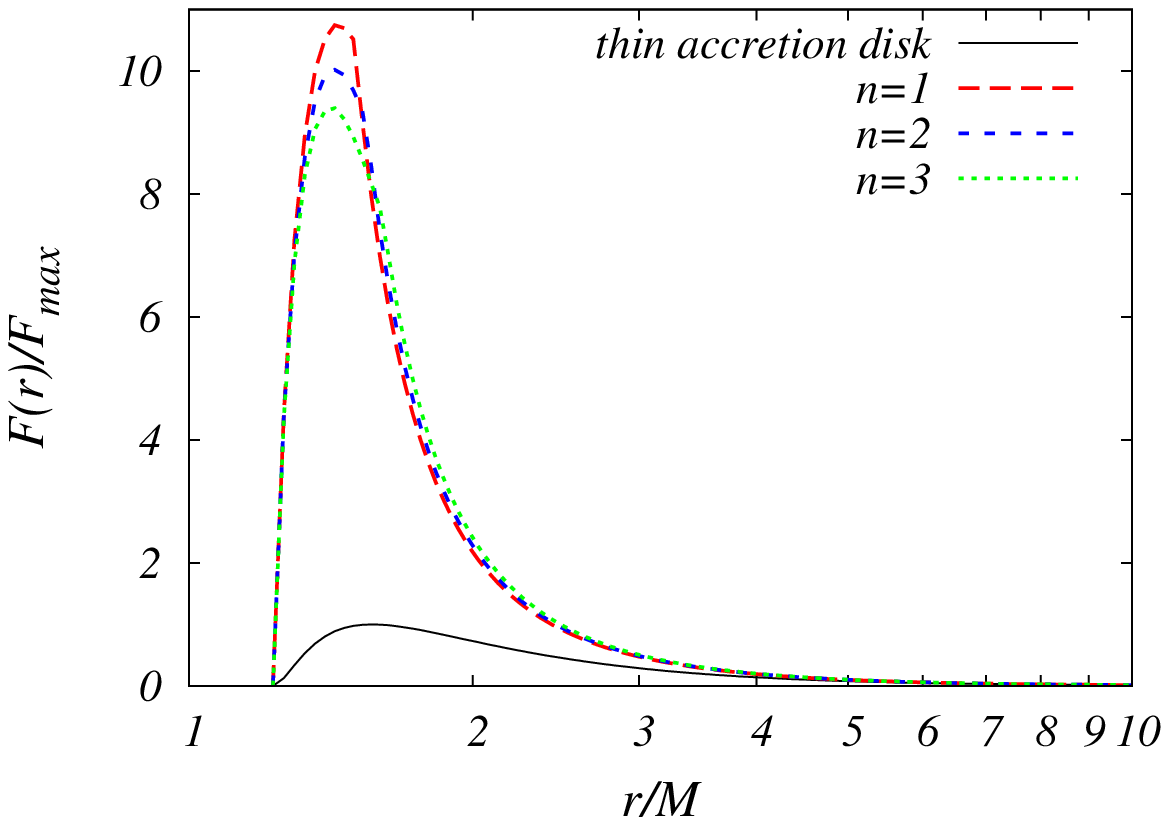}
\caption{The time-averaged flux $F(r)$ radiated by an accretion disk around
a rotating black hole ($a_{\ast }=0.9982$) for the standard accretion disk
and accretion disk with a magnetic torque. The angular boundary $\protect%
\theta _{max}$ is set to $\protect\pi /6$ (top), $\protect\pi /4$ (middle)
and $\protect\pi /3$ (bottom). The parameter $n$ takes the values 1,2 and 3. 
$F(r)$ is normalized with $F_{max}=0.00498\dot{M}_{0}/M^{2}$. }
\label{F_mc1}
\end{figure}

\subsubsection{Special case: magnetosphere with no jet}

Considering solely the magnetic torque exerted on a Keplerian disk by the
spinning black hole (no jet, $r_{in}=r_{ms}$), the expressions (\ref{dLDdt2}%
), (\ref{PD2}) inserted in Eqs. (\ref{cL}), (\ref{cE}) give $c_{L}=\Omega
c_{E}$. Then Eq. (\ref{ftot}) leads to%
\begin{eqnarray}
F_{tot} &=&F+\frac{\dot{M}_{0}}{4\pi r}\frac{4}{\dot{M}_{0}}\frac{-\Omega
_{,r}}{(\widetilde{E}-\Omega \widetilde{L})^{2}}\frac{M^{2}B_{H}^{2}}{\sqrt{%
\mathscr A(x_{ms})}}  \notag \\
&&\times \int_{r_{ms}}^{\mathrm{min}(r,r_{max})}(\widetilde{E}-\Omega 
\widetilde{L})(\Omega ^{H}-\Omega )  \notag \\
&&\times \left( \frac{r}{r_{ms}}\right) ^{1-n}\sqrt{\frac{\mathscr A}{%
\mathscr D}}\frac{r_{+}^{2}\sin ^{2}\theta }{2M-r_{-}\sin ^{2}\theta }{d}r\;.
\label{ftotmc}
\end{eqnarray}%
The radial dependence of $\sin ^{2}\theta $ is given by Eq. (\ref{sin2theta}%
). In the expression (\ref{ftotmc}) the photon flux depends on the magnetic
field $B_{H}$ and the parameter $n$ used in in the power law (\ref{BDBH}).
In Fig. \ref{F_mc1} we plotted the radial distribution of the photon flux
determined by Eq. (\ref{ftotmc}). In the plots the parameter $n$ is set to $%
1 $, $2$ and $3$ whereas $\theta _{max}$ takes the values $\pi /6$, $\pi /4$
and $\pi /3$. This case differs from the discussion of Uzdensky (2005) by
limiting the allowed range of $\theta _{max}$ to values $>0$, so providing
an additional free parameter in the boundary condition; this is the topology
required by a jet. With increasing values of $\theta _{max}$ we decrease the
angular boundary of the closed magnetic field lines, which decreases the
contribution of the torque conveyed by the flux lines to the emitted energy
of the disk. For larger values of $\theta _{max}$ the flux profiles fall
more steeply at higher radii. The increment of the parameter $n$ decreases
the radial fall-off of the magnetic field strength. Then the fall-off of the
flux profile at higher radii is somewhat less steeper, as seen in the
uppermost plot.

\section{BLACK HOLE EVOLUTION AND ENERGY CONVERSION EFFICIENCY \label%
{massspinevol}}

\subsection{Black hole mass and spin evolution equations}

\subsubsection{Canonical black hole}

In Bardeen's accretion model the time derivatives of the total mass and
angular momentum of the black hole are proportional to the specific energy $%
\widetilde{E}_{ms}$ and specific angular momentum $\widetilde{L}_{ms}$,
respectively, of the particles falling in from the marginally stable orbit,
the proportionality factor being in both cases the rate of change of the
rest mass. Page \& Thorne (1974) modified these relations by including the
radiation emitted by the disk and captured by the black hole, to get: 
\begin{eqnarray}
dM/dt &=&\dot{M}_{0}\widetilde{E}_{ms}+(dM/dt)_{rad}\;,  \label{dMdt} \\
dJ/dt &=&\dot{M}_{0}\widetilde{L}_{ms}+(dJ/dt)_{rad}\;.  \label{dJdt}
\end{eqnarray}%
From Eqs. (\ref{tildeE}) and (\ref{tildeL}) evaluated at $r_{ms}$ one
obtains $\widetilde{E}_{ms}$ and $\widetilde{L}_{ms}$. The time averaged
rates $(dM/dt)_{rad}$ and $(dJ/dt)_{rad}$ at which photons carry energy and
angular momentum into the black hole were computed by integrating the photon
flux $F$ over the disk surface in the Kerr potential (Thorne 1974), given
here as Eqs. (\ref{dMdtrad}) and (\ref{dJdtrad}) of Appendix A. The
evolution equations of the spin parameter $a_{\ast }$ and of the total black
hole mass $M$ therefore can be rewritten as 
\begin{eqnarray}
\frac{da_{\ast }}{d\ln \left( M/M_{1}\right) } &=&\frac{1}{M}\frac{%
\widetilde{L}_{ms}+\dot{M}_{0}^{-1}(dJ/dt)_{rad}}{\widetilde{E}_{ms}+\dot{M}%
_{0}^{-1}(dM/dt)_{rad}}-2a_{\ast }\;,  \label{dastardlnM} \\
\frac{dM}{dM_{0}} &=&\widetilde{E}_{ms}+\dot{M}_{0}^{-1}(dM/dt)_{rad}\;,
\label{dMdM02}
\end{eqnarray}%
where $M_{1}$ represents a mass scale, chosen as unity in what follows.
Despite appearances, the right hand side of Eq. (\ref{dastardlnM}) does not
depend on either the black hole mass $M$ or the accretion rate $\dot{M}_{0}$%
. Indeed, from Eqs. (\ref{tildeE}) and (\ref{tildeL}) $\widetilde{L}%
_{ms}\propto M$, while $\widetilde{E}_{ms}$ does not depend on either $M$ or 
$\dot{M}_{0}$. Since $F(r)\propto \dot{M}_{0}/M^{2}$, according to the
integrals (\ref{dMdtrad}) and (\ref{dJdtrad}) $(dM/dt)_{rad}\propto \dot{M}%
_{0}$ and $(dJ/dt)_{rad}\propto \dot{M}_{0}M$ hold, respectively. Then the
right hand side of Eq. (\ref{dastardlnM}) depends solely on $a_{\ast }$, the
differential equation becoming separable. Also, the extremum $da_{\ast
}/d\ln M=0$ does not depend on either $M$ or $\dot{M}_{0}$.

Further, as Eq. (\ref{dastardlnM}) does not depend on the mass accretion
rate, neither does the evolution $a_{\ast }(M)$. However, the evolution in
time depends on $\dot{M}_{0}$, cf. Eqs. (\ref{dMdt})-(\ref{dJdt}). As black
holes of few solar masses and supermassive galactic black holes exhibit huge
differences in $\dot{M}_{0}$, both their mass and spin evolutions occur at
different characteristic time scales.

A similar analysis of Eq. (\ref{dMdM02}) shows that $dM/dM_{0}$ does not
depend on $M$ or $\dot{M}_{0}$ either, but only on $a_{\ast }$.

\subsubsection{Black hole spin reversal due to accretion in Bardeen's model}

We show explicitly here, how the spin evolves as function of mass in the
simplest, Bardeen approximation (without corrections due to photon capture,
magnetic fields and truncation of radiation). For this, we integrate Eq. (%
\ref{dastardlnM}), with the radiative contributions left apart. By employing
the definitions (\ref{tildeE}) and (\ref{tildeL}) we get

\begin{equation}
\frac{\pm da_{\ast }}{d\ln M}=\frac{x_{ms}\mp 2a_{\ast }\pm 2a_{\ast
}x_{ms}^{-2}-a_{\ast }^{2}x_{ms}^{-3}}{1-2x_{ms}^{-2}\pm a_{\ast }x_{ms}^{-3}%
}\;,  \label{aux1}
\end{equation}%
with the upper (lower) sign referring to corotating (counterrotating)
accreting particles on the innermost marginally stable circular geodesic
orbit.

\begin{figure}
\centering\includegraphics[width=.4\textwidth]{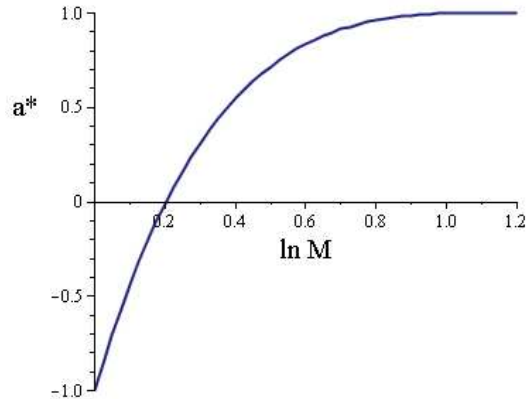}
\caption{The variation of the spin parameter $a_{\ast }$ due to standard
accretion only in terms of the mass (taken in units of the initial mass).
For convenience the counterrotation is represented by negative values of $%
a_{\ast }$. The graph is for initially counterrotating accreted particles.
This slows down, then stops the rotation of the black hole, after which, in
a corotating configuration the black hole spins up again in the opposite
sense. The initial radius of the innermost stable circular geodesic orbit is 
$r_{ms}=9M$ for counterrotating particles in the extreme Kerr geometry. When
the rotation stops, $r_{ms}=6M$ (Schwarzschild case). Finally when the black
hole reaches maximal rotation in the opposite sense, $r_{ms}=M$. In the
process the mass of the black hole is trebled (and it is $22\%$ over the
initial mass when it passes through the Schwarzschild state). Further
accretion increases only the mass, not the dimensionless spin.}
\label{Fig_accretion}
\end{figure}

The numerical analysis of the system (\ref{aux1}), (\ref{aux2}) shows that
counterrotating particles first reduce fast the rotation parameter to zero
(the mass gain being $22\%$), after which the rotation is again increased by
the corotating accretion up to the maximally rotating, extreme Kerr limit
(the mass being trebled). This is shown on Fig. \ref{Fig_accretion}. 
This evolution is slightly changed by the various effects considered in the
present paper.

\subsubsection{Symbiotic model}

In the generic case we take into account a) the torque $d\Delta L^{D}/dt$,
b) the extracted power $P^{D}$ due to the the magnetic coupling of the disk
and the hole, c) the loss of angular momentum $d\Delta L^{H}/dt$, and d) the
energy loss $P^{H}$ due to the BZ mechanism. For the time averaged rates $%
(dJ/dt)_{rad}$ and $(dM/dt)_{rad}$ we use the flux integral (\ref{ftot})
 and write the evolution equations in the form
\begin{eqnarray}
dM/dt &=&\dot{M}_{0}\widetilde{E}_{ms}+(dM/dt)_{rad}  \notag \\
&&-P^{H}-P^{D}\;,  \label{dMdt2} \\
dJ/dt &=&\dot{M}_{0}\widetilde{L}_{ms}+(dJ/dt)_{rad}  \notag \\
&&+d\Delta {L}^{H}/dt+d\Delta {L}^{D}/dt\;.  \label{dJdt2}
\end{eqnarray}%
Similar evolution equations were
given by Park \& Vishniac (1988) and Wang, Xiao \& Lei (2002). However Park
\& Vishniac considered a black hole magnetosphere with only open field
lines; while the photon capture was ignored by both Park \& Vishniac and
Wang, Xiao \& Lei. We obtain the evolution equations 
\begin{eqnarray}
\frac{da_{\ast }}{d\ln M} &=&\frac{1}{M}\frac{\widetilde{L}_{ms}+N_{J}}{%
\widetilde{E}_{ms}+N_{M}}-2a_{\ast }\;,  \label{dastardlnM2} \\
\frac{dM}{dM_{0}} &=&\widetilde{E}_{ms}+N_{M}\;,  \label{dMdM03}
\end{eqnarray}%
with 
\begin{eqnarray}
N_{J} &\equiv &\dot{M}_{0}^{-1}[(dJ/dt)_{rad}+d\Delta {L}^{H}/dt+d\Delta {L}%
^{D}/dt]\;, \\
N_{M} &\equiv &\dot{M}_{0}^{-1}[(dM/dt)_{rad}-P^{H}-P^{D}]\;.
\end{eqnarray}%
These equations contain the jet parameter $r_{in}$, and the parameters $n$, $%
\theta _{max}$ of the magnetosphere. The spin limit of the generalized
equation (\ref{dastardlnM2}) is independent of the mass scale and the
accretion rate as in the case of the evolution equation (\ref{dastardlnM})
of $a_{\ast }$. This can be concluded from the relations $d\Delta
L^{H}/dt\sim \dot{M}_{0}M$, $d\Delta L^{D}/dt\sim \dot{M}_{0}M$, $P^{H}\sim 
\dot{M}_{0}$ and $P^{D}\sim \dot{M}_{0}$ which give $N_{J}\sim M$ and render 
$N_{M}$ independent of $M$ and $\dot{M}_{0}$.

\subsection{Radiative truncation of the disk}

We present the evolution of the black hole mass and spin in the
disk-jet-hole symbiotic models without magnetosphere in Fig. \ref%
{a_M_M0_r_jet}, for the sequence of nozzle radii $r_{in}=4M$, $8M$ and $12M$%
, both for accretion disks with isotropic emission of radiation and electron
scattering atmosphere.

For zero initial value of $a_{\ast }$ the mass accretion with $0\leq
(M_{0}-M_{0i})/M_{i0}\lesssim $ $1.3$ increases both the total mass and the
spin of the black hole and the mass evolution is essentially independent of
the photon capture. Here $M_{0}$ represents the rest mass of the black hole,
which includes the accreted mass. For $(M_{0}-M_{0i})/M_{i0}\gtrsim 1.3$ the
black hole has already reached the spin limit and thus can gain no
additional rotational energy by accretion. However, the accreting rest mass
further increases the total mass $M$ of the black hole. At the spin limit $%
r_{ms}$ and $\widetilde{E}_{ms}$ reach minimal values, as seen in Fig. \ref%
{r_ms} and Eq. (\ref{tildeE}), therefore the slope of the upper part of the
mass evolution curves decrease. The photon capture $(dM/dt)_{rad}$ is
becoming less important with increasing the truncation radius of the
radiation from the disk and the slope of the curve will further decrease.

If the initial value of the spin parameter is $1$, its value will change
only slightly until the limit is reached. The radius $r_{ms}$ and specific
energy $\tilde{E}_{ms}$ also decrease somewhat before reaching their
limiting values and the photon capture becomes more important. Thus in the
mass regime $(M_{0}-M_{0i})/M_{i0}\gtrsim 1.3$ all evolution curves of the
mass for $a_{\ast i}=1$ have identical slopes with the corresponding (same $%
r_{in}$) curves of the initially non-rotating configurations. 
\begin{figure*}
\centering
\includegraphics[width=.48\textwidth]{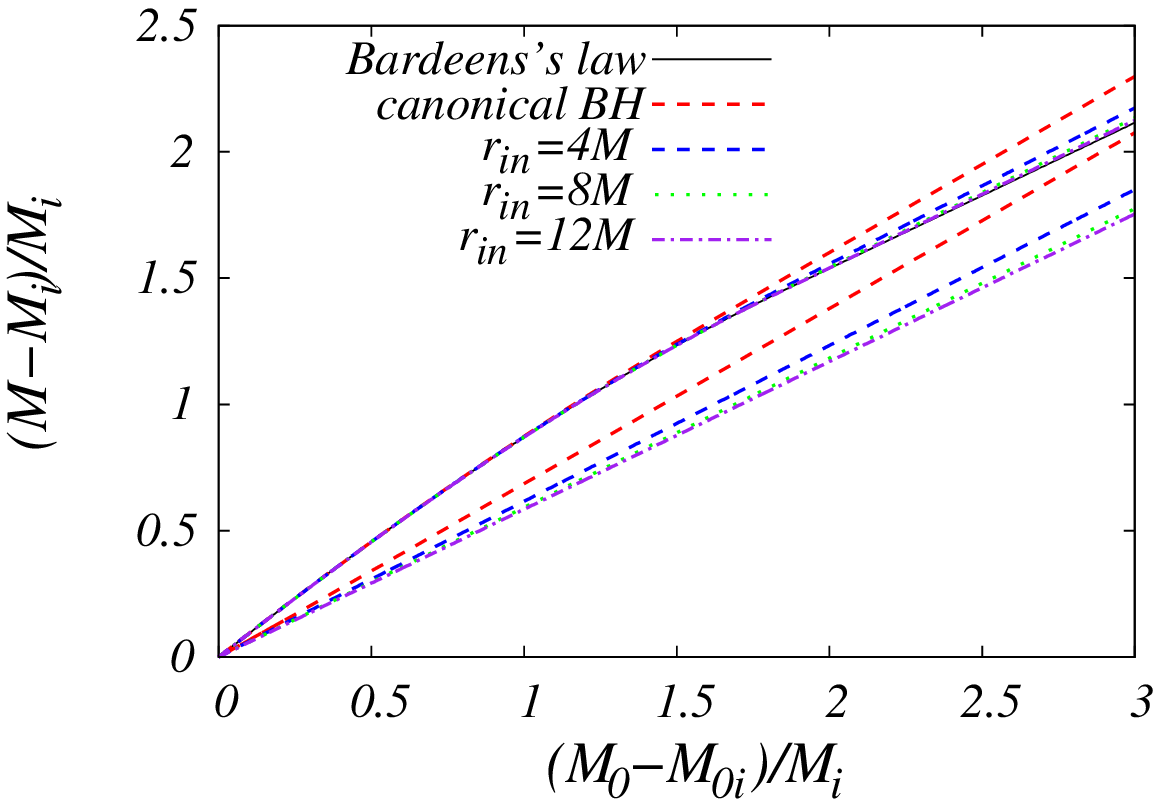} %
\includegraphics[width=.48\textwidth]{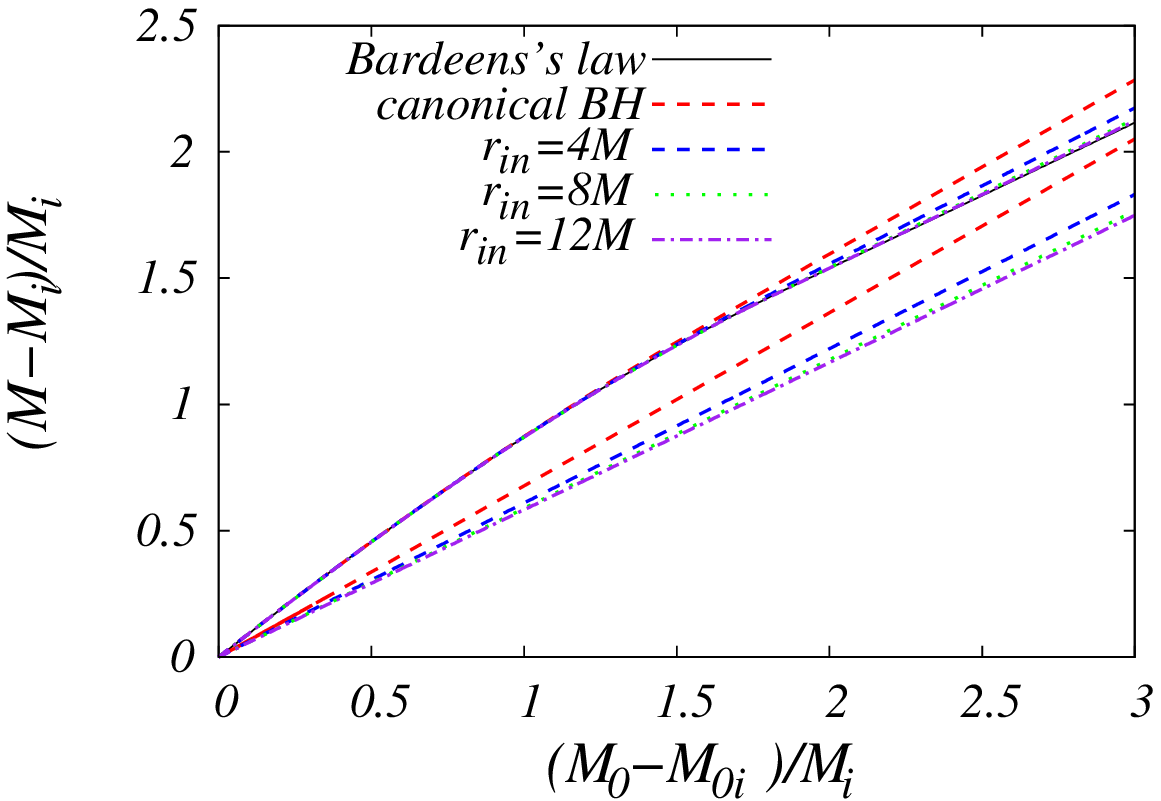}\newline
\includegraphics[width=.48\textwidth]{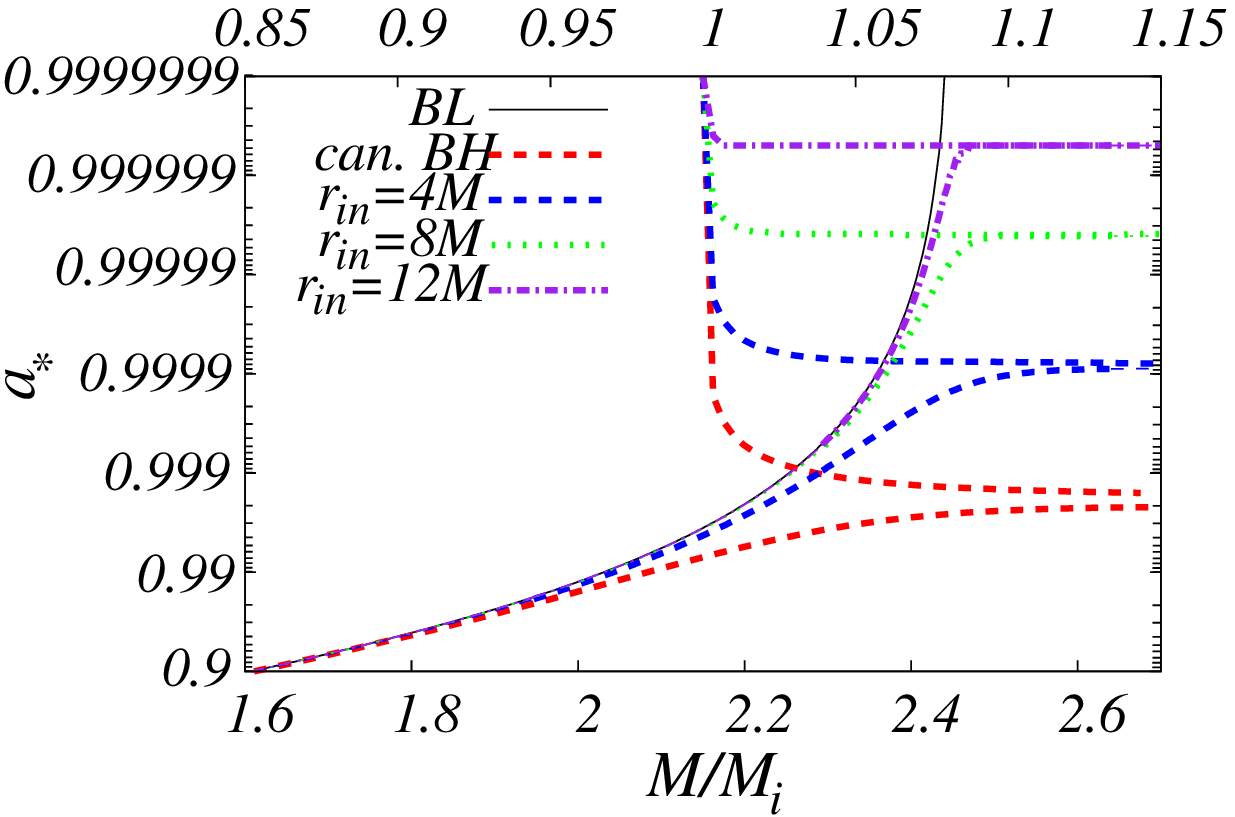} %
\includegraphics[width=.48\textwidth]{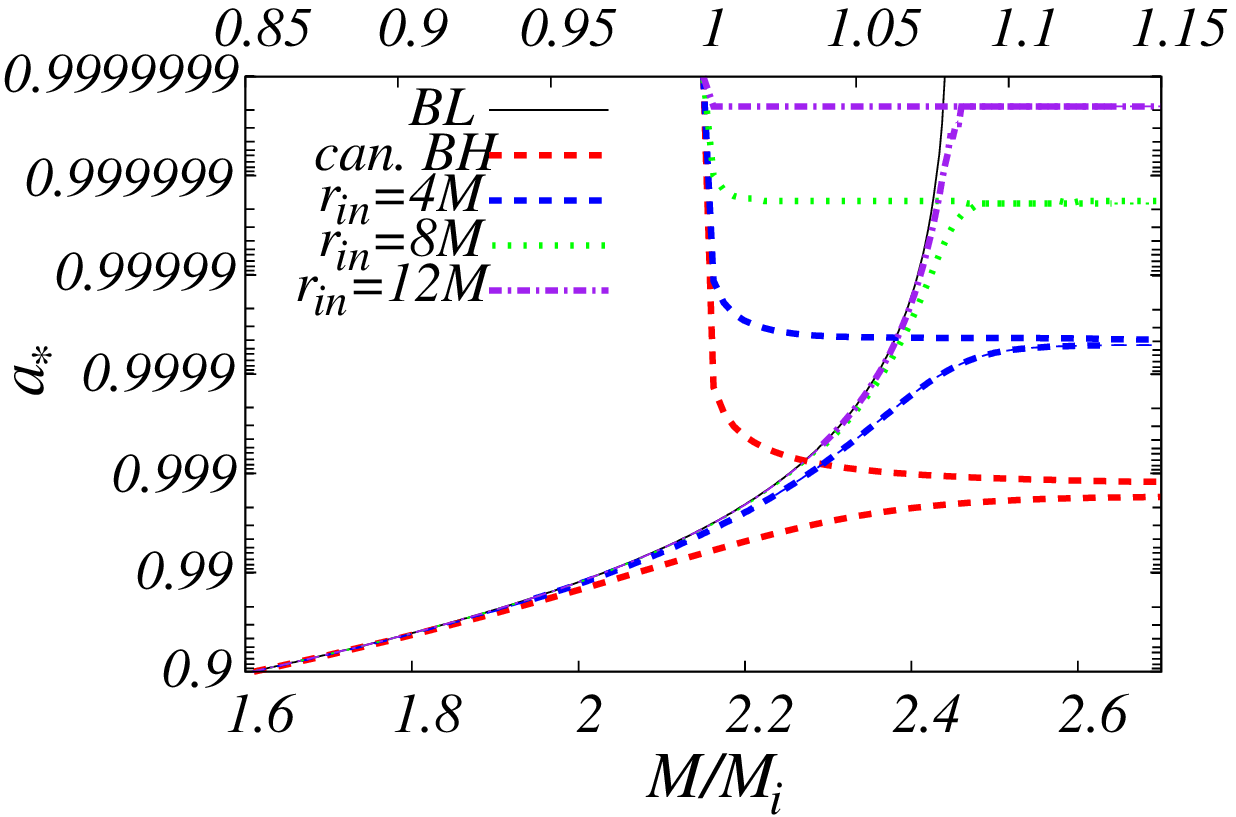}
\caption{The effect of the jets (of radiative truncation) on the evolutions
of the total mass-energy $M$ (upper plots) and spin parameter $a_{\ast }$
(lower plots) in the absence of the magnetosphere (generalization of Figs. 1
and 2 of Thorne 1974). $M_{i}$ and $M_{i0}$ are inital black hole total and
rest masses. Left panel models are with isotropic disk emission; right panel
models are with electron scattering atmosphere. Black curves represent
Bardeen evolutions (no photon capture, initially non-rotating black holes).
The two limiting cases of canonical black holes (allowing accretion and
photon capture) are the dashed red curves (lower curve on mass evolution:
initially extreme Kerr $a_{\ast i}=1$; upper curve on mass evolution:
initially non-rotating $a_{\ast i}=0$ ). Similar evolutions are depicted for
the radiative truncation radii $r_{in}=4M$, $8M$ and $12M$. All evolutions
with $a_{\ast i}=0$ exhibit an initially faster increase of mass than in the
Bardeen case, while mass increase for $a_{\ast i}=1$ is slower. The rotation
parameter on the vertical axis is represented on the $log_{10}(0.1/(1-a_{%
\ast }))$ scale; there are two mass scales indicated on the horizontal axis,
one for each of the initially non-rotating and extreme Kerr configurations.
The jets (the truncation of radiation) renders the spin limit closer to the
extreme Kerr case. }
\label{a_M_M0_r_jet}
\end{figure*}

The spin evolution is also represented in Fig. \ref{a_M_M0_r_jet}. The
surface of the disk and the corresponding energy flux decrease by increasing
the radiative truncation radius. $(dM/dt)_{rad}$ becoming smaller, renders
the evolution of the total black hole mass more close to Bardeen's law. For
the highest radiative truncation radius considered, $r_{in}=12M$ the spin
limit is almost $1$, similarly to the extreme Kerr holes without photon
capture. This is an indication that only a negligible number of photons is
captured by the hole, or otherwise stated, confirms the common-sense
expectation, that the majority of the captured photons are emitted quite
close to the gravitational radius. The limiting values of $a_{\ast }$ are
given in Table \ref{q_m_r_jet} for both the isotropic and electron
scattering models. 
\begin{table}
\caption{The upper limit of the spin parameter $a_{\ast }$ and the
efficiency $\protect\epsilon $ for a symbiotic model with a radiative
truncation radius $r_{in}$\textbf{.} The values refer to isotropic emission
(I) and electron scattering atmosphere (ES).}
\label{q_m_r_jet}
\begin{center}
\begin{tabular}{lccccc}
\hline
$r_{in}$ [$M$] & $a_{\ast }$ (I) & $\epsilon $ (I) & $a_{\ast }$ (ES) & $%
\epsilon $ (ES) &  \\ \hline
4 & 0.9999106 & 0.383 & 0.9999480 & 0.389 &  \\ 
8 & 0.9999959 & 0.408 & 0.9999981 & 0.411 &  \\ 
12 & 0.9999995 & 0.415 & 0.9999998 & 0.417 &  \\ 
&  &  &  &  & 
\end{tabular}%
\end{center}
\end{table}

The efficiency for converting accreted mass into outgoing radiation is 
\begin{eqnarray}
\epsilon &=&1-\dot{M}_{0}^{-1}dM/dt  \notag \\
&=&1-\widetilde{E}_{ms}-\dot{M}_{0}^{-1}(dM/dt)_{rad}\;.  \label{eps}
\end{eqnarray}%
Canonical black holes for isotropic emission and electron scattering
atmosphere have $\epsilon $\thinspace $=0.302$ and $\epsilon $\thinspace $%
=0.308$, respectively (Thorne 1974). In Table \ref{q_m_r_jet} we give the
values of $\epsilon $, considerably increased due to the jet. For $r_{in}=4M$
\ the efficiency is close to $0.400$, the value characterizing extreme Kerr
black holes with photon capture. The explanation for this is that the
specific energy $\widetilde{E}_{ms}$ in Eq. (\ref{eps}) is evaluated at a
high spin limit. Increasing the radiative truncation radius $\epsilon $
exceeds $0.400$, approaching $0.423$, which is the efficiency of the extreme
black Kerr hole without photon capture.

\subsection{Magnetosphere models}

In what follows we study the spin evolution for standard accretion disk
models in the presence of a black hole magnetosphere. If the inner edge of
disk is always at the marginally stable orbit then the only effects on the
mass accretion are caused by the BZ mechanism and the torque conveyed from
the hole to the disk via the magnetic field lines. These two effects have
been studied separately by Park \& Vishniac (1988) and Wang, Xiao \& Lei
(2002), respectively. We investigate numerically their combined effect and
concentrate on the high spin regime.

In the evolution equations (\ref{dastardlnM2}) and (\ref{dMdM03}) the terms $%
d\Delta L^{H}/dt$, $d\Delta L^{D}/dt$, $P^{H}$ and $P^{D}$ of the torque and
the extracted power are given by Eqs. (\ref{dLHdtanal}), (\ref{Panal}), (\ref%
{dLDdt2}) and (\ref{PD2}) and the terms $(dJ/dt)_{rad}$ and $(dM/dt)_{rad}$
are determined by applying the flux integral (\ref{ftotmc}). The efficiency
for the conversion of the accreting mass into outgoing radiation defined by
Eq. (\ref{eps}) takes the form 
\begin{equation}
\epsilon =1-\widetilde{E}_{ms}-N_{M}\;.  \label{eps2}
\end{equation}

\subsubsection{Open magnetic field lines from the horizon}

\begin{figure*}
\centering
\includegraphics[width=.48\textwidth]{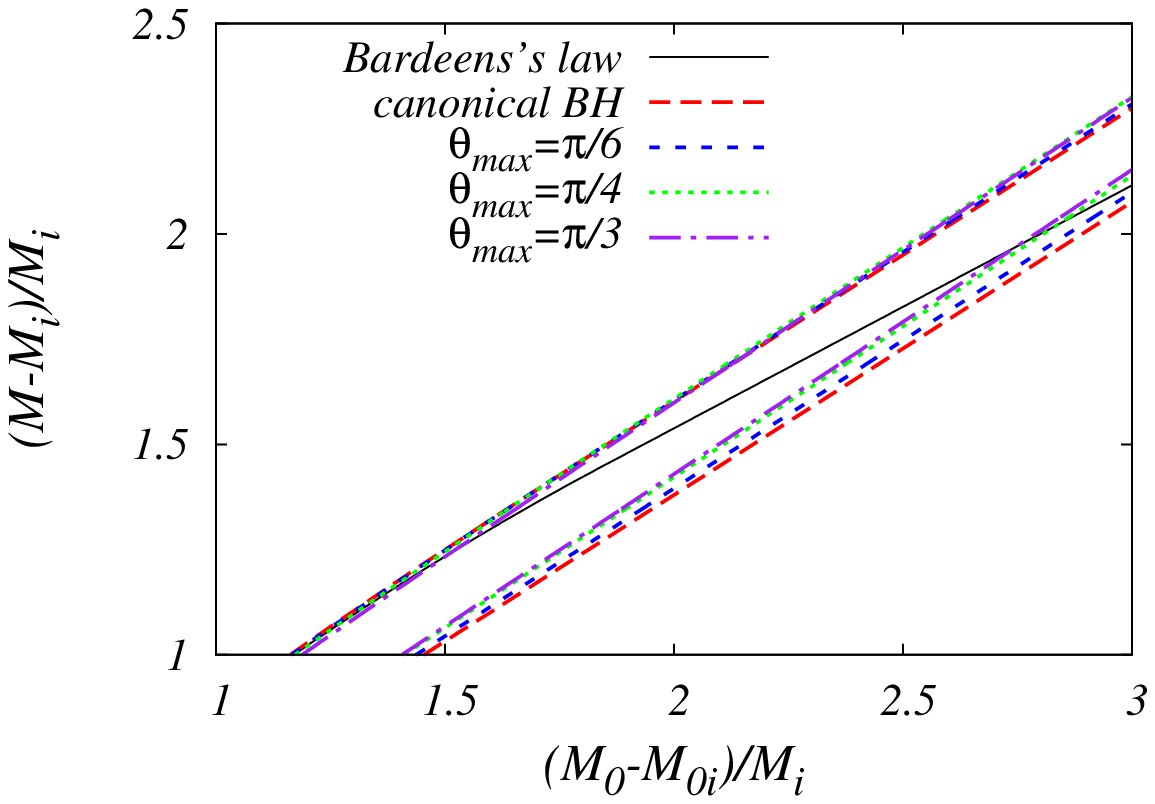} %
\includegraphics[width=.48\textwidth]{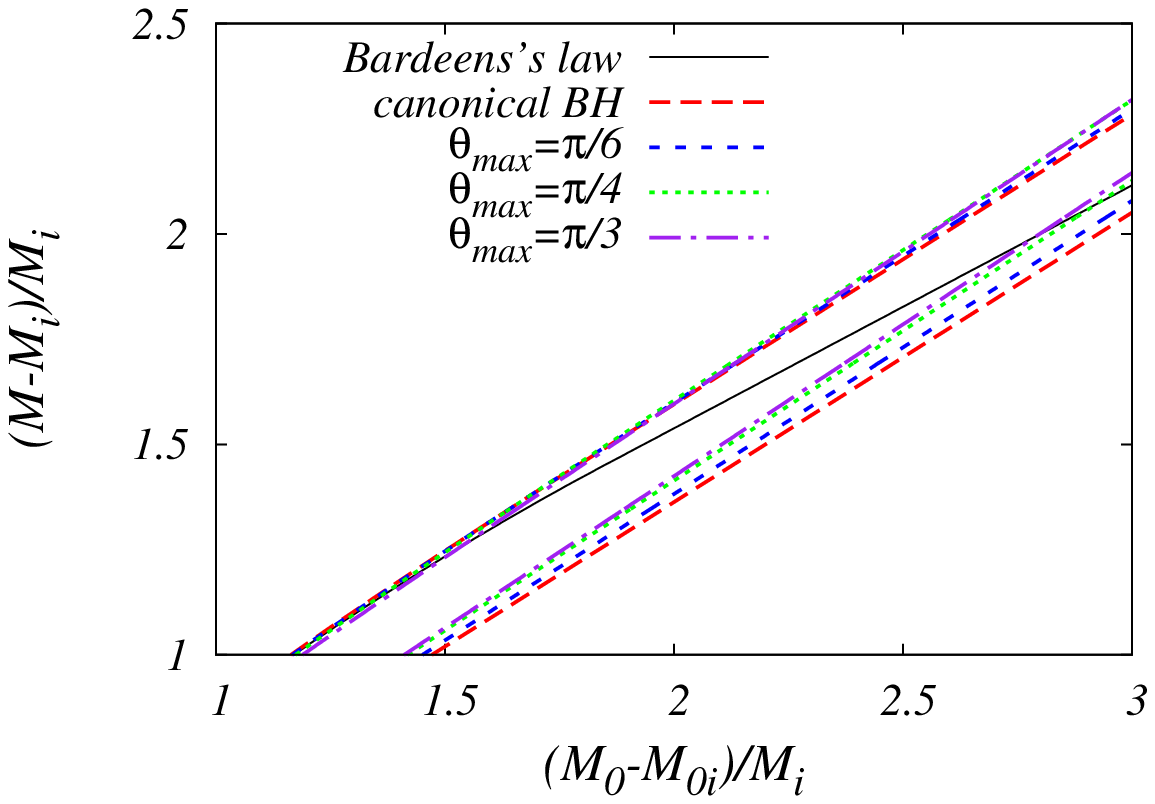}\newline
\includegraphics[width=.48\textwidth]{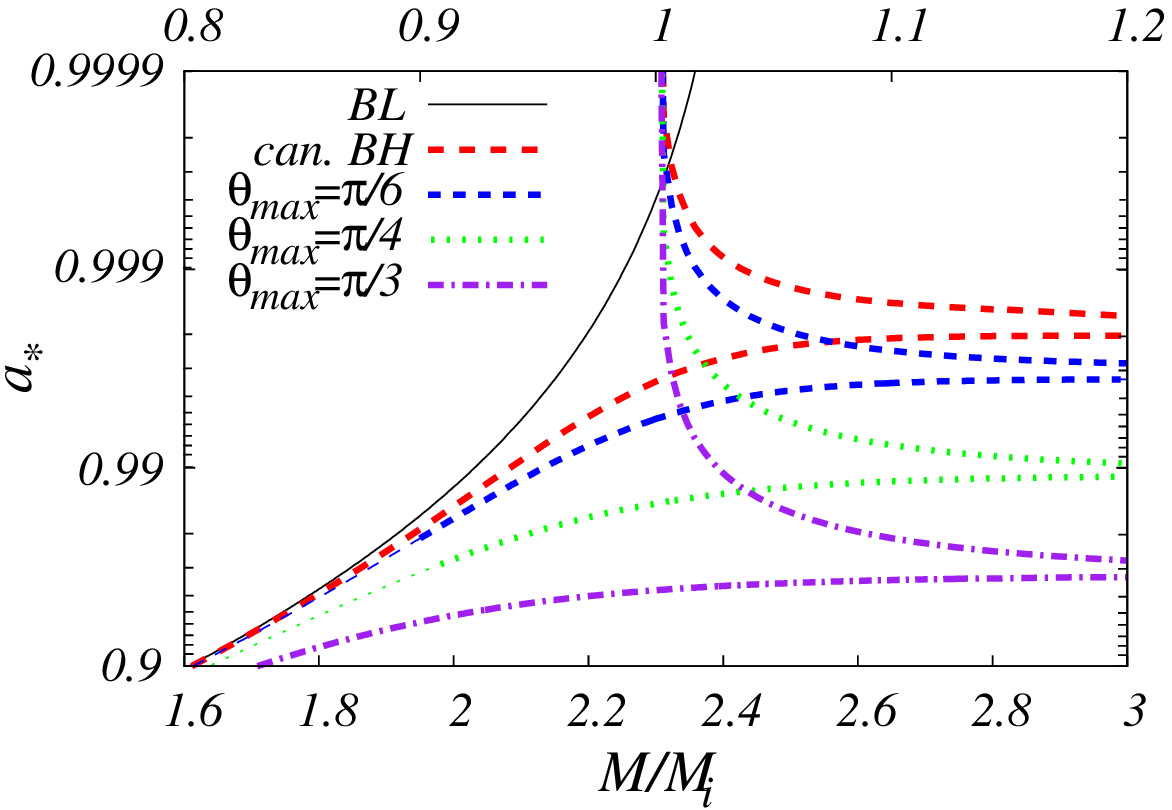} %
\includegraphics[width=.48\textwidth]{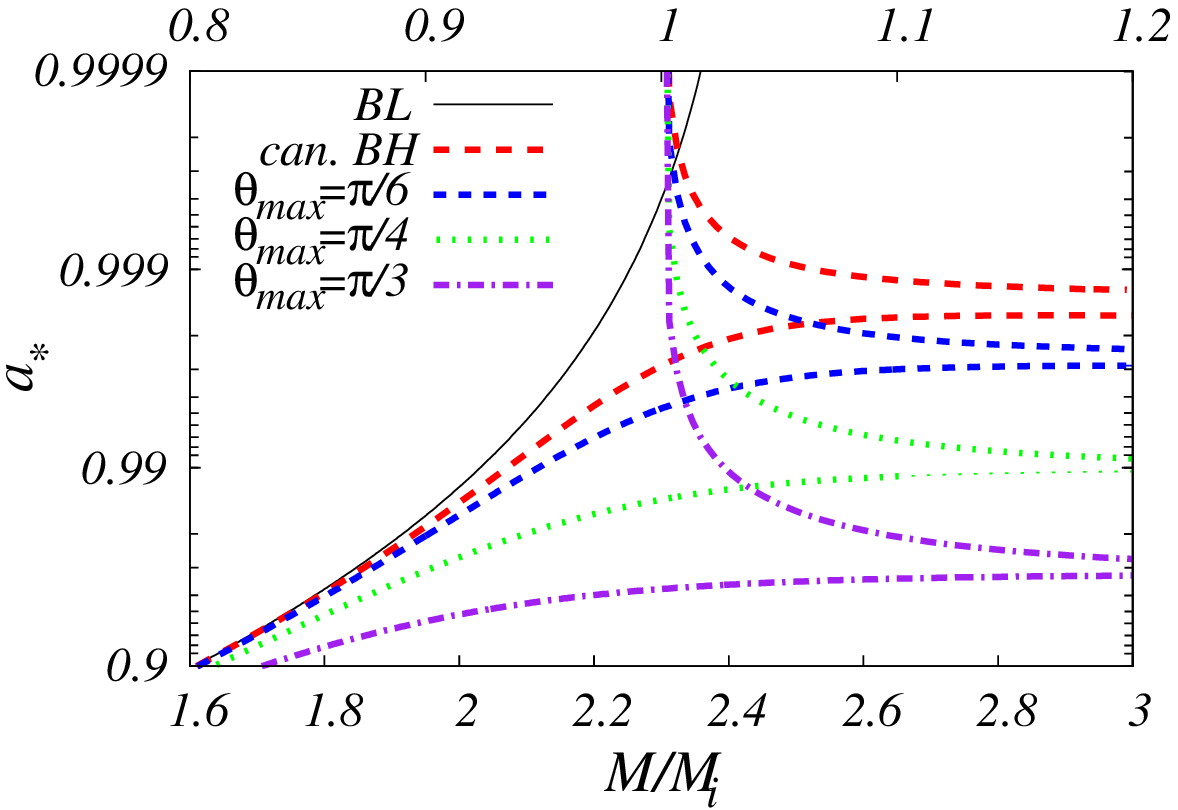}
\caption{The evolution of the total mass-energy $M$ (upper plots) and spin
parameter $a_{\ast }$ (lower plots) for a black hole with the inclusion of
the BZ mechanism. Left hand panels show $M$ and $a_{\ast }$ for isotropic
disk emission; right hand panels display the same plots for the electron
scattering atmosphere.}
\label{a_M_BZ}
\end{figure*}

If we consider the BZ mechanism and no closed magnetic field lines ($d\Delta
L^{D}/dt=0$ and $P^{D}=0$), the only parameter left is the boundary angle $%
\theta _{max}$ of the open lines on the event horizon. By increasing $\theta
_{max}$ we also increase the total area of the flux tube formed by the open
field lines on the horizon. As a result, the torque exerted on the flux tube
by the black hole raises and more rotational energy is extracted from the
hole in the BZ process. This reduces the spin limit, which is presented in
Table \ref{theta_max} and in Fig. \ref{a_M_BZ}. The extracted power $P^{H}$
takes larger values when we increase the horizontal area of the open field
lines even at smaller values of the spin limit. For lower values of spin
parameter the mass-energy flux $(dM/dt)_{rad}$ becomes smaller as well and
these effects decrease the term $N_{M}$ in the mass evolution equation.
Nevertheless, it is compensated by the higher values of $\widetilde{E}_{ms}$
taken at the lower spin limits. The net effect is that the slopes of the
mass evolution curves are steeper with the increasing boundary angle $\theta
_{max}$ and more total mass is produced by the same amount of the accreted
rest mass, as seen in Fig. \ref{a_M_BZ}. The efficiency $\epsilon $ given in
Table \ref{theta_max} also exhibits a behaviour similar to the one of the
spin limit. With the increasing area of the flux tube formed by the open
magnetic field lines the BZ process reduces the conversion efficiency of the
accreted mass into radiation. Park \& Vishniac (1988) applying a maximally
efficient BZ mechanism and a constant accretion rate in their analysis
demonstrated that $a_{\ast }$ is decreasing in time and the black hole will
spin down from the initial value $a_{\ast i}\approx 1$. Comparing their
findings with our result, we can state that the extremely rapidly rotating
black holes with $a_{\ast }\approx 1$ will slow down to the spin limits
shown in Table \ref{theta_max} for the given boundary angles $\theta _{max}$%
. 
\begin{table}
\caption{The limit of the spin parameter $a_{\ast }$ and the efficiency $%
\protect\epsilon $ of the accretion process in the presence of open magnetic
field lines (BZ mechanism) for different boundary angles $\protect\theta %
_{max}$.}
\label{theta_max}
\begin{center}
\begin{tabular}{lcccc}
\hline
$\theta_{max}$ & $a_{*}$ (I) & $\epsilon$ (I) & $a_{*}$ (ES) & $\epsilon$
(ES) \\ \hline
$\pi/6$ & 0.9964 & 0.300 & 0.9969 & 0.307 \\ 
$\pi/4$ & 0.9891 & 0.286 & 0.9897 & 0.290 \\ 
$\pi/3$ & 0.9651 & 0.276 & 0.9657 & 0.281 \\ 
&  &  &  & 
\end{tabular}%
\end{center}
\end{table}

\subsubsection{Closed magnetic field lines}

\begin{figure*}
\centering
\includegraphics[width=.48\textwidth]{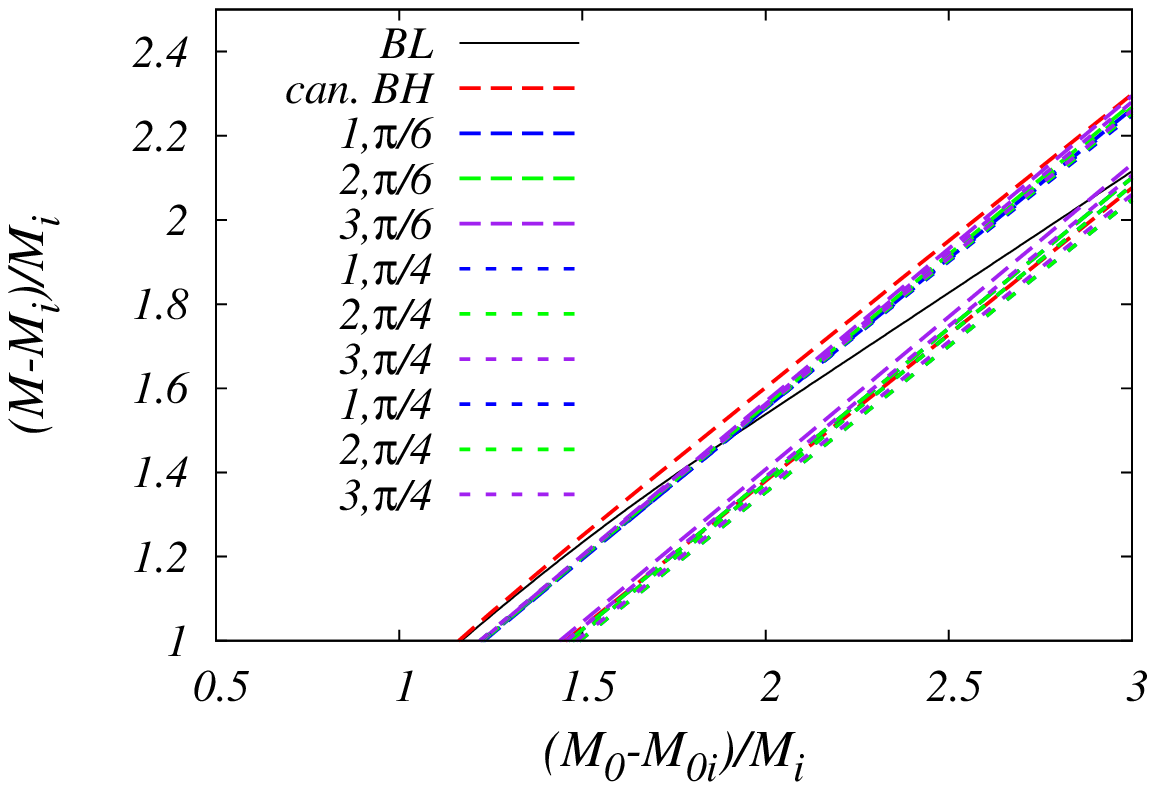} %
\includegraphics[width=.48\textwidth]{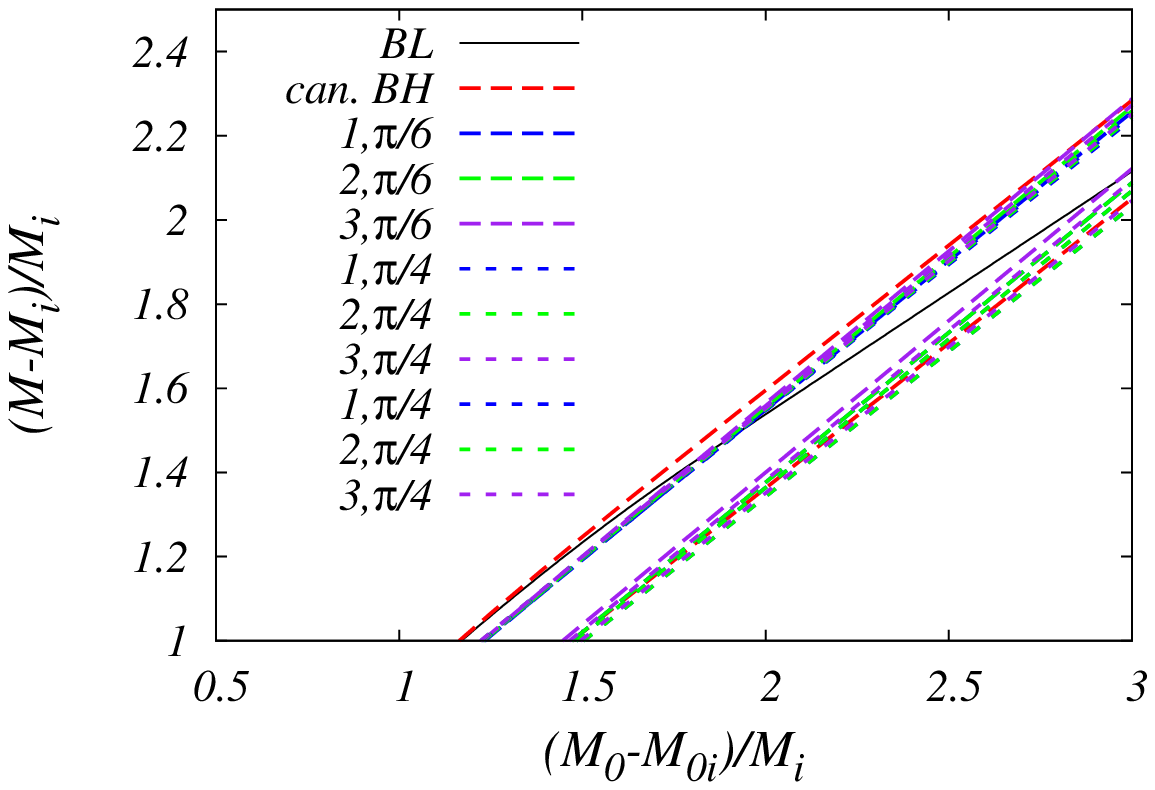}\newline
\includegraphics[width=.48\textwidth]{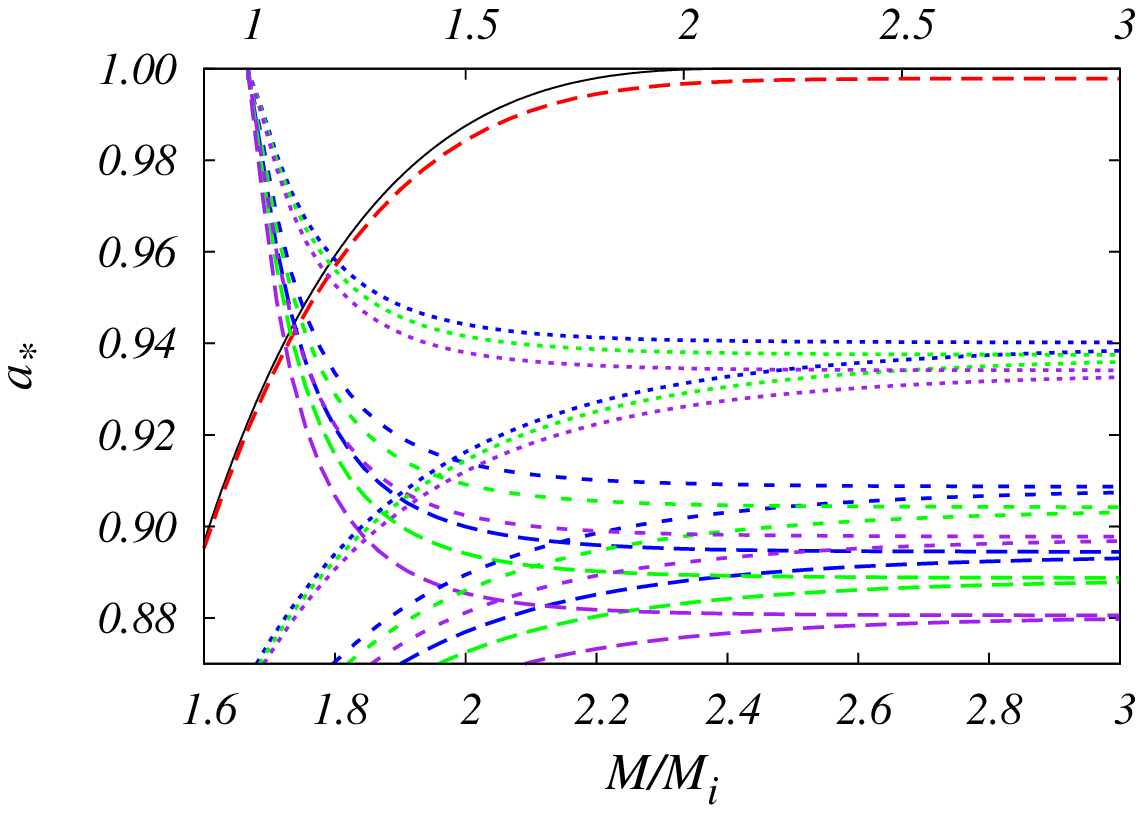} %
\includegraphics[width=.48\textwidth]{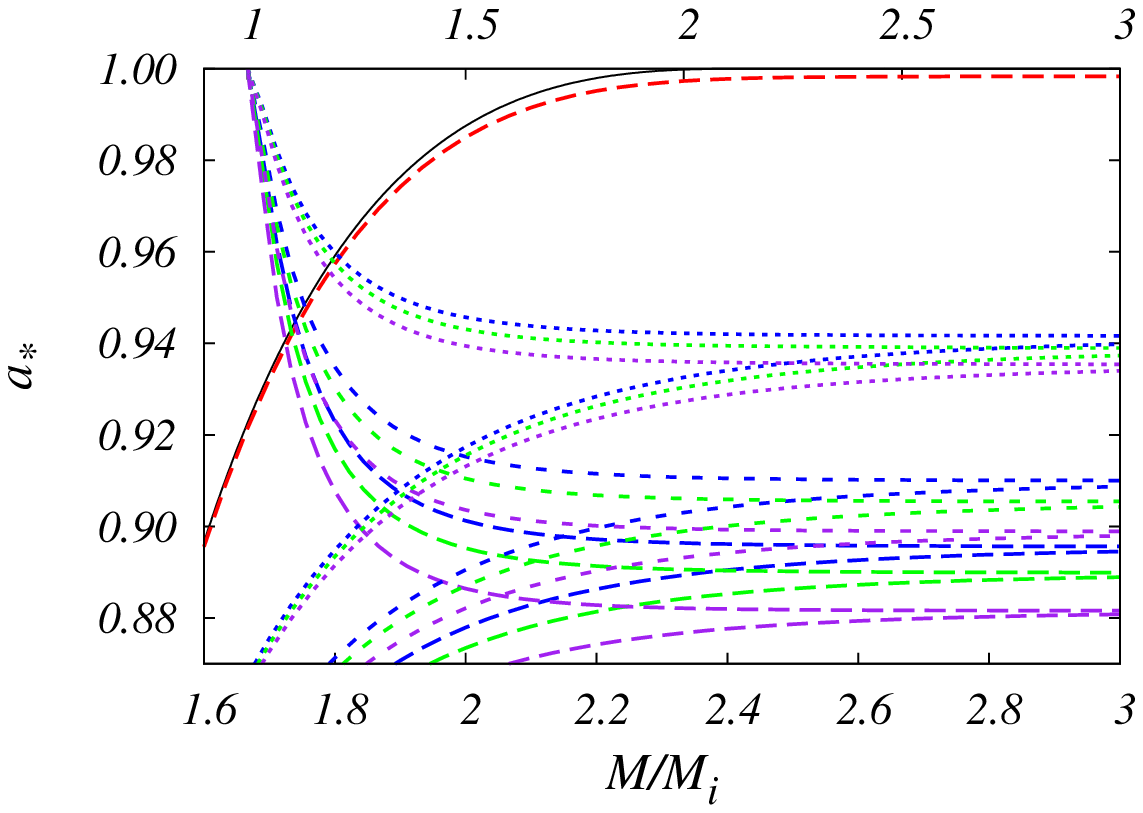}
\caption{The evolution of the the total mass-energy $M$ (upper plots) and
the spin parameter $a_{\ast }$ (lower plots) for a black hole, with the
inclusion of the magnetic torque exerted on the disk by the hole. The
integer numbers refer to $n$, the angles to $\protect\theta _{\max }$. Left
hand panels show $M$ and $a_{\ast }$ for isotropic disk emission, right hand
ones display the same plots for electron scattering atmosphere. Linestyles
in the lower plots are as for the upper plots. }
\label{a_M_mc}
\end{figure*}

By switching the BZ process off ($d\Delta L^{H}/dt=0$ and $P^{H}=0$), we can
study the effects of the magnetic torque exerted on the accretion disk by
the black hole. The area of flux tube formed on the horizon by the closed
field lines connecting the disk to the hole is determined by the angle $%
\theta _{max}$ again but it is now decreasing with the increasing boundary
angle. The effect of the magnetic coupling between the hole and the disk
becomes smaller for greater values of $\theta _{max}$, as opposed to the BZ
process. This can be seen in Table \ref{theta_max_na} containing the spin
limits and the efficiency for different values of $\theta _{max}$ and $n$.
As we increase $\theta _{max}$ for a fixed $n$ we obtain higher spin limits
approaching the limiting values of the canonical black holes. The highest
values derived for $\theta _{max}=\pi /3$ are still somewhat smaller than
0.9978 (isotropic emission) and 0.9982 (electron scattering atmosphere). The
efficiency also follows the same tendency: for $\theta _{max}=\pi /6$ the
efficiency $\epsilon $ is smaller than 0.302 (isotropic emission) and 0.308
(electron scattering atmosphere) obtained for the standard accretion disks
but approaches and somewhat exceeds those values with increasing $\theta
_{max}$. Although for a fixed $\theta _{max}$ the maximal flux emitted by
the disk and the number of the captured photons are reduced if we increase
the exponent $n$, the quantities $d\Delta L^{D}/dt$ and $P^{D}$ compensate
the diminishing effects of $(dJ/dt)_{rad}$ and $(dM/dt)_{rad}$. As a result,
the spin limits decrease for higher values of $n$, deviating more from those
of the canonical black hole. Fig. \ref{a_M_mc} presenting the mass and spin
evolution of the black hole demonstrates this kind of dependence of the
limiting values on the parameters of the magnetic fields. For $\theta
_{max}=\pi /6$ and $n=3$ the mass evolution is faster than in the case of
the canonical black hole because of the relatively high value of $\widetilde{%
E}_{ms}$ at the lower spin limit but the magnetic coupling of the hole and
the disk becomes more dominant with decreasing boundary angle and reduces
the mass evolution rate of the hole. Any decrease in the exponent $n$ also
causes a reduction in the mass evolution rate. Wang, Xio \& Lei (2002, 2003)
already analyzed the black hole evolution in the magnetically coupled model
for the parameters $0<\theta _{max}<\pi /2$ and $1<n<6$. They obtained the
same trends of the spin limit by varying $n$ for fixed values of $\theta
_{max}$, even if they neglected the photon capture effect decreasing the
limiting value of $a_{\ast }$. 
\begin{table}
\caption{The spin limits $a_{\ast }$ and the efficiency $\protect\epsilon $
for the accretion process in black hole magnetosphere with closed field
lines for different boundary angles $\protect\theta _{max}$ and magnetic
field strength power law exponents $n$.}
\label{theta_max_na}
\begin{center}
\begin{tabular}{lccccc}
\hline
$\theta _{max}$ & $n$ & $a_{\ast }$ (I) & $\epsilon $ (I) & $a_{\ast }$ (ES)
& $\epsilon $ (ES) \\ \hline
$\pi /6$ & 1 & 0.8944 & 0.291 & 0.8956 & 0.294 \\ 
& 2 & 0.8887 & 0.284 & 0.8899 & 0.288 \\ 
& 3 & 0.8807 & 0.275 & 0.8817 & 0.278 \\ \hline
$\pi /4$ & 1 & 0.9086 & 0.296 & 0.9100 & 0.300 \\ 
& 2 & 0.9042 & 0.291 & 0.9053 & 0.295 \\ 
& 3 & 0.8980 & 0.284 & 0.8991 & 0.287 \\ \hline
$\pi /3$ & 1 & 0.9400 & 0.308 & 0.9416 & 0.313 \\ 
& 2 & 0.9374 & 0.305 & 0.9389 & 0.309 \\ 
& 3 & 0.9342 & 0.301 & 0.9356 & 0.305%
\end{tabular}%
\end{center}
\end{table}

\subsubsection{Combined open and closed magnetic field line topology}

\begin{figure*}
\centering
\includegraphics[width=.48\textwidth]{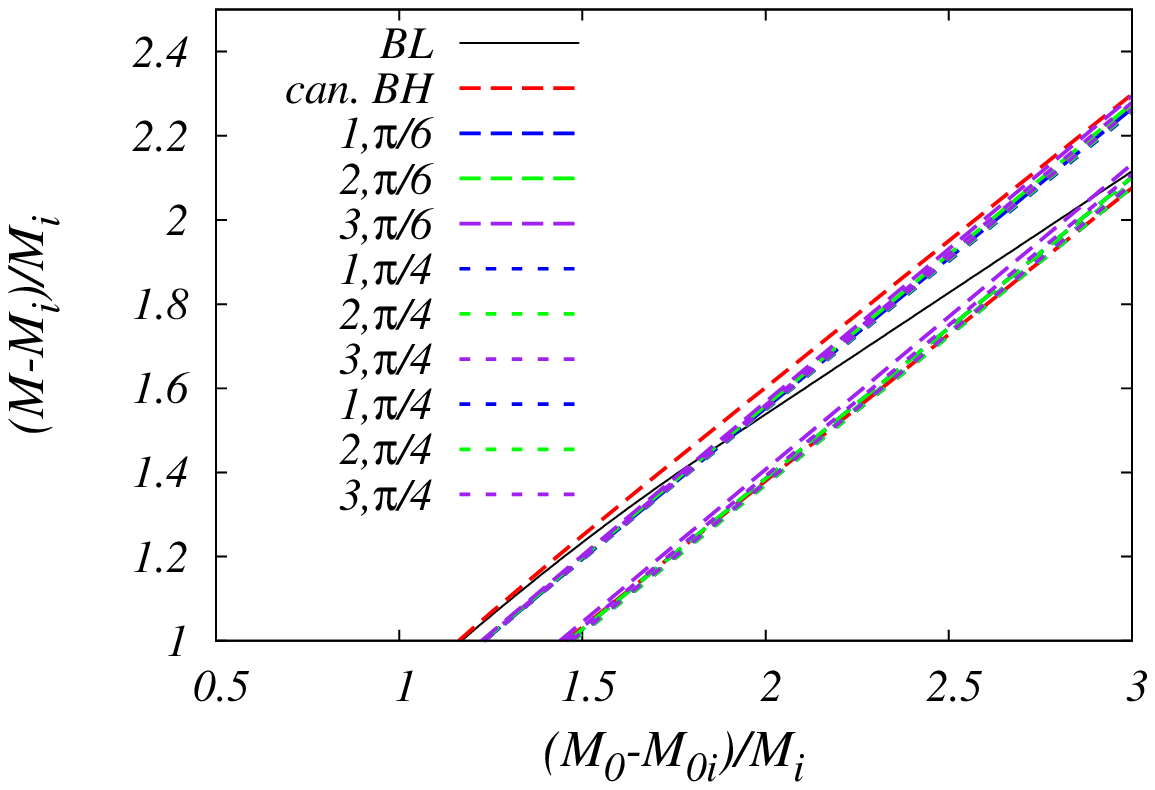} %
\includegraphics[width=.48\textwidth]{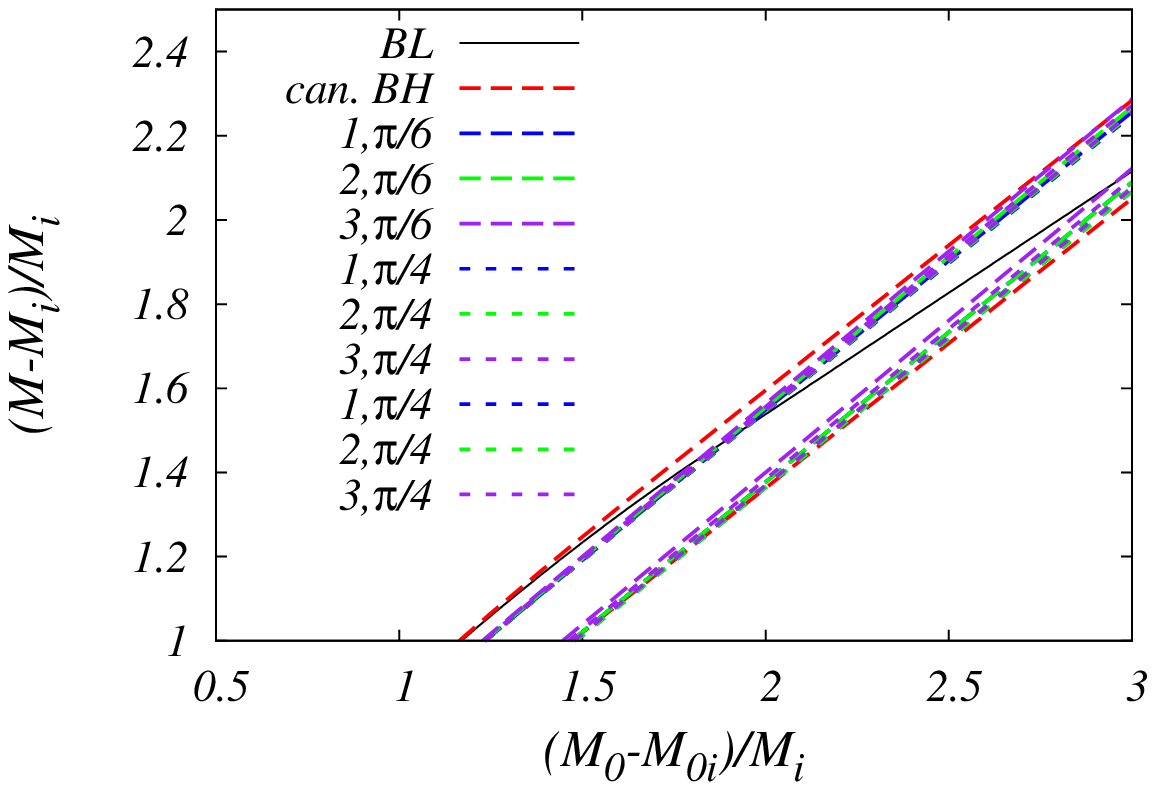}\newline
\includegraphics[width=.48\textwidth]{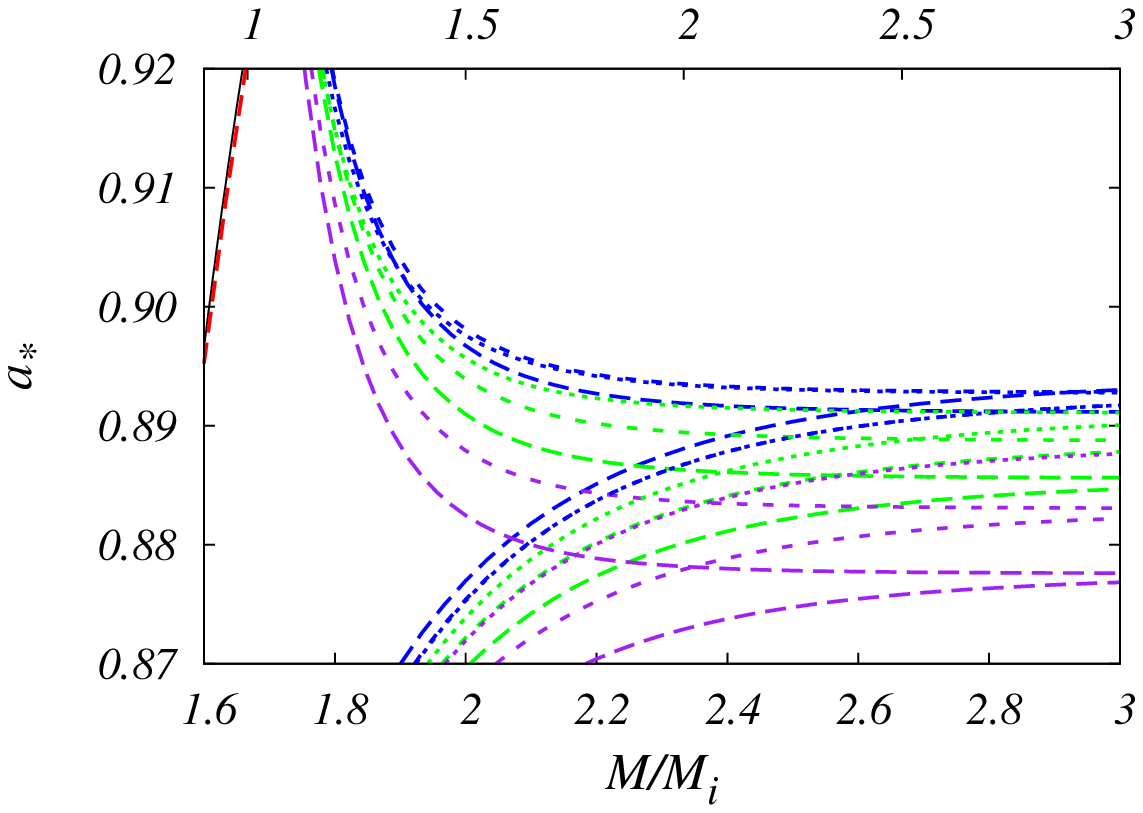} %
\includegraphics[width=.48\textwidth]{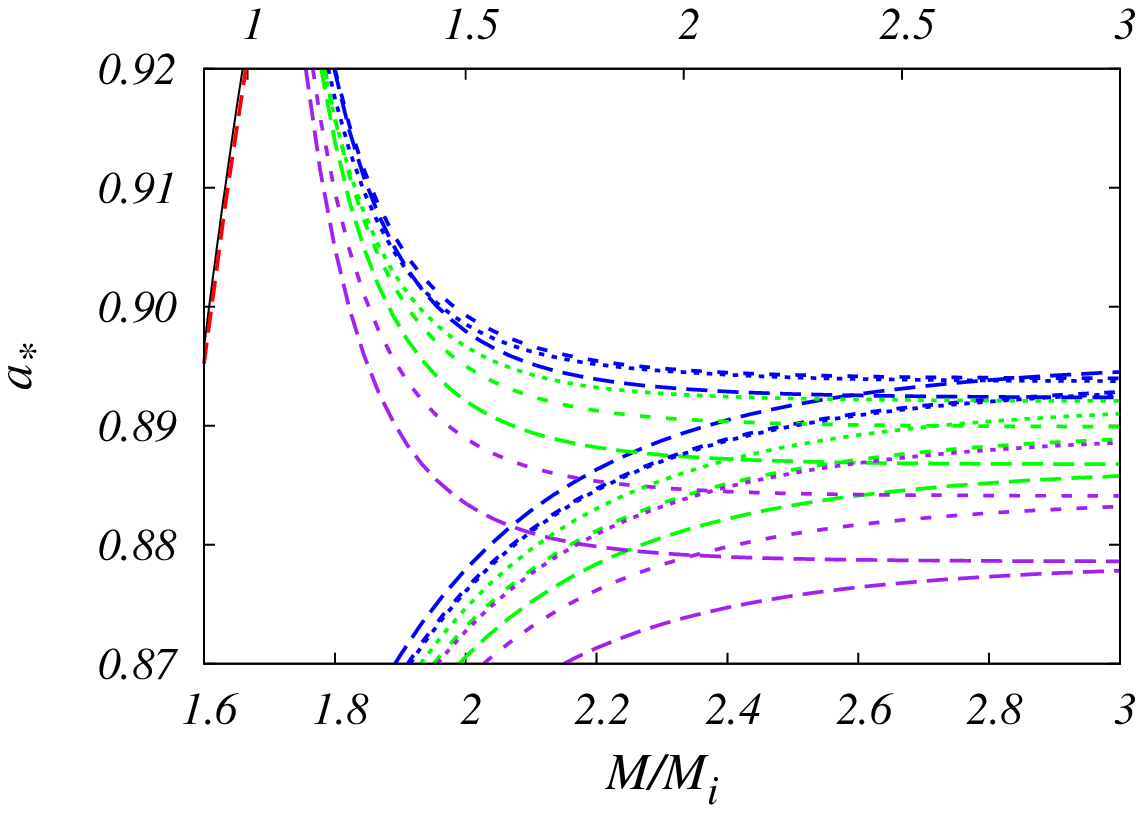}
\caption{The evolution of the the total mass-energy $M$ (the upper plots)
and the spin parameter $a_{\ast }$ (the lower plots) for a black hole with
the inclusion of the BZ mechanism and the magnetic torque exerted on the
disk by the hole. Left panel has $M$ and $a_{\ast }$ for an isotropic disk
emission; right panel for the electron scattering atmosphere. Linestyles in
the lower plots are as for the upper plots. }
\label{a_M_mc_BZ}
\end{figure*}

A complete black hole magnetosphere model contains both the open and the
close flux lines emanating from the event horizon. Hence both the BZ
mechanism and the torque produced by the hole on flux lines anchored in the
accretion disk affect on the evolution characteristics of the rotating black
hole. In the second part of Table \ref{theta_max_nb} we present the spin
limits and the efficiency for the combined effect of the BZ mechanism and
the magnetic torque. Here we see that the limiting values of $a_{\ast }$
deviate more from the values obtained for canonical black holes than without
the BZ mechanism included. Since the variation of $\theta _{max}$ increases
the area of the flux tube with open (closed) field line topology and the
same time decreases the area of the flux tube with the closed (open) one,
the separation angle $\theta _{max}$ determines if the BZ process or the
magnetic coupling of the black hole to the disk has a more dominant role.
Therefore, the BZ mechanism produces only a weak effect for $\theta
_{max}=\pi /6$ and a relatively small decrease in the spin limit compared
with the values for the magnetic coupling. For $\theta _{max}=\pi /3$ the
contribution of the BZ mechanism is stronger and so is the reduction in the
limiting values of the spin. An increasing parameter $n$ reduces the spin
limit and the efficiency, an effect which increases with the value of $%
\theta _{max}$ (largest for $\theta _{max}=\pi /6$). Fig. \ref{a_M_mc_BZ}
shows the evolution of the black hole parameters. The trends in the mass
evolution are the same as for the magnetic coupling case but the differences
in the steepness of the mass evolution curves are moderated by the BZ
mechanism. 
\begin{table}
\caption{The spin limit $a_{\ast }$ and the efficiency $\protect\epsilon $
for the accretion process in black hole magnetosphere with combined open and
closed magnetic field topology, for various boundary angles $\protect\theta %
_{max}$ and magnetic field strength power law exponents $n$.}
\label{theta_max_nb}
\begin{center}
\begin{tabular}{lccccc}
\hline
$\theta _{max}$ & $n$ & $a_{\ast }$ (I) & $\epsilon $ (I) & $a_{\ast }$ (ES)
& $\epsilon $ (ES) \\ \hline
$\pi /6$ & 1 & 0.8911 & 0.290 & 0.8923 & 0.294 \\ 
& 2 & 0.8856 & 0.284 & 0.8867 & 0.287 \\ 
& 3 & 0.8777 & 0.274 & 0.8787 & 0.278 \\ \hline
$\pi /4$ & 1 & 0.8928 & 0.293 & 0.8939 & 0.296 \\ 
& 2 & 0.8887 & 0.288 & 0.8899 & 0.292 \\ 
& 3 & 0.8832 & 0.282 & 0.8842 & 0.285 \\ \hline
$\pi /3$ & 1 & 0.8927 & 0.296 & 0.8937 & 0.298 \\ 
& 2 & 0.8911 & 0.293 & 0.8920 & 0.296 \\ 
& 3 & 0.8887 & 0.291 & 0.8897 & 0.293 \\ 
&  &  &  &  & 
\end{tabular}%
\end{center}
\end{table}

\subsection{Symbiotic model with magnetosphere}

In this subsection we combine the effect of the open and closed magnetic
field topology with the radiative truncation of the disk emerging from the
symbiotic model. 
\begin{figure*}
\centering
\includegraphics[width=.48\textwidth]{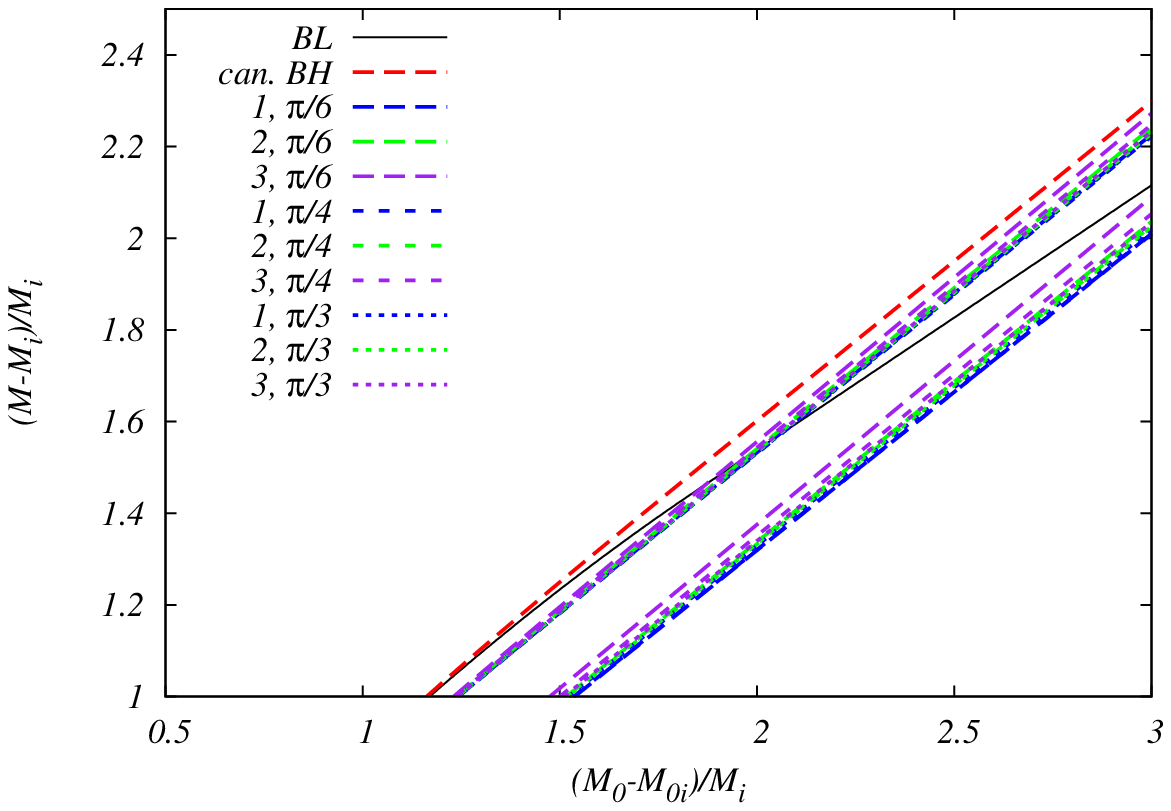} %
\includegraphics[width=.48\textwidth]{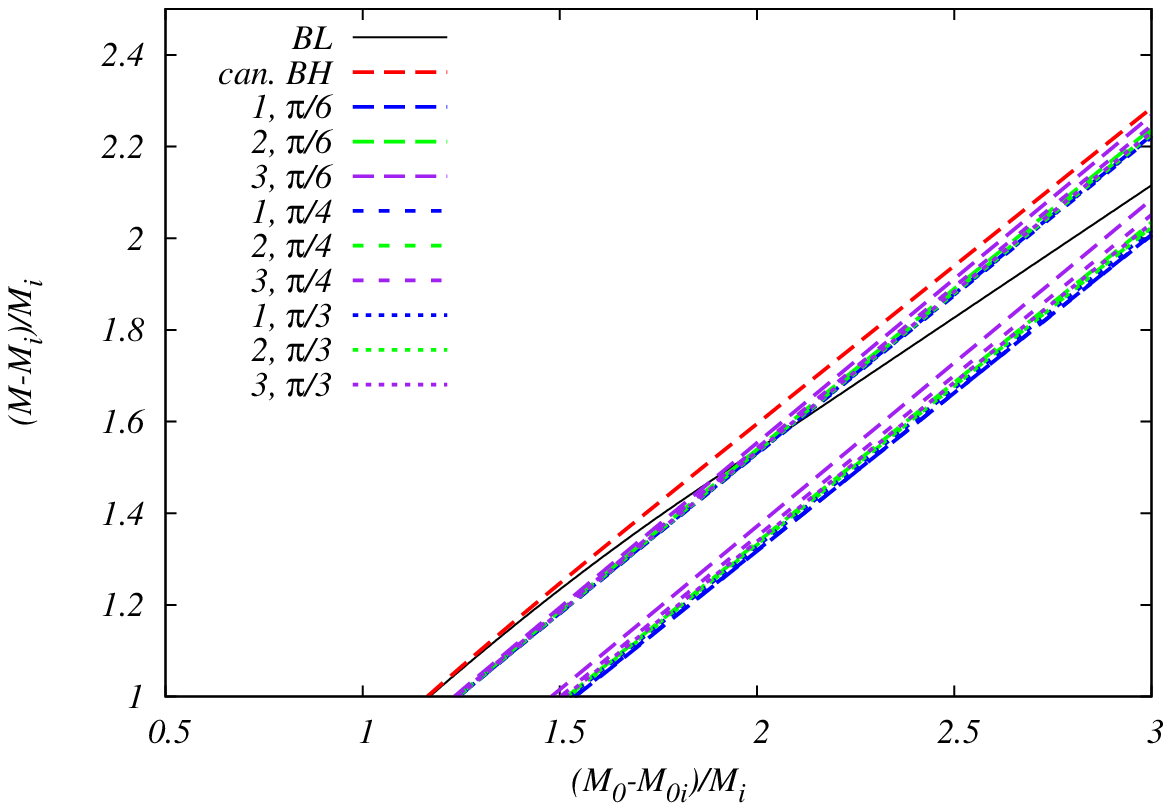}\newline
\includegraphics[width=.48\textwidth]{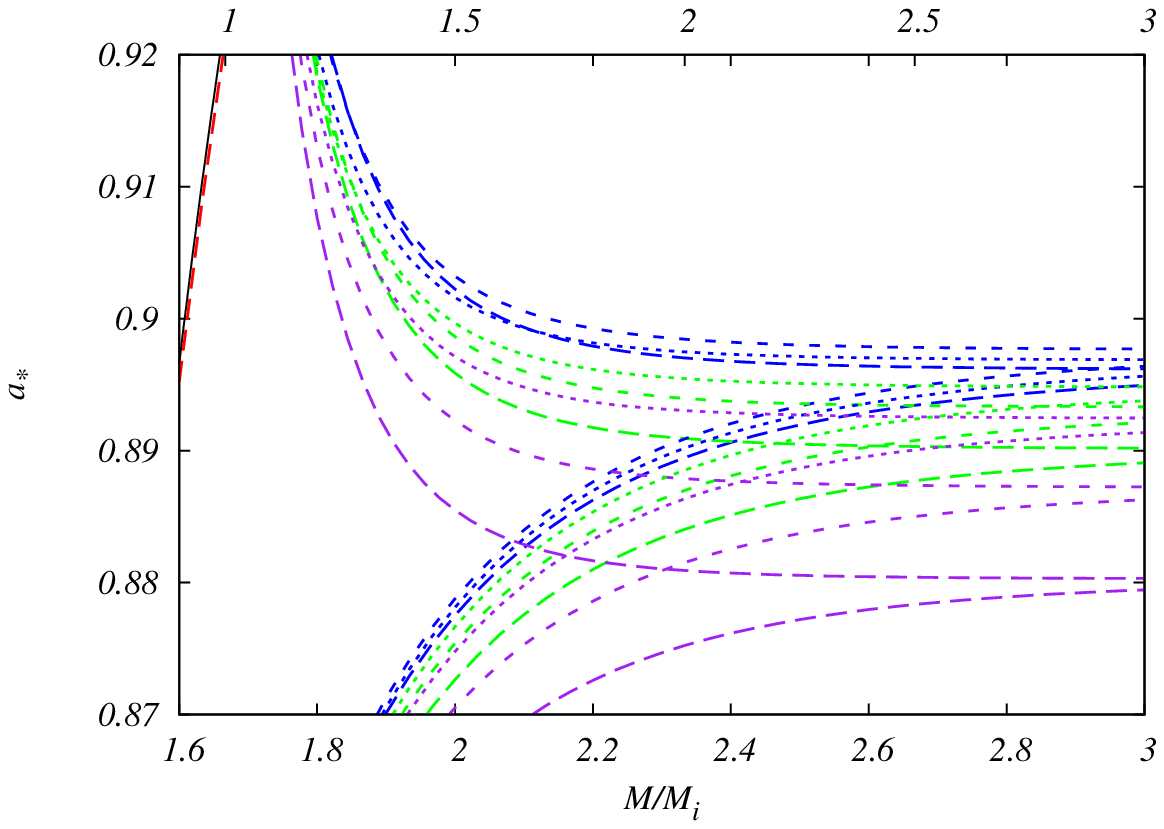} %
\includegraphics[width=.48\textwidth]{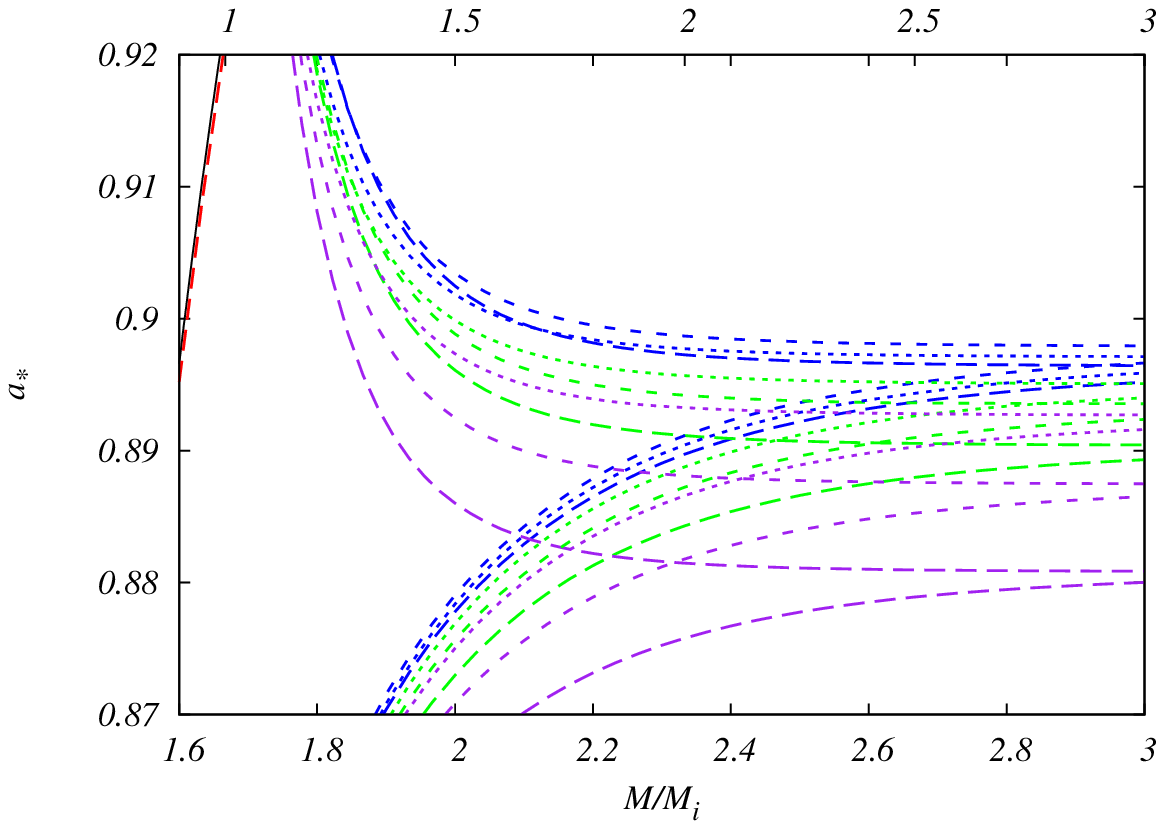}
\caption{The evolution of the the total mass-energy $M$ (upper plots) and
the spin parameter $a_{\ast }$ (lower plots) for the symbiotic model with
black hole magnetosphere and $r_{in}=4M$. For the curves running above the
black curve indicating Bardeens's law (BL) the initial value of $a_{\ast }$
is zero. With the initial value of 1 we obtained the curves running below
the black curve. Left panel has $M$ and $a_{\ast }$ for an isotropic disk
emission; right panel for the electron scattering atmosphere. }
\label{a_M_M0_r_in_4_mc_BZ}
\end{figure*}
\begin{table}
\caption{The spin limit $a_{\ast }$ and the efficiency $\protect\epsilon $
of the accretion process in symbiotic system with black hole magnetosphere
for different boundary angles $\protect\theta _{max}$ and magnetic field
strength power law exponents $n$. The first set of nine lines shows $a_{\ast
}$ and $\protect\epsilon $ when the radius $r_{in}$ is set to $4M$; the
second set of nine lines displays these quantities when $r_{in}\in (8M,12M)$%
. }
\label{theta_max_n_all}
\begin{center}
\begin{tabular}{lccccc}
\hline
$\theta_{max}$ & $n$ & $a_{*}$ (I) & $\epsilon$ (I) & $a_{*}$ (ES) & $%
\epsilon$ (ES) \\ \hline
$\pi/6$ & 1 & 0.8961 & 0.309 & 0.8964 & 0.309 \\ 
& 2 & 0.8901 & 0.301 & 0.8904 & 0.302 \\ 
& 3 & 0.8803 & 0.285 & 0.8808 & 0.267 \\ \hline
$\pi/4$ & 1 & 0.8976 & 0.312 & 0.8979 & 0.312 \\ 
& 2 & 0.8933 & 0.306 & 0.8935 & 0.306 \\ 
& 3 & 0.8872 & 0.298 & 0.8874 & 0.298 \\ \hline
$\pi/3$ & 1 & 0.8969 & 0.310 & 0.8971 & 0.311 \\ 
& 2 & 0.8948 & 0.308 & 0.8950 & 0.309 \\ 
& 3 & 0.8924 & 0.305 & 0.8927 & 0.306 \\ \hline\hline
$\pi/6$ & 1 & 0.8968 & 0.311 & 0.8969 & 0.311 \\ 
& 2 & 0.8901 & 0.302 & 0.8909 & 0.303 \\ 
& 3 & 0.8822 & 0.291 & 0.8825 & 0.292 \\ \hline
$\pi/4$ & 1 & 0.8983 & 0.313 & 0.8984 & 0.314 \\ 
& 2 & 0.8939 & 0.308 & 0.8939 & 0.308 \\ 
& 3 & 0.8879 & 0.300 & 0.8879 & 0.300 \\ \hline
$\pi/3$ & 1 & 0.8975 & 0.312 & 0.8976 & 0.312 \\ 
& 2 & 0.8955 & 0.311 & 0.8955 & 0.311 \\ 
& 3 & 0.8931 & 0.307 & 0.8931 & 0.307 \\ 
&  &  &  &  & 
\end{tabular}%
\end{center}
\end{table}

The truncation of the radiation from the disk (determined by $r_{in}$)
reduces the effect of the photon capture by the black hole and brings the
accretion process closer to Bardeen's evolution law. However, the BZ
mechanism and the extraction of mass-energy and rotational energy of the
hole via the closed field lines are more important effects. In particular,
the magnetic torque of the closed magnetic field lines typically reduce the
spin limit to values at about 0.89, as is seen on Fig. \ref{a_M_M0_r_in_4_mc_BZ}.
Table \ref{theta_max_n_all} shows the new numbers, and the difference with
respect to the numbers of Table \ref{theta_max_nb} is but a slight increase
due to the radiative truncation.

For a powerful jet such a low value of the spin is forbidden by arguments on
the pion decay. As the severe decrease in the final spin is due to the
torque of the closed magnetic field lines coupling to the very inner region
of the disk, one could speculate why the contribution of the closed magnetic
field lines is disfavoured by observations. One possibility is that the
inner region of the disk (hence the associated closed magnetic field torque)
is missing. This could arise for example if the central Kerr black hole is
replaced by a binary black hole merger (Liu and Shapiro 2010), such that
tidal torques acting on the gaseous accretion disk around the binary create
a gap in the disk. The closed magnetic field lines could also decouple from
the disk, slipping along the accreting plasma and exerting a much smaller
torque on the disk, as discussed in McKinney and Narayan (2007). Or
alternatively the inner disk can decouple from the outer disk, being spun up
by the black hole and then swallowed whole, following which a new inner disk
is built up (similarly to the Biermann \& Hall (1973) model, where a
temporary storage of angular momentum in a disk ring was employed). The most
radical option would be to forbid the closed magnetic field lines connecting
the horizon to the disk, all magnetic field lines (emanating from both the
horizon and the disk) being open.

We present the mass accretion and the spin limit for the symbiotic model
with only open magnetic field line contribution on Fig. \ref%
{a_M_M0_r_jet_mc_BZ} and Table \ref{r_jet_theta_max}. 
\begin{figure*}
\centering
\includegraphics[width=.48\textwidth]{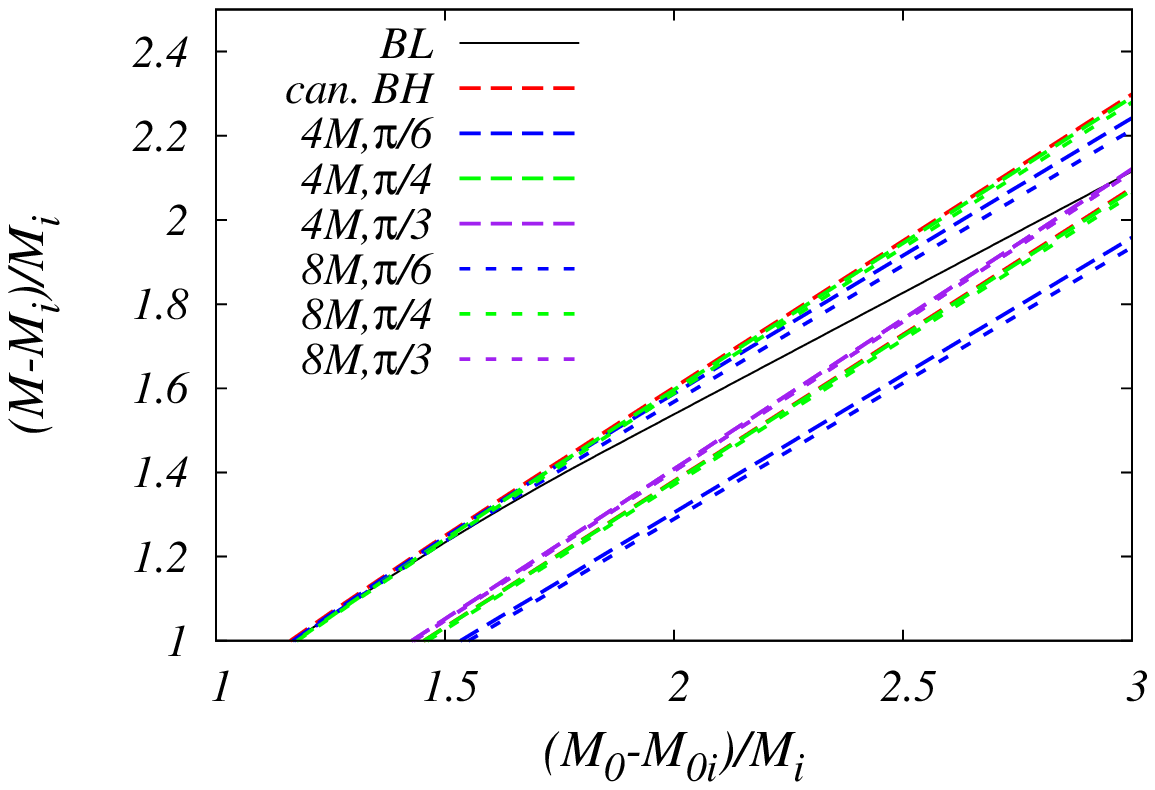} %
\includegraphics[width=.48\textwidth]{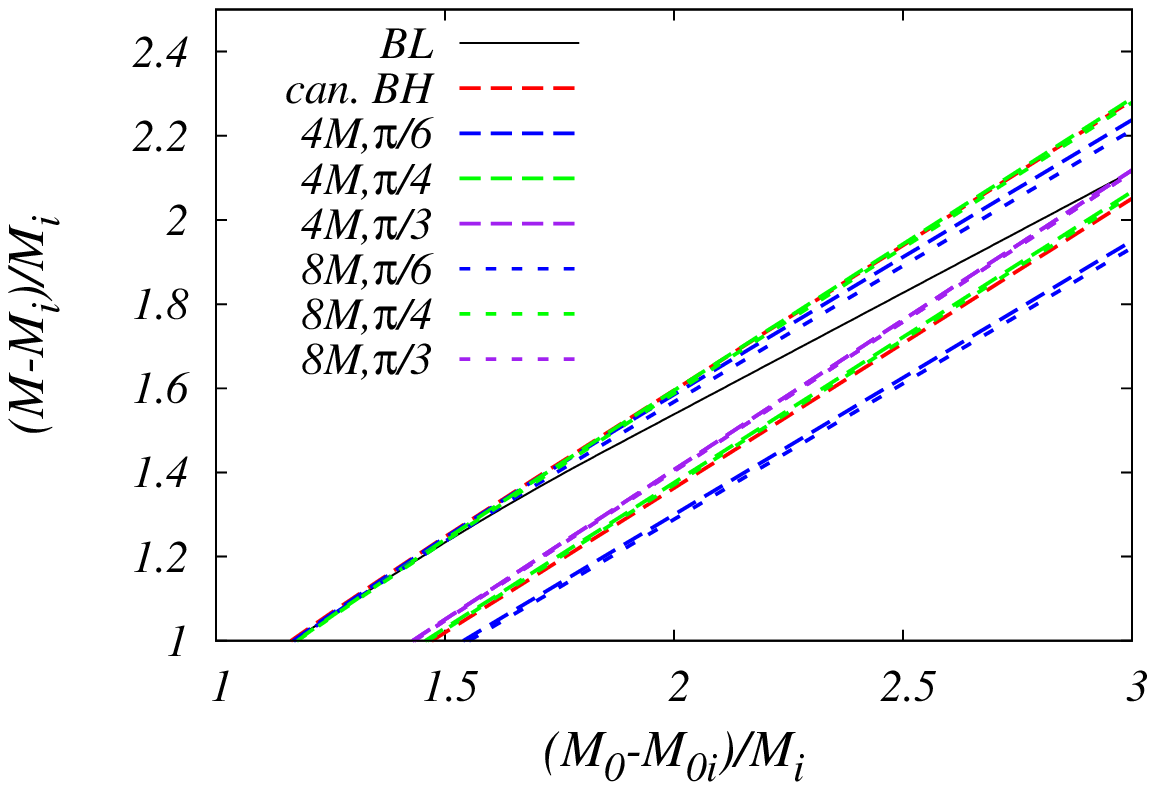}\newline
\includegraphics[width=.48\textwidth]{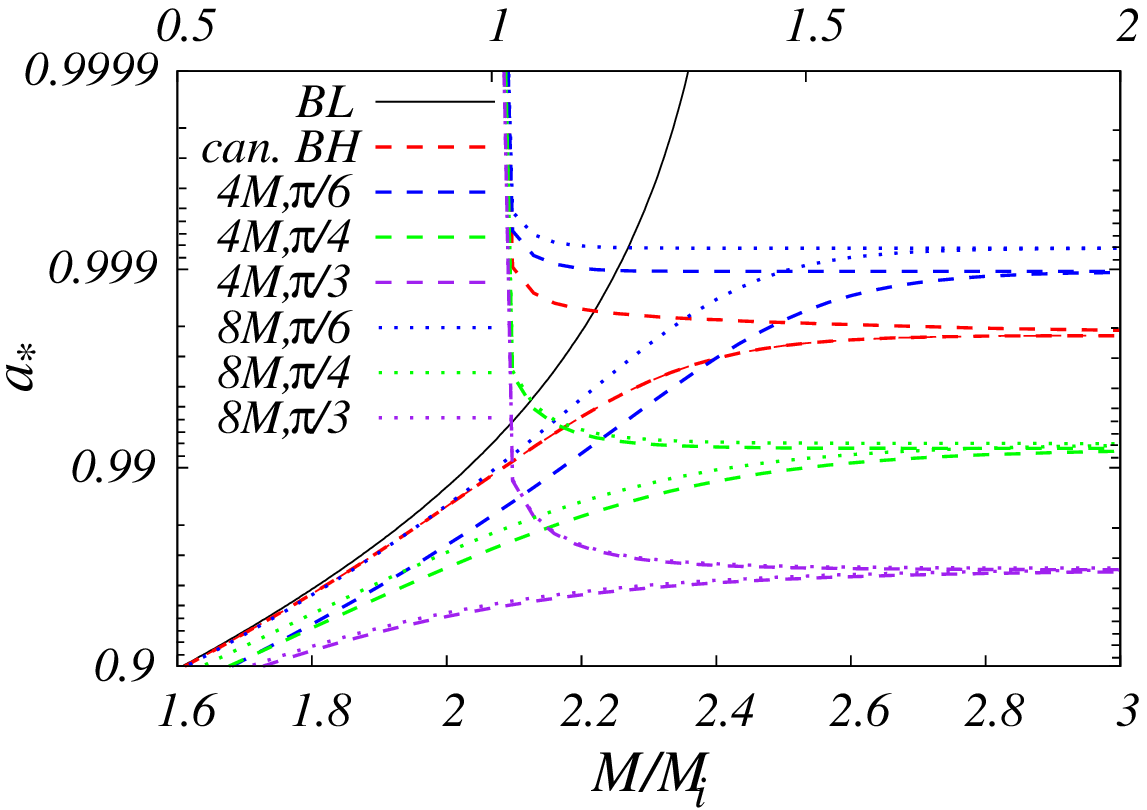} %
\includegraphics[width=.48\textwidth]{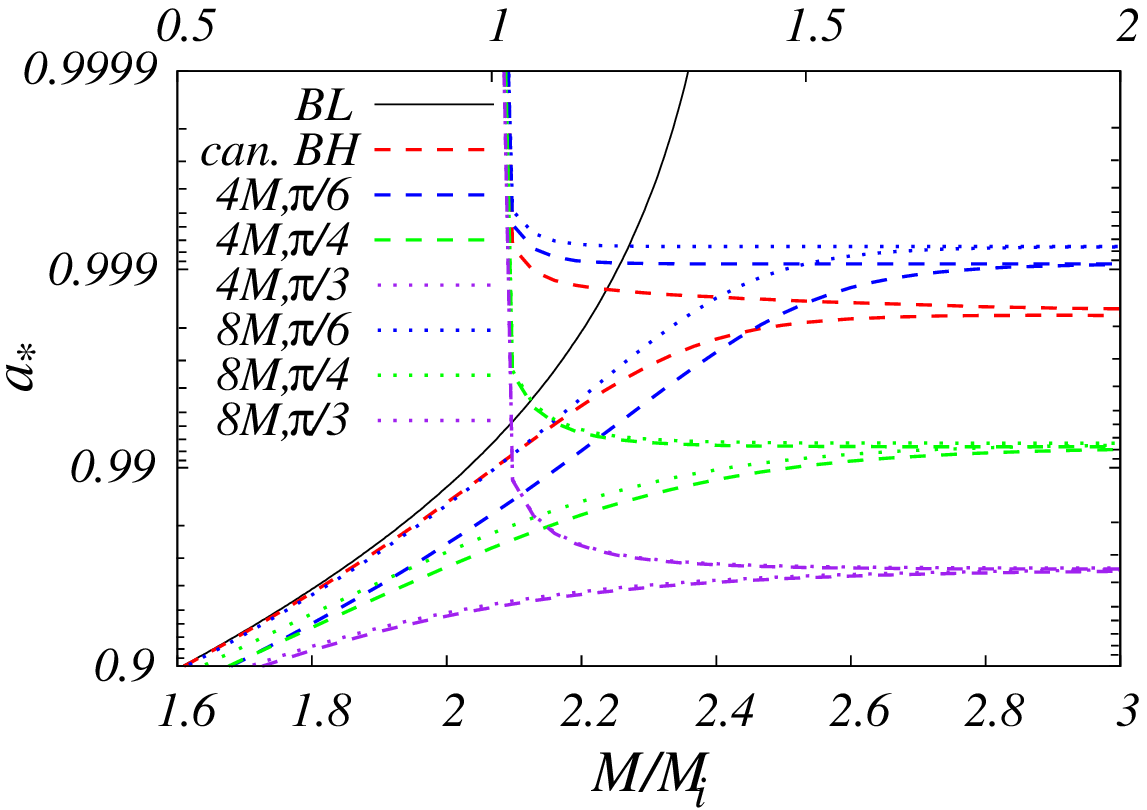}
\caption{The evolution of the the total mass-energy $M$ (upper plots) and
the spin parameter $a_{\ast }$ (lower plots) for the symbiotic model with
open magnetosphere topology and radiative truncation radius $r_{in}=4M$ and $%
8M$. For the curves running above the black curve representing Bardeens's
law (BL) the initial value of $a_{\ast }$ is zero. With the initial value of
1 we obtain the curves running below the black curve. Left panel has $M$ and 
$a_{\ast }$ for an isotropic disk emission; right panel for the electron
scattering atmosphere.}
\label{a_M_M0_r_jet_mc_BZ}
\end{figure*}
\begin{table}
\caption{The upper limit of the spin parameter and the efficiency $\protect%
\epsilon $ for the accretion process in a symbiotic system with black hole
magnetosphere, but no coupling of the closed magnetic field lines to the
disk below $r_{in}$. The notations are as for Table 2. The value of $r_{in}$%
, also representing the truncation radius of the radiation from the
accretion disk is set to $4M$, $8M$ and $12M$. The parameter $\protect\theta %
_{max}$ runs over the values $\protect\pi /6$, $\protect\pi /4$ and $\protect%
\pi /3$. }
\label{r_jet_theta_max}
\begin{center}
\begin{tabular}{lccccc}
\hline
$r_{in}[M]$ & $\theta _{max}$ & $a_{\ast }$ (I) & $\epsilon $ (I) & $a_{\ast
}$ (ES) & $\epsilon $ (ES) \\ \hline
4 & $\pi /6$ & 0.9990 & 0.347 & 0.9991 & 0.350 \\ 
& $\pi /4$ & 0.9920 & 0.307 & 0.9922 & 0.309 \\ 
& $\pi /3$ & 0.9673 & 0.287 & 0.9675 & 0.288 \\ \hline
8 & $\pi /6$ & 0.9992 & 0.354 & 0.9992 & 0.353 \\ 
& $\pi /4$ & 0.9925 & 0.311 & 0.9925 & 0.311 \\ 
& $\pi /3$ & 0.9679 & 0.290 & 0.9679 & 0.290 \\ \hline
12 & $\pi /6$ & 0.9992 & 0.354 & 0.9992 & 0.354 \\ 
& $\pi /4$ & 0.9925 & 0.311 & 0.9925 & 0.311 \\ 
& $\pi /3$ & 0.9680 & 0.290 & 0.9680 & 0.290 \\ 
&  &  &  &  & 
\end{tabular}%
\end{center}
\end{table}
The numbers are close to those given in Table \ref{theta_max} (referring to
the open magnetic field lines, with the whole disk radiating), but slightly
increased by the radiative truncation, as expected.

Since the effect of the closed field lines is sensitive to the parameter $n$
only in the very inner region (which was decoupled), the results are
independent of the model parameter $n$. Then only the variation of the
angular boundary $\theta _{max}$ of the open field lines, affects the
limiting values of $a_{\ast }$. These spin limits are given in Table \ref%
{r_jet_theta_max}. For $r_{in}=8M$ and $12M$ they tend to be identical,
indicating that the photon capture from the regions closer to the horizon is
more important. With increasing $\theta _{max}$ we strengthen the effects
due to the BZ process and decrease the spin limits, also indicated by the
numbers of Table \ref{theta_max}.

For wider solid angle of the open field lines the mass evolution are
steeper. Nonetheless, the mass evolution curves with increasing $\theta
_{max}$ are approaching the evolution profile of the canonical black hole
because the increase of the radiative truncation radius compensates the
effect of the BZ mechanism. For higher radii $r_{in}$ the mass evolution
curves are also steeper.

The efficiency $\epsilon $ shown in Table \ref{r_jet_theta_max} is slightly
decreasing for higher values of $\theta _{max}$ as more and more rotational
energy is extracted by the BZ process. The trends characterizing the pure
symbiotic model are also present where the increasing radiative truncation
radius causes in the value of $\epsilon $ a mild increase. Thus in symbiotic
models combined with black hole magnetosphere we can produce conversion
efficiencies higher than those obtained for the canonical black holes if the
angle $\theta _{max}$ is small enough. For greater boundary angles, such as $%
\theta _{max}=\pi /3$, the canonical black holes are still more efficient in
the mass-to-radiation conversion mechanism. The truncation effect here is
not so important, though $\epsilon $ can be increased somewhat by applying
greater values for $r_{in}$.

\section{CONCLUSION}

In this paper we explored a black hole - accretion disk - magnetosphere -
jet symbiotic model, consistent with the observations on the accretion disk
and on jet. The thin, steady-state accretion disk feeds the black hole in
both angular momentum and mass. Observations of the source GRS1915+105
indicate that the innermost region of the disk does not radiate when the jet
is powerful. (The initial Lorentz factors of jet flows can be very large, of
order $\gtrsim 50$.) Hence we assume no radiation below a certain radius, as
opposed to the exterior luminous part of the disk. The surplus of energy and
angular momentum transported via accretion into the black hole can be
regarded as a source of the strong jet.

The jet is driven purely electromagnetically by the black hole rotation. We
assumed that the jet starts as a Poynting flux, carrying no baryonic matter,
which is added into the jet only far up from the black hole by intersecting
clouds, stars, stellar winds and from shear and boundary instabilities.
Thus, there is no direct connection between the disk and jet in this model,
in agreement with observations on the long lifetime of jets. Due to the
extremely long lifetime of a relativistic jet and its related inefficient
spin-down, we could neglect the backreaction of the jet onto the black hole.

The magnetosphere consists of an open magnetic field region with field lines
emanating from the polar regions of the black hole, exhibiting the
Blandford-Znajek mechanism, responsible for powering the jets in AGN's and
GRB's; a closed magnetic field region with field lines emanating from the
horizon under a polar angle $\theta >\theta _{max}$ and anchored to the
disk, thus exerting a torque on it; and finally an exterior open magnetic
field line configuration, extending from $r>r_{\max }$ (the radius where the
closed field lines emanating at $\theta _{max}$ are anchored), with no
direct coupling to the black hole.

Under these assumptions we have studied the spin evolution of the black hole
in its symbiosis with the accretion disk, jet and the black hole
magnetosphere. The topology of the magnetic field has by far the most
prominent role: it drastically modifies the number of the captured photons
and in turn the spin evolution of the black hole. While for open field line
topology the limiting values of the spin parameter and the conversion
efficiency are slightly decreased as compared to the values derived for
canonical black holes, the inclusion of the closed magnetic field
configuration lowers the spin limit considerably, from $\symbol{126}0.998$
to $\symbol{126}0.89$. This happens in spite of the size of the closed
magnetic field shrinking with increasing spin. Thus the topological
properties of the magnetic field strongly correlate with the spin limit.

Therefore, assuming that observations suggest that spins larger than 0.95
are common we confirm Hirose et al. (2004) and McKinney (2005) that the
magnetic field configuration strongly connecting black hole and disk should
be uncommon.

Therefore observations on black hole spins could favour or disfavour the
existence (or the coupling to the disk) of the closed magnetic field line
region. In particular, a high black hole spin, as would be inferred from the
low energy cutoff of the electron energy distribution in the low radio
frequency spectrum of the jet indirectly forbids the existence of such a
closed magnetic field.

As for the jet contribution, we find that narrow opening angles led to
larger spin limits, than for canonical black holes. The radiative truncation
of the disk suggested by the presence of the jet further increases the spin
limit. We conclude that collimated jets slightly increase both the spins and
the efficiency.

\section*{Acknowledgments}

The authors would like to thank Alina-C\u{a}t\u{a}lina Donea, Zolt\'{a}n
Keresztes and Li-Xin Li for discussions, and for G\'{a}bor Paragi and
Tiberiu Harko for assistance in running the numerical codes in the early
phases of this work. The collaboration between the University of Szeged and
the University of Bonn was via an EU Sokrates/Erasmus contract and between
the University of Szeged and Hong Kong University via the GRF grant No.
701808P of the government of the Hong Kong SAR. L\'{A}G was supported by
COST Action MP0905 "Black Holes in a Violent Universe", the Hungarian
Scientific Research Fund (OTKA) grants nos. 69036 and 81364. Support for PLB
was coming from the AUGER membership and theory grant 05 CU 5PD 1/2 via
DESY/BMBF and VIHKOS.

\appendix

\section{Basic formulae of the photon capture}

The time averaged rates $(dM/dt)_{rad}$ and $(dJ/dt)_{rad}$ at which the
photons leaving the disk carry energy and angular momentum to the event
horizon are given by the integrals (Page \& Thorne 1974) 
\begin{eqnarray}
\left( \frac{dM}{dt}\right) _{rad} &=&-4\pi \int_{r_{ms}}^{\infty
}\int_{0}^{\pi /2}\int_{0}^{2\pi }n_{t}(r,\Theta ,\Phi )C(\Theta ,\Phi ) 
\notag \\
&&\times S(\Theta )\cos \Theta \sin \Theta rF(r){d}\Phi {d}\Theta {d}r\;,
\label{dMdtrad} \\
\left( \frac{dJ}{dt}\right) _{rad} &=&4\pi \int_{r_{ms}}^{\infty
}\int_{0}^{\pi /2}\int_{0}^{2\pi }n_{\phi }(r,\Theta ,\Phi )C(\Theta ,\Phi )
\notag \\
&&\times S(\Theta )\cos \Theta \sin \Theta rF(r){d}\Phi {d}\Theta {d}r\;.
\label{dJdtrad}
\end{eqnarray}%
In the above integrands $n_{t}$ and $n_{\phi }$ are components of the
renormalized photon momentum:%
\begin{eqnarray*}
-n_{t} &=&\mathscr C^{-1/2}(\mathscr G+x^{-1}\mathscr D^{1/2}\sin \Theta
\sin \Phi )\;, \\
n_{\phi } &=&M\mathscr C^{-1/2}(x\mathscr F+x\mathscr B\mathscr D^{1/2}\sin
\Theta \sin \Phi )\;,
\end{eqnarray*}%
$C$ is the photon capture function of the photons radiated in the direction $%
(\Theta ,\Phi )$, taking the value $1$ if the photons are captured by the
hole and $0$ if they escape to infinity. $S=1/\pi $ stands for isotropic
emission and $S=(3/7\pi )(1+2\cos \Theta )$ for electron scattering
atmosphere. The photon flux was expressed analytically (Page \& Thorne 1974)
as 
\begin{eqnarray}
F(x) &=&-\dot{M}_{0}/(4\pi r(x))\frac{3}{2M}\frac{1}{x^{2}(x^{3}-3x+2a_{\ast
})}  \notag \\
&&\times \left[ x-x_{ms}+\frac{3}{2}a_{\ast }\ln \left( \frac{x}{x_{ms}}%
\right) \right.  \notag \\
&&-\frac{3(x_{1}-a_{\ast })^{2}}{x_{1}(x_{1}-x_{2})(x_{1}-x_{3})}\ln \left( 
\frac{x-x_{1}}{x_{ms}-x_{1}}\right)  \notag \\
&&-\frac{3(x_{2}-a_{\ast })^{2}}{x_{2}(x_{2}-x_{1})(x_{2}-x_{3})}\ln \left( 
\frac{x-x_{2}}{x_{ms}-x_{2}}\right)  \notag \\
&&\left. -\frac{3(x_{3}-a_{\ast })^{2}}{x_{3}(x_{3}-x_{1})(x_{3}-x_{2})}\ln
\left( \frac{x-x_{3}}{x_{ms}-x_{3}}\right) \right] \;.  \label{Fanal}
\end{eqnarray}%
where $x_{ms}=\sqrt{r_{ms}/M}$, given by Eq. (\ref{aux2}) and $x_{1},$ $%
x_{2} $ and $x_{3}$ are the roots of the equation $x^{3}-3x+2a_{\ast }=0$.


\begin{thebibliography}{99}
\bibitem{} Abramowicz M., Jaroszynski M. \& Sikora M., \textit{Astron.
Astrophys.}, \textbf{63}, 221 (1978)

\bibitem{} Agol E. \& Krolik J. H., \textit{Astrophys. J.} \textbf{528}, 161
(2000)

\bibitem{} Afshordi N. \& Paczy\'{n}ski B. \textit{Astrophys. J.} \textbf{592%
}, 354 (2003)

\bibitem{} Araudo A. T., Bosch-Ramon V., Romero G. E., \textit{Astron. \&
Astrophys. }\textbf{522}, A97 (2010)%

\bibitem{} Armitage Ph. J. \& Natarajan P., \textit{Astrophys. J.}, \textbf{%
523}, L7 (1999)%

\bibitem{} Bardeen J. M., \textit{Nature} \textbf{226}, 64 (1970)

\bibitem{} Bardeen J. M., Press W. H., \& Teukolsky S. A., \textit{%
Astrophys. J.} \textbf{178}, 347 (1972)

\bibitem{} Berti E. \& Volonteri M., \textit{Astrophys. J.} \textbf{684},
822 (2008) 

\bibitem{} Biermann P. L. \& Hall D., \textit{Astron. \& Astrophys. }\textbf{%
27}, 249 (1973)

\bibitem{} Blandford R. D., \textit{Month. Not. Roy. Astr. Soc.} \textbf{176}%
, 465 (1976)

\bibitem{} Blandford R. D. \& Znajek R. L., \textit{Month. Not. Roy. Astr.
Soc.} \textbf{179}, 433 (1977) 

\bibitem{} Blandford R. D. \& K\"{o}nigl A., \textit{Astrophys. J.} \textbf{%
232}, 34 (1979)

\bibitem{} Blandford R. D. \& Payne D. G. \textit{Month. Not. Roy. Astr. Soc.%
} \textbf{199}, 883 (1982)

\bibitem{} Camenzind M., \textit{Astron. \& Astrophys.} \textbf{156}, 137, 
 \textit{Astron. \& Astrophys.}  \textbf{162}, 32 (1986)

\bibitem{} Camenzind M., \textit{Astron. \& Astrophys.} \textbf{184}, 341
(1987)

\bibitem{} Chini R., Kreysa E., Biermann P. L., \textit{Astron. \& Astrophys.%
} \textbf{219}, 87 (1989)

\bibitem{} Done C., Madejski G. M., \& \.{Z}ycki P. T., \textit{Astrophys. J.%
} \textbf{536}, 213 (2000)

\bibitem{} Donea A. C. \& Biermann P. L., \textit{Astron. \& Astrophys.} 
\textbf{316}, 43 (1996) 

\bibitem{} Du\c{t}an I., arXiv:1001.5434 (2010) 


\bibitem{} Eikenberry S. S., \textit{Astrophys. J. Lett.} \textbf{494}, L61
(1998) 

\bibitem{} Faber S. M., Tremaine S., Ajhar E. A., et al. \textit{Astron. J.} 
\textbf{114}, 1771 (1997) 

\bibitem{} Falcke H, in \textquotedblleft Jets from Stars and Galactic
Nuclei", Ed. W. Kundt, \textit{Springer Lecture Notes} \textbf{471}, 19
(1996); astro-ph/9512093 

\bibitem{} Falcke H. \& Biermann P. L., \textit{Astron. \& Astrophys.} 
\textbf{293}, 665 (1995) 
%

\bibitem{} Falcke H., Malkan M. A. \& Biermann P. L., \textit{Astron. \&
Astrophys.} \textbf{298}, 375 (1995) 

\bibitem{} Falcke H. \& Biermann P. L., \textit{Astron. \& Astrophys.} 
\textbf{308}, 321 (1996) 

\bibitem{} Falcke H. \& Biermann P. L., \textit{Astron. \& Astrophys.} 
\textbf{342}, 49 (1999) 
%

\bibitem{} Falcke H., Nagar N. M., Wilson A. S. \& Ulvestad J. S., \textit{%
Astrophys. J.} \textbf{542}, 197 (2000) 

\bibitem{} Fender R. P., \textit{Month. Not. Roy. Astr. Soc.} \textbf{343},
.L99 (2003) 


\bibitem{} Gammie C. F., \textit{Astrophys. J.} \textbf{522}, L57 (1999)

\bibitem{} Ghisellini G., Tavecchio F., \textit{Month. Not. Roy. Astr. Soc.} 
\textbf{386}, L28 (2008) 

\bibitem{} Gopal-Krishna, Biermann P. L., \& Wiita P. J., \textit{Astrophys.
J. Lett.} \textbf{603}, L9 (2004) 

\bibitem{} Hirose, S., et al., textit{Astrophys. J.} \textbf{606}, 1083
(2004) 

\bibitem{} Hirotani K., Takahashi M., Nitta S., \& Tomimatsu A., \textit{%
Astrophys. J.} \textbf{386}, 455 (1992)


\bibitem{} Hughes S. A. \& Blandford, R. D. , \textit{Astrophys. J.} \textbf{%
585}, L101 (2003)

\bibitem{} Janiuk, A., Czerny, B., \textit{Month. Not. Roy. Astr. Soc.} 
in press, E-print: arXiv:1102.3257 (2011)

\bibitem{} Kormendy J. \& Richstone D., \textit{Annual Rev. of Astron. \&
Astrophys.} \textbf{33}, 581 (1995) 

\bibitem{} Krolik J. H., \textit{Astrophys. J.} \textbf{515}, L73 (1999)

\bibitem{} Li L. X., \textit{Astrophys. J.}, \textbf{533}, L115 (2000)

\bibitem{} Li L. X., \textit{Phys. Rev.} \textbf{D} \textbf{68}, 024022
(2003)

\bibitem{} Liang E. T. P. \& Price R. H., \textit{Astrophys. J.} \textbf{218}%
, 247 (1977)

\bibitem{} Liu Y. T. \& Shapiro S. L., \textit{Accretion Disks Around Binary
Black Holes: A Quasistationary Model}, E-print: arXiv:1011.0002 (2010)

\bibitem{} Macdonald D. A. \& Thorne K. S., \textit{Month. Not. Roy. Astr.
Soc.} \textbf{198}, 345 (1982)

\bibitem{} McKinney, J.C., \textit{Astrophys. J. Lett.} \textbf{630}, L5
(2005) 

\bibitem{} McKinney, J.C., Narayan, R., \textit{Month. Not. Roy. Astr. Soc.} 
\textbf{375}, 513, 531 (2007)

\bibitem{} Mahadevan R., \textit{Nature} \textbf{394}, 651 (1998) 

\bibitem{} Mannheim K., Schulte M. \& Rachen J., \textit{Astron. \&
Astrophys.}, \textbf{303}, L41 (1995) 

\bibitem{} Markoff S., Falcke H., \& Fender R., \textit{Astron. \& Astrophys.%
} \textbf{372}, L25 (2001) 

\bibitem{}  Mizuno Y., Nishikawa K.I., Koide S., Hardee P., Fishman G.J., Proceedings of the VI Microquasar Workshop: Microquasars and Beyond. September 18-22, 2006, Como, Italy., p.45.1,  astro-ph/0609344

\bibitem{} Moderski R., Sikora M., \textit{Month. Not. Roy. Astr. Soc.} 
\textbf{283}, 854 (1996)


\bibitem{} Nagar N. M., Falcke H., Wilson A. S. \& Ho L. C., \textit{%
Astrophys. J.} \textbf{542}, 186 (2000) 

\bibitem{} Nagar N. M., Wilson A. S. \& Falcke H., \textit{Astrophys. J.} 
\textbf{559}, L87 (2001) 

\bibitem{} Nagar N. M., Falcke H., Wilson A. S. \& Ulvestad J. S., \textit{%
Astron. \& Astrophys.} \textbf{392}, 53 (2002) 

\bibitem{} Nagar N. M., Falcke H. \& Wilson A. S., \textit{Astron. \&
Astrophys.} \textbf{435}, 521 (2005) 

\bibitem{} Nitta S., Takahashi M. \& Tomimatsu A., \textit{Phys. Rev.} 
\textbf{D} \textbf{44}, 2295 (1991)

\bibitem{} Novikov I. D. \& Thorne K. S., 1973 in Black Holes, ed. C. DeWitt
\& B. DeWitt (New York: Gordon \& Breach) p. 343

\bibitem{} Page D. N. \& Thorne K. S., \textit{Astrophys. J.}, \textbf{191},
499 (1974)

\bibitem{} Park S. J. \& Vishniac E. T., \textit{Astrophys. J.}, \textbf{332}%
, 135 (1988)

\bibitem{} Perez-Fournon I. \& Biermann P. L. \textit{Astron. \& Astroph} 
\textbf{130}, L13 (1984)

\bibitem{} Shakura N. I., \textit{Astronomicheskii Zhurnal} \textbf{49}, 921
(1972) 

\bibitem{} Shakura N. I. \& Sunyaev R. A., \textit{\ Astron. \& Astroph} 
\textbf{24}, 33 (1973)

\bibitem{} Shapiro S. L., Lightman A. P., Eardley D. M., \textit{Astrophys.
J.} \textbf{204}, 187 (1976)

\bibitem{} Shibata M. \& Sasaki M., \textit{Phys. Rev.} \textbf{D} \textbf{58%
}, 104011 (1998)

\bibitem{} Sikora M, Begelman M. C., Madejski G. M., Lasota J.-P., \textit{%
Astrophys. J.}, \textbf{265}, 272 (2005)




\bibitem{} Thorne K. S., \textit{Astrophys. J.} \textbf{191}, 507 (1974)

\bibitem{} Thorne K. S. \& Price R. H. \textit{Astrophys. J.} \textbf{195},
L101 (1975)

\bibitem{} Thorne K. S., Price R. H. \& Macdonald D. A. 1986 Black Holes:
The Membrane Paradigm. (New Heaven: Yale Univ. Press)

\bibitem{} Toomre A., Toomre, J \textit{Astrophys. J.} \textbf{178}, 623
(1972)

\bibitem{} Uzdensky D. A. \textit{Astrophys. J.} \textbf{620}, 889 (2005)

\bibitem{} Wang D. X., Xiao K. \& Lei W. H., \textit{Month. Not. Roy. Astr.
Soc.} \textbf{335}, 655 (2002)

\bibitem{} Wang D. X., Lei W. H. \& Ma R. Y., \textit{Month. Not. Roy. Astr.
Soc.} \textbf{342}, 851 (2003)

\bibitem{} Whysong D. \& Antonucci R., \textit{New Astronomy Reviews} 
\textbf{47}, 219 (2003) 

\bibitem{} Windhorst R. A., van Heerde G. M. \& Katgert P., \textit{Astron.
Astrophys. Suppl. Ser.} \textbf{58}, 1 (1984) 


%

\bibitem{} Yuan F., Markoff S., \& Falcke H., \textit{Astron. Astrophys.} 
\textbf{383}, 854 (2002) 

\bibitem{} Yuan F., Markoff S., Falcke H., \& Biermann P. L., \textit{%
Astron. Astrophys.} \textbf{391}, 139 (2002) 

\bibitem{} Znajek R. L., \textit{Month. Not. Roy. Astr. Soc.} \textbf{179},
457 (1977)
\end{thebibliography}
\end{document}